\pdfoutput=1
\documentclass[colorlinks=true,linkcolor=refcolor,citecolor=refcolor,urlcolor=refcolor,accepted]{uai2025} 

\usepackage[american]{babel}
\usepackage{xcolor}
\usepackage{ dsfont } 
\usepackage{natbib} 

\usepackage{mathtools} 
\usepackage{booktabs} 
\usepackage{tikz} 
\usepackage{ifthen}

\usepackage{comment}

\usepackage{mathtools}
\usepackage{amsmath}
\usepackage{amssymb}
\usepackage{amsthm}
\usepackage{algorithm}
\usepackage{algorithmicx}
\usepackage{algpseudocode}
\usepackage{booktabs}       
\usepackage{amsfonts}       
\usepackage{float}

\usepackage{pgf}
\usetikzlibrary{arrows.meta}
\usetikzlibrary{calc,math}
\usepackage{ulem}

\theoremstyle{definition}
\newtheorem{theorem}{Theorem}

\usepackage{bm}

\usepackage{array}
\usetikzlibrary{shapes.geometric}
\definecolor{blueMain}{RGB}{66,133,244}

\definecolor{PurpleGMiddle}{RGB}{152,78,163}

\definecolor{GreenGMiddle}{RGB}{77,175,74}

\definecolor{OrangeGMiddle}{RGB}{255,127,0} 

\definecolor{BlueGMiddle}{RGB}{55,126,184} 

\definecolor{RedGMiddle}{RGB}{228,26,28}

\definecolor{PinkGMiddle}{RGB}{223,101,176}

\definecolor{mustardGLight}{RGB}{230,171,2} 

\definecolor{orangeSand}{RGB}{241,105,19} 

\definecolor{OrangeGLight}{RGB}{251,128,114}
\definecolor{BlueGLight}{RGB}{128,177,211}
\definecolor{BlueGDark}{RGB}{31,120,180}
\definecolor{GreenGDark}{RGB}{0,104,55} 
\definecolor{RedGDark}{RGB}{165,15,21} 
\definecolor{BrownGDark}{RGB}{166,86,40} 
\definecolor{YellowGDark}{RGB}{255,255,51} 
\definecolor{mustardGDark}{RGB}{166,118,29} %

\definecolor{PinkGDLight}{RGB}{247,129,191}
\definecolor{MyGray}{RGB}{128, 128, 128} 

\definecolor{MyGrayLight}{RGB}{217,217,217} 

\definecolor{MyGrayDark}{RGB}{90,90,90} 

\definecolor{MyCyan}{RGB}{28,144,153} 
\definecolor{NavyBlueG}{RGB}{28,28,132} 

\definecolor{colorMain1}{RGB}{117,112,179} 
\definecolor{colorMain2}{RGB}{102,166,30} 
\definecolor{colorMain3}{RGB}{217,95,2} 
\definecolor{colorMain4}{RGB}{231,41,138} 
\definecolor{colorMain5}{RGB}{27,158,119} 

\newcommand{\CausalArrow}{\kern-0.25em\mathrel{\tikz[baseline=-0.5ex] \draw[-{Triangle}, thick] (0,0) -- (1.5em,0);}\kern-0.25em}
\newcommand{\CausalArrowColor}[1]{\kern-0.25em\mathrel{\tikz[baseline=-0.5ex] \draw[-{Triangle}, thick,#1] (0,0) -- (1.5em,0);}\kern-0.25em}
\newcommand{\CausalArrowColorText}[2]{%
    \kern-0.0em\mathrel{\raisebox{-0.25ex}{
        \tikz[baseline=-0.5ex] {
            \draw[-{Triangle}, thick, #1] (-0.15em,0) -- (1.5em,0);
            \node[anchor=south east] at (1.3em,-0.2em) {\fontsize{5pt}{5pt}\selectfont \smash{\llap{#2}}}; 
        }
    }}\kern-0.0em%
}

\theoremstyle{definition}

\newcommand{\blocktheorem}[1]{%
  \csletcs{old#1}{#1}
  \csletcs{endold#1}{end#1}
  \RenewDocumentEnvironment{#1}{o}
    {\par\addvspace{1.5ex}
     \noindent\begin{minipage}{\textwidth}
     \IfNoValueTF{##1}
       {\csuse{old#1}}
       {\csuse{old#1}[##1]}}
    {\csuse{endold#1}
     \end{minipage}
     \par\addvspace{1.5ex}}
}

\blocktheorem{theo}
\blocktheorem{defi}
\blocktheorem{lem}
\blocktheorem{rem}
\blocktheorem{col}

\newcommand{\CausalArrowColorNamePath}{\CausalArrowColorText{PathNameColor}{\PathName}}
\newcommand{\CausalArrowColorNameHub}{\CausalArrowColorText{HubNameColor}{\HubName}}
\newcommand{\CausalArrowColorNameOld}{\CausalArrowColorText{OldNameColor}{\OldName}}
\newcommand{\CausalArrowColorNameMid}{\CausalArrowColorText{MidNameColor}{\MidName}}
\newcommand{\CausalArrowColorNameFar}{\CausalArrowColorText{FarNameColor}{\FarName}}

\newcommand{\CausalArrowColorNameSelf}{\CausalArrowColorText{SelfNameColor}{\SelfName}}

\colorlet{colorLinks}{NavyBlueG}
\colorlet{refcolor}{NavyBlueG}

\colorlet{TitleColor}{PinkGDLight}

\usepackage{enumitem}
\setlist[itemize]{label=\textcolor{colorLinks}{\textbullet}}
\setlist[enumerate]{label=\textcolor{colorLinks}{\arabic*}}

\colorlet{TransitiveColor}{GreenGMiddle} 
\newcommand{\colorTransitive}[1]{\textcolor{TransitiveColor}{#1}}
\newcommand{\transitive}{\textcolor{TransitiveColor}{\textsc{\mbox{Path}}}}

\colorlet{HubColor}{BlueGMiddle} 
\newcommand{\colorHub}[1]{\textcolor{HubColor}{#1}}
\newcommand{\hub}{\textcolor{HubColor}{\textsc{\mbox{Hub}}}}

\colorlet{ForwardColor}{PinkGMiddle}
\newcommand{\colorForward}[1]{\textcolor{ForwardColor}{#1}}
\newcommand{\forward}{\textcolor{ForwardColor}{\textsc{\mbox{Old}}}}

\colorlet{BackwardColor}{PurpleGMiddle}
\newcommand{\colorBackward}[1]{\textcolor{BackwardColor}{#1}}
\newcommand{\backward}{\textcolor{BackwardColor}{\textsc{\mbox{New}}}}

\colorlet{BidirectedColor}{PinkGMiddle}

\colorlet{UndirectedColor}{PurpleGMiddle} %

\colorlet{NonLocalInteriorColor}{RedGMiddle}
\newcommand{\colorNonLocalInterior}[1]{\textcolor{NonLocalInteriorColor}{#1}}
\newcommand{\nonLocalInterior}{\textcolor{NonLocalInteriorColor}{\textsc{\mbox{Near}}}}

\colorlet{NonLocalExteriorColor}{mustardGLight}
\newcommand{\colorNonLocalExterior}[1]{\textcolor{NonLocalExteriorColor}{#1}}
\newcommand{\nonLocalExterior}{\textcolor{NonLocalExteriorColor}{\textsc{\mbox{Far}}}}

\colorlet{NonLocalInterfaceColor}{orangeSand}
\newcommand{\colorNonLocalInterface}[1]{\textcolor{NonLocalInterfaceColor}{#1}}
\newcommand{\nonLocalInterface}{\textcolor{NonLocalInterfaceColor}{\textsc{\mbox{Mid}}}}

\colorlet{SelfArrowColor}{refcolor}
\newcommand{\colorSelfArrow}[1]{\textcolor{SelfArrowColor}{#1}}
\newcommand{\selfArrow}{\textcolor{SelfArrowColor}{\textsc{\mbox{Self}}}}

\def\HubName{\colorHub{\hub}}
\def\PathName{\colorTransitive{\transitive}}
\def\OldName{\colorForward{\forward}}
\def\NewName{\colorBackward{\backward}}
\def\FarName{\colorNonLocalExterior{\nonLocalExterior}}
\def\MidName{\colorNonLocalInterface{\nonLocalInterface}}
\def\NearName{\colorNonLocalInterior{\nonLocalInterior}}

\colorlet{HubNameColor}{HubColor}
\colorlet{PathNameColor}{TransitiveColor}
\colorlet{OldNameColor}{ForwardColor}
\colorlet{NewNameColor}{BackwardColor}
\colorlet{FarNameColor}{NonLocalExteriorColor}
\colorlet{MidNameColor}{NonLocalInterfaceColor}
\colorlet{NearNameColor}{NonLocalInteriorColor}
\colorlet{SelfNameColor}{SelfArrowColor}

\def\SelfName{\colorSelfArrow{\selfArrow}}

\colorlet{rjColor}{BackwardColor} 
\colorlet{djColor}{ForwardColor} 
\colorlet{irColor}{HubColor} 
\colorlet{diColor}{TransitiveColor} 
\colorlet{ddColor}{NonLocalExteriorColor} 
\colorlet{drColor}{NonLocalInterfaceColor} 
\colorlet{rrColor}{NonLocalInteriorColor}

\newcommand{\avgdegree}{\ensuremath{\big\langle d{(n)} \big\rangle}}

\newcommand{\emphWord}[1]{\textit{\textbf{#1}}}

\newcommand{\din}{d^{\text{in}}}
\newcommand{\dout}{d^{\text{out}}}

\newcommand{\thetain}{\colorHub{\theta_{\text{in}}}}
\newcommand{\thetaout}{\colorTransitive{\theta_{\text{out}}}}
\newcommand{\thetaold}{\colorForward{\theta_{\text{old}}}}

\newcommand{\alphaP}{\protect\ensuremath{\colorSelfArrow{\alpha}}}
\newcommand{\betaP}{\protect\ensuremath{\colorSelfArrow{\beta}}}

\newcommand{\bigO}[1]{\mathcal{O}\big( #1 \big)}
\newcommand{\biggerO}[1]{\mathcal{O}\Big( #1 \Big)}
\newcommand{\littleO}[1]{o\big( #1 \big)}
\newcommand{\bigTh}[1]{\Theta\big( #1 \big)}

\newcommand{\polyColor}{black}
\newcommand{\constColor}{black}
\newcommand{\logColor}{black}

\newcommand*\ONode[1]{%
  \tikz \node [draw, circle, inner sep=0pt, minimum size=9pt, yshift=0pt] {\raisebox{-2pt}[4pt][0pt]{\smash{$#1$}}};%
}

\newcommand*\NoONode[1]{%
  \tikz \node [circle, inner sep=0pt, minimum size=9pt, yshift=0pt] {\raisebox{-2pt}[4pt][0pt]{\smash{$#1$}}};%
}

\newcommand*\OSNode[1]{%
  \tikz \node [draw, circle, inner sep=0pt, minimum size=6pt, yshift=0pt] {\raisebox{-2pt}[3pt][0pt]{\smash{$#1$}}};%
}
\newcommand*\OSSNode[1]{%
  \tikz \node [draw, circle, inner sep=0pt, minimum size=4pt, yshift=0pt] {\raisebox{-2pt}[3pt][0pt]{\smash{$#1$}}};%
}
\newcommand*\NoOSNode[1]{%
  \tikz \node [circle, inner sep=0pt, minimum size=6pt, yshift=0pt] {\raisebox{-2pt}[3pt][0pt]{\smash{$#1$}}};%
}

\newcommand*\ONodeOne[1]{%
  \tikz \node [draw=black, circle, fill=white, inner sep=0pt, minimum size=9pt, yshift=0pt] {\raisebox{-2pt}[4pt][0pt]{\smash{$#1$}}};%
}
\newcommand*\OSNodeOne[1]{%
  \tikz \node [draw=black, circle, fill=white, inner sep=0pt, minimum size=6pt, yshift=0pt] {\raisebox{-2pt}[3pt][0pt]{\smash{$#1$}}};%
}

\newcommand*\ONodeTwo[1]{%
  \tikz \node [draw=black, circle, fill=white,  inner sep=0pt, minimum size=9pt, yshift=0pt] {\raisebox{-2pt}[4pt][0pt]{\smash{$#1$}}};%
}
\newcommand*\OSNodeTwo[1]{%
  \tikz \node [draw=black, circle, fill=white, inner sep=0pt, minimum size=6pt, yshift=0pt] {\raisebox{-2pt}[3pt][0pt]{\smash{$#1$}}};%
}

\newcommand*\BoxNode[1]{%
  \kern1pt\ONode{#1}\kern1pt
}
\newcommand*\BoxNodeS[1]{%
  \kern1pt\OSNode{#1}\kern1pt
}
\newcommand*\BoxNodeSS[1]{%
  \kern1pt\OSSNode{#1}\kern1pt
}

\newcommand*\BoxNodeOne[1]{%
  \kern1pt\ONodeOne{#1}\kern1pt
}
\newcommand*\BoxNodeOneS[1]{%
  \kern1pt\OSNodeOne{#1}\kern1pt
}

\newcommand*\BoxNodeTwo[1]{%
  \kern1pt\ONodeTwo{#1}\kern1pt
}
\newcommand*\BoxNodeTwoS[1]{%
  \kern1pt\OSNodeTwo{#1}\kern1pt
}

\newcommand*\UnBoxNode[1]{%
  \kern1pt\NoONode{#1}\kern1pt
}
\newcommand*\UnBoxNodeS[1]{%
  \kern1pt\NoOSNode{#1}\kern1pt
}
\newcommand{\DyadPairCircleNodes}{\dyadpair{\BoxNodeS{}}{\BoxNodeS{}}}

\newcommand{\eqspace}[1][]{\kern1pt #1 \kern1pt}
\newcommand{\Zero}{\kern1pt\circ\kern0.5pt}
\newcommand{\Star}{\kern1.4pt\star\kern1.4pt}

\usepackage[noabbrev]{cleveref}

\newcommand{\spacebefsection}{\vspace{-1pt}}
\newcommand{\spacebefsubsection}{\vspace{-2pt}}
\newcommand{\spacebefsubsubsection}{\vspace{-3pt}}

\newcommand{\spaceendsection}{\vspace{-1pt}}
\newcommand{\spaceendsubsection}{\vspace{-1pt}}
\newcommand{\spaceendsubsubsection}{\vspace{-1pt}}

\newcommand{\spaceafterfivedyadsfig}{\vspace{100pt}}

\newcommand{\ie}{\protect{i.e.}}
\newcommand{\eg}{\protect{e.g.}}

\newcommand{\textNodeRelationNode}[1]{\protect{\text{ #1}}}
\newcommand{\textNodeRelationNodeType}[1]{\protect{\text{``#1''}}}

\newcommand{\thmlabel}[1]{\textit{\textbf{#1}}}

\newcommand{\dyadpair}[2]{\ensuremath{(#1#2)}}

\newcommand{\refTypeCausalArrows}{Table~\ref{Def:TypesOfCausalArrows}}

\newcommand{\ppaName}{Distributed Affine Preferential Attachment} 
\newcommand{\ppaAcron}{DAPA}

\newcommand{\thmNamePPA}{\thmlabel{Phase transitions in the average degree of the \protect{\ppaAcron} model}}
\newcommand{\thmNamePPAAppendix}{\textit{(P\MakeLowercase{hase transitions in the asymptotic growth rate of the average degree of the} \protect{\ppaAcron} \MakeLowercase{model})}}

\newcommand{\thmNamePPAdegdist}{\thmlabel{Power-law degree distributions of the \protect{\ppaAcron} model}}
\newcommand{\thmNamePPAAppendixdegdist}{\textit{(P\MakeLowercase{ower-law degree distributions of the} \protect{\ppaAcron} \MakeLowercase{model})}}

\newcommand{\thmNameNodeDel}{\thmlabel{\mbox{Deletion-invariant} causal \mbox{meta-DAGs}}}

\newcommand{\thmNameHasse}{\thmlabel{Deletion and marginalization-invariant causal meta-DAGs}}

\newcommand{\spacestartappendixsection}{\vspace{0pt}}

\newcommand{\potentiallyImprove}[1]{\textcolor{black}{#1}} 

\newcommand{\potentiallyImproveN}[1]{\textcolor{black}{#1}} 

\newcommand{\potentiallyRemove}[1]{\textcolor{black}{#1}} 

\newcommand{\novo}[1]{\textcolor{black}{#1}}

\newcommand{\paperTitle}{\huge{Causal Models for Growing Networks}}
\newcommand{\paperTitleAppendix}{\huge{APPENDIX --- \protect{\paperTitle}}}
\title{
\paperTitle}
\author[1,a]{\href{mailto:<gecia.bravo@gmail.com>?Subject=CausalModelsForGrowingNetworks2025}{Gecia \mbox{Bravo-Hermsdorff}}{}}
\author[2,b]{Lee M.~Gunderson}
\author[2,c]{Kayvan Sadeghi}
\affil[1]{%
    School of Informatics\\
    University of Edinburgh\\
    Edinburgh, Scotland, UK  
}
\affil[2]{%
    Department of Statistical Science\\
    University College London\\
    London, England, UK 
}
\affil[a,b,c]{%
    \{\href{mailto:<gecia.bravo@gmail.com>?Subject=CausalModelsForGrowingNetworks2025}{\text{gecia.bravo@gmail.com}},
    \hspace{1pt}  
    \href{mailto:<l.gunderson@ucl.ac.uk>?Subject=CausalModelsForGrowingNetworks2025}{\text{l.gunderson@ucl.ac.uk}}, \hspace{1pt}
    \href{mailto:<k.sadeghi@ucl.ac.uk>?Subject=CausalModelsForGrowingNetworks2025}{\text{k.sadeghi@ucl.ac.uk}}\}
}
  
\begin{document}
\maketitle

\begin{abstract}
Real-world networks grow over time; statistical models based on node exchangeability are not appropriate.  
Instead of constraining the structure of the \textit{distribution} of edges, 
we propose that the relevant symmetries refer to the \mbox{\textit{causal structure}} between them.  
We first enumerate the 96 causal directed acyclic graph (DAG) models over pairs of nodes (dyad variables) in a growing network with finite ancestral sets that are invariant to node deletion. 
We then partition them into 21 classes with ancestral sets that are closed under node marginalization. 
Several of these classes are remarkably amenable to distributed and asynchronous evaluation.  
As an example, we highlight a simple model that exhibits flexible power-law degree distributions and emergent phase transitions in sparsity, which we characterize analytically.  
With few parameters and much conditional independence, our proposed framework provides natural baseline models for causal inference in relational data. 
\end{abstract}

\input{002_0TikZNature}

\section{Causality and Networks}
\label{sec:motivation}
\spacebefsection
The importance of causal and relational reasoning cannot be understated.  
Causality is fundamentally about explaining how interventions influence observations and extrapolating from such explanations  \citep{deutsch2011beginning}. 
Individuals can act on their environment, learning from the result of their interventions.  
Multiple individuals in the same environment naturally lead to networks of interactions.  
Individuals themselves are complex systems, composed of many interacting parts.  
Collective changes of these interactions allow the individual to learn, which in turn shapes these webs of interactions \citep{chazelle2019iterated, chazelle2012dynamics, bravo2023quantifyinghuman}. 
From the microcosm of an individual to the macrocosm of many,\footnote{Such as the ants and their Aunt Hillary \citep{hofstadter1999godel}.}  
learning is an intrinsically compositional, temporal, and collective process \citep{coecke2023compositionality,Rovelli2021-ROVHMS}.

A wide range of systems can be described as growing temporal networks: 
trades \citep{adamic2017trading}, 
financial transactions \citep{arnold2024insights}, 
citations \citep{radicchi2011citation}, 
media interactions \citep{goglia2024structure}, 
etc. 
A guiding principle for useful null models is to exploit the (often approximate) symmetries of the system \citep{villar2023towards}.  
For network data, 
the classical notion of exchangeability of graph distributions implies invariance to relabeling of the nodes \citep{orbanz2017subsampling}. 

However, node-exchangeable models of networks, 
such as graphons \citep{lovasz2012large,gunderson2024graph} and Exponential Random Graph Models (ERGMs) \citep{harris2013introduction,lauritzen2018random}, 
frequently have difficulty describing real-world networks. 
For example, 
they tend to struggle to describe sparse networks, 
essentially treating them all as equivalent to the network without edges \citep{orbanz2014bayesian}. 
While various modifications have been suggested to cope with these issues, 
many of the hallmarks of real-world networks do not sit comfortably in this framework.

Perhaps the reason (at least partially) is that real-world networks do not typically pop into existence fully-developed.  
For growing networks, the order in which the nodes arrive is arguably their most basic ``feature''.  
With this perspective, 
instead of invariance of the \textit{distribution} of edges 
(with respect to node permutation and subsampling), 
we ask for invariance of the causal mechanisms \textit{generating} the edges 
(with respect to node deletion and marginalization).  

By systematically enumerating causal models with these properties, 
we find statistically-streamlined models for growing networks that exhibit 
emergent features characteristic of real-world networks, 
and offer a baseline framework for causal inference in relational data.

We highlight one model in particular, which we call {\ppaName} ({\ppaAcron}),      
and analytically characterize its asymptotic degree distribution, 
which exhibits a flexible power-law and a striking phase transition in the growth rate of the average degree. 
The choice of name is due to its surprisingly parallelizable causal structure.  
Indeed, most naturally-occurring growing networks are distributed and asynchronus systems, 
so it stands to reason that the causal models describing their generation might be more amenable to distributed and asynchronous computation.

\paragraph{Outline (see Figure~\ref{fig:FirstNatureFigure}).}
%
In Section~\ref{sec:framework}, we present our framework in detail. 
In Section~\ref{sec:parttwoppa}, we showcase a simple model with surprisingly rich behavior. 
In Section~\ref{sec:extensions}, we discuss applications of our framework to problems of generalization and causal inference. 
In Section~\ref{sec:discussion}, we discuss some promising sequels and conclude. 
%

\spaceendsection
\section{Our framework}
\label{sec:framework}
\spacebefsection
We are concerned with describing causal models for growing networks. 
Both of these structures can be interpreted as graphs,
so to avoid confusion, we refer to the data being generated as the \emphWord{growing network} 
and the causal structure describing its generation as the \emphWord{causal DAG} or \emphWord{meta-DAG}. 
%
For instance,  
in Figs.~\ref{Fig:SchemeWithLocalInfluences} and \ref{Fig:SchemeWithNonInfluences} 
the white nodes and black edges represent the growing network and the 
(\colorHub{c}\colorTransitive{o}\colorForward{l}\colorBackward{o}\colorNonLocalExterior{r}\colorNonLocalInterface{f}\colorNonLocalInterior{u}l)
directed arrows between these black edges represent the meta-DAG. 
\spaceendsection
\subsection{The Growing Network}
\label{sec:GrowingNetworkDescription}
\spacebefsubsection
The \emphWord{nodes} in the growing network are indexed by the natural numbers $\mathbb{N}$ with the standard $\leq$ ordering.   
Intuitively, we can think of the nodes as ``arriving'' in that order, 
and 
then 
deciding with other nodes to connect to.
We refer to specific node indices with lower-case letters 
(sometimes $i$ and $j$, and sometimes $a$, $b$, $c$, $d$), 
with ordering implied lexicographically.  
Unspecified nodes are represented as open circles $\BoxNodeS{}$, 
with ordering implied by position (such as in \refTypeCausalArrows).  

The random variables in our model are indexed 
pairs of distinct nodes \dyadpair{i}{j},  
which we refer to as \emphWord{dyad variables}, or simply \emphWord{dyads}.  
That is, we take the node ordering of the growing network as given and model distributions over their dyadic connections. 
While the set of nodes is infinite, 
one can equivalently think of the growing network as a random process defining an infinite 
family of probability distributions 
over the $n \choose 2$ dyad variables between the first $n$ nodes,   
for each \mbox{$n \in \mathbb{N}$}.
The dyads variables could in principle take values from any set of outcomes,  
but in  our example model in Section~\ref{sec:parttwoppa} 
they are binary (for presence or absence of an edge). 
%
\spaceendsubsection
\subsection{The Causal Meta-DAG}
\label{ref:TheMetaDAGDescription}
\spacebefsubsection
The  meta-DAG is a directed acyclic graph (DAG) that represents 
the causal relationships between the dyad variables of the growing network 
(see Appendix~\ref{appendix:AppendixMetaDAGS} Figs.~\ref{Fig:ExampleGraphicalModelLocalDirect}, \ref{Fig:ExampleGraphicalModelNonLocalDirect}, and \ref{Fig:ExampleGraphicalModelLocalDirectSeveral} for several examples).
It refers to the generative model for the growing network. 
As such, the dyad variables of the growing network
are represented as the 
vertices of the meta-DAG. 
For instance,  
in Figs.~\ref{Fig:SchemeWithLocalInfluences} and \ref{Fig:SchemeWithNonInfluences} 
the white nodes and black edges represent the growing network and the (\colorHub{c}\colorTransitive{o}\colorForward{l}\colorBackward{o}\colorNonLocalExterior{r}\colorNonLocalInterface{f}\colorNonLocalInterior{u}\colorSelfArrow{l}) directed arrows between these black 
edges represent the meta-DAG. 
To avoid confusion, 
we refer to the vertices of the meta-DAG as ``dyads'', 
using 
the word ``nodes'' only to 
refer to the nodes of the growing network. 

We refer to these directed edges as \emphWord{causal arrows}.  
A causal arrow \mbox{$\dyadpair{a}{b}\hspace{2pt} \CausalArrow \hspace{2pt}\dyadpair{c}{d}$} from a parent dyad \dyadpair{a}{b} to a child dyad \dyadpair{c}{d} 
indicates that the 
outcome of the parent dyad variable \dyadpair{a}{b} 
\textit{can affect} 
the outcome of the child dyad \mbox{variable \dyadpair{c}{d}}, 
whereas the outcome of the child dyad variable \dyadpair{c}{d} \textit{cannot affect} the outcome of the parent dyad variable \dyadpair{a}{b}. 
This is typically phrased in terms of performing interventions on variables, 
such as in the do-calculus of ``hard'' interventions \citep{pearl1994probabilistic}, 
but also applies to various notions of ``soft'' interventions \citep{lorenz2023causal, bravo2024intervention}.  

%
\spaceendsubsection
\subsection{Invariances of the Causal Meta-DAGs}
\label{ref:InvarianceCausalModel}
\spacebefsubsection
One of our reasons for defining the nodes to be countably infinite is that 
our notions of invariance  
and symmetries
({\refTypeCausalArrows}, 
\Cref{thm:NodeDeletionInvariantCausalModels,thm:21InvariantCausalModels}) 
are more natural to state.

Deleting a node $\BoxNodeS{}$ from the growing network leaves a set of nodes that is isomorphic to the original; 
there is a unique order-preserving map $\varphi$ from the remaining nodes to the original nodes:  
\[\quad \varphi(i) = \begin{cases} i-1 \quad & \text{if}\quad \BoxNodeS{}<i \\ i \quad & \text{if}\quad i<\BoxNodeS{} \\\end{cases} \]
Deleting a node also deletes the dyads  containing that node.  
This relabeling induces a map from the remaining dyads to the original dyads:
\vspace{-1pt}
\begin{align*}
    \varphi\big(\dyadpair{i}{j}\big) = \dyadpair{\varphi(i)}{\varphi(j)} 
\end{align*}
Similarly, for each deleted dyad, any causal arrows referencing it (either as a parent or as a child) are also deleted.  
And the relabeling maps the remaining causal arrows:
\begin{align*}
    \varphi\Big(\dyadpair{i}{j}\hspace{1pt} \CausalArrow \hspace{1pt}\dyadpair{k}{l} \Big) = \Big(\varphi\big(\dyadpair{i}{j}\big)\hspace{1pt} \CausalArrow  \hspace{1.2pt} \varphi\big(\dyadpair{k}{l}\big)\Big) 
\end{align*}
A meta-DAG is a set of causal arrows between dyad variables.  We want to classify all meta-DAGs that are invariant to this action of node deletion and relabeling.  
That is, what sets of causal arrows are isomorphic to their image under the map $\varphi$: 
\vspace{-5pt}
\begin{align*}
    \varphi\Big(\big\{\text{causal arrows}\big\} \Big) \cong \big\{\text{causal arrows}\big\} 
\end{align*}
To answer this question, 
note that the only property that is preserved 
by $\varphi$ is the relative ordering of the nodes in the growing network.  
Indeed, if we define a set of causal arrows that makes reference to nodes that are any specific number of steps away 
(such as ``immediate predecessor''), 
deleting nodes changes this property, and the set of causal arrows will not be invariant.  

With this in mind, consider a generic dyad \dyadpair{i}{j} between two nodes $i<j$.  
There are five ways that an arbitrary node, which we denote by an open circle $\BoxNodeS{}$, 
can relate to the nodes $i$ and $j$:
\def\TempSpacingFirst{\kern1.25em}
\def\TempSpacingSecond{\kern-0.25em}
\vspace{-5pt}
\begin{align*}
\vspace{-1pt}
    \renewcommand{\arraystretch}{1.35}
    \begin{array}{crc}
    \BoxNode{} \UnBoxNode{i} \UnBoxNode{} \UnBoxNode{j} \UnBoxNode{} 
  \TempSpacingFirst  & \textNodeRelationNodeType{distant}  \textNodeRelationNode{nodes}\hphantom{\textNodeRelationNode{node}} & \TempSpacingSecond \BoxNodeS{}<i<j\\
    \UnBoxNode{} \BoxNode{i} \UnBoxNode{} \UnBoxNode{j} \UnBoxNode{}
    \TempSpacingFirst  & \textNodeRelationNodeType{past} \textNodeRelationNode{node}\hphantom{\textNodeRelationNode{nodes}} & \TempSpacingSecond \BoxNodeS{}=i<j\\ 
    \UnBoxNode{} \UnBoxNode{i} \BoxNode{} \UnBoxNode{j} \UnBoxNode{}
    \TempSpacingFirst  & \textNodeRelationNodeType{recent} \textNodeRelationNode{nodes}\hphantom{\textNodeRelationNode{node}} & \TempSpacingSecond  i<\BoxNodeS{}<j\\  
    \UnBoxNode{} \UnBoxNode{i} \UnBoxNode{} \BoxNode{j} \UnBoxNode{}
    \TempSpacingFirst  & \textNodeRelationNodeType{current} \textNodeRelationNode{node}\hphantom{\textNodeRelationNode{nodes}}  & \TempSpacingSecond  i<\BoxNodeS{}=j\\
    \UnBoxNode{} \UnBoxNode{i} \UnBoxNode{} \UnBoxNode{j} \BoxNode{}
   \TempSpacingFirst  & \textNodeRelationNodeType{future} \textNodeRelationNode{nodes}\hphantom{\textNodeRelationNode{node}}  & \TempSpacingSecond i<j<\BoxNodeS{}
   \end{array}
    \vspace{-8pt}
\end{align*}
Note that the set of ``future'' nodes is infinite, 
whereas the other four sets of node types are finite (with two of them having a single element). 

To enumerate the relationships between \textit{dyads}, 
\ie, pairs of nodes,
consider all the ways that \textit{two} arbitrary nodes can relate to \dyadpair{i}{j}.  
These two arbitrary nodes denote the parent dyad with a causal arrow to the child dyad: \mbox{$\DyadPairCircleNodes\hspace{2pt} \CausalArrow \hspace{2pt}\dyadpair{i}{j}$}.  
There are 12 types of deletion-invariant causal arrows\footnote{Plus the identity arrow from a dyad to itself makes 13.} in total. 
However, 5 of those options contain  ``future'' node(s) in the parent dyad, 
leading to a causal meta-DAG in which child dyads have infinitely many parents.

To a 
avoid issues associated with infinitely many variables we require that 
our 
causal meta-DAGs
have finite ancestral sets, 
where an ancestral set contains the parents of all variables in the set \citep{lauritzen1996graphical}.  
Thereby, we exclude these 5 options in which the parent dyad has at least one ``future'' node.
This ensures that the probability distribution resulting from a model following one such causal meta-DAG   
is unique \citep{peters2021causal}, 
and  also makes the model straightforward to sample from.

The remaining 7 types of arrows (defined in \refTypeCausalArrows) constitute the choices one has for constructing deletion-invariant 
meta-DAGs with finite ancestral sets; 
including any causal arrow of a given type requires one to include \textit{all} arrows of that type. 
\vspace{-6pt}
\def\KernTemp{\kern-0.15cm}
\def\KernTempRight{\kern1em}
\def\KernTempSpaceOne{\kern-0.1cm}
\def\KernTempSpaceTwo{\kern-0.0cm}
\def\PreHorizontalLineGap{-9pt}
\def\HorizontalLineTemp{\kern-0.2cm\rule{8cm}{1pt}\kern-8cm\\[-4pt]}
\renewcommand{\arraystretch}{1.35}
\begin{table}[ht]
  \centering
  \begin{tabular}{l l l}
    \HorizontalLineTemp
    \KernTemp\BoxNodeOne{} \UnBoxNode{} \BoxNodeTwo{i} \UnBoxNode{} \UnBoxNode{} \UnBoxNode{j} \UnBoxNode{} \UnBoxNode{} & 
      \KernTempSpaceOne\colorHub{\hub} & 
      \KernTempSpaceTwo\kern6pt 
      $\mathmakebox[24pt][r]{\dyadpair{a}{b}}
      \mathmakebox[24pt][c]{\CausalArrowColor{HubColor}} 
      \mathmakebox[24pt][l]{\dyadpair{a}{c}}\KernTempRight$ \\
    \KernTemp\UnBoxNode{} \UnBoxNode{} \BoxNodeOne{i} \BoxNodeTwo{} \UnBoxNode{} \UnBoxNode{j} \UnBoxNode{} \UnBoxNode{} & 
      \KernTempSpaceOne\colorTransitive{\transitive} & 
      \KernTempSpaceTwo\kern6pt 
      $\mathmakebox[24pt][r]{\dyadpair{a}{b}}
      \mathmakebox[24pt][c]{\CausalArrowColor{TransitiveColor}}
      \mathmakebox[24pt][l]{\dyadpair{b}{c}}\KernTempRight$ \\
    \KernTemp\BoxNodeOne{} \UnBoxNode{} \UnBoxNode{i} \UnBoxNode{} \UnBoxNode{} \BoxNodeTwo{j} \UnBoxNode{} \UnBoxNode{} & 
      \KernTempSpaceOne\colorForward{\forward} & 
      \KernTempSpaceTwo\kern6pt 
      $\mathmakebox[24pt][r]{\dyadpair{a}{c}} 
      \mathmakebox[24pt][c]{\CausalArrowColor{ForwardColor}}
      \mathmakebox[24pt][l]{\dyadpair{b}{c}}\KernTempRight$ \\
    \KernTemp\UnBoxNode{} \UnBoxNode{} \UnBoxNode{i} \BoxNodeOne{} \UnBoxNode{} \BoxNodeTwo{j} \UnBoxNode{} \UnBoxNode{} & 
      \KernTempSpaceOne\colorBackward{\backward} & 
      \KernTempSpaceTwo\kern6pt  
      $\mathmakebox[24pt][r]{\dyadpair{b}{c}}
      \mathmakebox[24pt][c]{\CausalArrowColor{BackwardColor}}
      \mathmakebox[24pt][l]{\dyadpair{a}{c}}\KernTempRight$ \\[\PreHorizontalLineGap]
    \HorizontalLineTemp
    \KernTemp\BoxNodeOne{} \BoxNodeTwo{} \UnBoxNode{i} \UnBoxNode{} \UnBoxNode{} \UnBoxNode{j} \UnBoxNode{} \UnBoxNode{} & 
      \KernTempSpaceOne\colorNonLocalExterior{\nonLocalExterior} & 
      \KernTempSpaceTwo\kern6pt 
      $\mathmakebox[24pt][r]{\dyadpair{a}{b}}
      \mathmakebox[24pt][c]{\CausalArrowColor{NonLocalExteriorColor}}
      \mathmakebox[24pt][l]{\dyadpair{c}{d}}\KernTempRight$ \\
    \KernTemp\BoxNodeOne{} \UnBoxNode{} \UnBoxNode{i} \BoxNodeTwo{} \UnBoxNode{} \UnBoxNode{j} \UnBoxNode{} \UnBoxNode{}  & 
      \KernTempSpaceOne\colorNonLocalInterface{\nonLocalInterface} & 
      \KernTempSpaceTwo\kern6pt 
      $\mathmakebox[24pt][r]{\dyadpair{a}{c}}
      \mathmakebox[24pt][c]{\CausalArrowColor{NonLocalInterfaceColor}}
      \mathmakebox[24pt][l]{\dyadpair{b}{d}}\KernTempRight$ \\
    \KernTemp\UnBoxNode{} \UnBoxNode{} \UnBoxNode{i} \BoxNodeOne{} \BoxNodeTwo{} \UnBoxNode{j} \UnBoxNode{} \UnBoxNode{} & 
      \KernTempSpaceOne\colorNonLocalInterior{\nonLocalInterior} & 
      \KernTempSpaceTwo\kern6pt 
      $\mathmakebox[24pt][r]{\dyadpair{b}{c}}
      \mathmakebox[24pt][c]{\CausalArrowColor{NonLocalInteriorColor}}
      \mathmakebox[24pt][l]{\dyadpair{a}{d}}\KernTempRight$ \\[\PreHorizontalLineGap]
    \HorizontalLineTemp
    \KernTemp\UnBoxNode{} \UnBoxNode{} \BoxNodeOne{i} \UnBoxNode{} \UnBoxNode{} \BoxNodeTwo{j} \UnBoxNode{} \UnBoxNode{} & 
      \KernTempSpaceOne\colorSelfArrow{\selfArrow} & 
      \KernTempSpaceTwo\kern6pt 
      $\mathmakebox[24pt][r]{\dyadpair{a}{b}}
      \mathmakebox[24pt][c]{\CausalArrowColor{SelfArrowColor}}
      \mathmakebox[24pt][l]{\dyadpair{a}{b}}\KernTempRight$ \\[\PreHorizontalLineGap]
    \HorizontalLineTemp
  \end{tabular}
  \caption{\textbf{The seven types of deletion-invariant  causal arrows with finite ancestral sets.} 
  The \textit{left} column represents the relative ordering of the parent dyad 
  {\protect\DyadPairCircleNodes} 
  and the child dyad {\protect\dyadpair{i}{j}}. 
  and the child dyad {\protect\dyadpair{i}{j}}. 
The \textit{middle} column are the names we have given to each type of causal arrow.  
 The \textit{right} column shows the causal arrows from the parent dyad to the child dyad in terms of lexicographic node indices \mbox{$a<b<c<d$}.  
 See Figs.~\ref{Fig:SchemeWithLocalInfluences}~and~\ref{Fig:SchemeWithNonInfluences} for the  representation of the causal arrows in the growing network and Appendix~\ref{appendix:AppendixMetaDAGS}   Figs.~\ref{Fig:ExampleGraphicalModelLocalDirect}~and~\ref{Fig:ExampleGraphicalModelNonLocalDirect}
 for their representation in the causal meta-DAGs. 
  }
  \label{Def:TypesOfCausalArrows}
\end{table}
\vspace{-10pt}
\spaceendsubsection
\subsection{Seven Types of Causal Arrows}
\label{sec:typesofcausalarrows}
\spacebefsubsubsection
Recall that the variables in our growing network models are the dyads; 
the nodes have no intrinsic properties other than their (relative ordering and) dyadic relationships with other nodes. %
At face value, this might appear overly simplistic.   
For example, in a growing network of citations between publications, 
this would mean that decisions about which publications to cite 
are determined solely by their interconnected bibliographies, 
and not explicitly on their content or quality. 
However, abstracting the notions of publication and citation 
to include notions like ``words/concepts/people'' and ``usage/reference/ideologies'',
it is not unreasonable to think of the content of such objects as 
encoded in the \potentiallyImproveN{structure of what they reference.}\footnote{
    This 
    relation-centric  
    perspective appears across many fields of mathematics.  
    To quote \citet{mazur2008one} on an essential tenet of category theory: 
    \textit{``Mathematical objects are determined by---and understood by---the network of relationships they enjoy with all the other objects of their species.''} 
}
\spaceendsection
\subsubsection{Arrows Between Dyads Sharing a Node} 
\label{sec:localcausalarrows}
\begin{figure}[h] 
\begin{tikzpicture}
    \newcommand{\LocLength}{1.75}
    \newcommand{\LocCosAngle}{0.92387953251} 
    \newcommand{\LocSinAngle}{0.38268343236} 
    \newcommand{\LocCosAngleTwo}{0.70710678118} 
    \newcommand{\TopNodeY}{3.0}
    \newcommand{\TopNodeX}{\TopNodeY*\LocSinAngle/\LocCosAngle}
    \newcommand{\LeftDashedLength}{3.0}
    \newcommand{\RightDashedLength}{5.0}
    \newcommand{\NewEdgeThickness}{2pt}
    \newcommand{\NewNodeSize}{13pt}
    \newcommand{\OtherEdgeThickness}{1pt}
    \newcommand{\DottedThickness}{1pt}
    \newcommand{\CausalThickness}{1pt}
    \newcommand{\OtherNodeSize}{12pt}
    \newcommand{\NodeBorderThickness}{1pt}
    \draw[line width=\NewEdgeThickness] (0,0) -- (\TopNodeX,\TopNodeY);
    \draw[line width=\DottedThickness,dotted] (-\LeftDashedLength,0) -- (\RightDashedLength,0); 
    \draw[line width=\DottedThickness,dotted] (-\LeftDashedLength,\TopNodeY) -- (\RightDashedLength,\TopNodeY); 
    \node[right, anchor=west,fill=white,draw=none] at (-\LeftDashedLength,0) {``current'' node}; 
    \node[right, anchor=east,fill=white,draw=none] at (\RightDashedLength,\TopNodeY) {``past'' node}; 
    \newcommand{\TiltedOffsetX}{-0.5}
    \newcommand{\TiltedOffsetY}{-0.5}
    \node [label={[rotate=0,align=center]``recent'' nodes\\ \mbox{$i<\BoxNodeS{}<j$}}] at (\RightDashedLength-1.8,\TopNodeY/2-1.5) {};
    \node [label={[rotate=0,align=center]``distant'' nodes\\ \mbox{$\BoxNodeS{}<i<j$}}] at (-\LeftDashedLength+2.0,\TopNodeY*3/2-0.5) {};
    \newcommand{\CausalRadius}{\LocLength/2}
    \draw[color=BackwardColor,line width=\CausalThickness, -latex] (\CausalRadius*\LocCosAngle,\CausalRadius*\LocSinAngle) arc (22.5:67.5:\CausalRadius);
    \node [label={[rotate=-22.5,anchor=south west]\NewName},inner sep=0] at (\CausalRadius*\LocSinAngle,\CausalRadius*\LocCosAngle) {};
    \draw[color=ForwardColor,line width=\CausalThickness, -latex] (-\CausalRadius*\LocSinAngle,\CausalRadius*\LocCosAngle) arc (90+22.5:90-22.5:\CausalRadius);
    \node [label={[rotate=-22.5,align=center,anchor=south east]\OldName},inner sep=0] at (\CausalRadius*\LocSinAngle,\CausalRadius*\LocCosAngle) {};
    \draw[color=HubColor,line width=\CausalThickness, -latex] (\TopNodeX+\CausalRadius*\LocSinAngle,\TopNodeY-\CausalRadius*\LocCosAngle) arc (-90+22.5:-90-22.5:\CausalRadius);
    \node [label={[rotate=-22.5,align=center,anchor=north west]\HubName},inner sep=0] at (\TopNodeX-\CausalRadius*\LocSinAngle,\TopNodeY-\CausalRadius*\LocCosAngle) {};
    \draw[color=TransitiveColor,line width=\CausalThickness, -latex] (\TopNodeX-\CausalRadius*\LocSinAngle,\TopNodeY+\CausalRadius*\LocCosAngle) arc (90+22.5:270-22.5:\CausalRadius);
    \node [label={[rotate=-22.5,align=center,anchor=north east]\PathName},inner sep=0] at (\TopNodeX-\CausalRadius*\LocSinAngle,\TopNodeY-\CausalRadius*\LocCosAngle) {};
    \draw[line width=\OtherEdgeThickness] (0,0) -- (-2.4*\LocLength*\LocSinAngle,2.4*\LocLength*\LocCosAngle);
    \node [fill=white, draw=black, circle, minimum size=\NewNodeSize, inner sep=0, line width=\NodeBorderThickness] at (-2.4*\LocLength*\LocSinAngle,2.4*\LocLength*\LocCosAngle) {};
    \draw[line width=\OtherEdgeThickness] (\TopNodeX,\TopNodeY) -- (\TopNodeX-\LocLength*\LocSinAngle,\TopNodeY+\LocLength*\LocCosAngle);
    \node [fill=white, draw=black, circle, minimum size=\NewNodeSize, inner sep=0, line width=\NodeBorderThickness] at (\TopNodeX-\LocLength*\LocSinAngle,\TopNodeY+\LocLength*\LocCosAngle) {};
    \draw[line width=\OtherEdgeThickness] (\TopNodeX+\LocLength*\LocSinAngle,\TopNodeY-\LocLength*\LocCosAngle) -- (\TopNodeX,\TopNodeY);
    \node [fill=white, draw=black, circle, minimum size=\NewNodeSize, inner sep=0, line width=\NodeBorderThickness] at (\TopNodeX+\LocLength*\LocSinAngle,\TopNodeY-\LocLength*\LocCosAngle) {};
    \draw[line width=\OtherEdgeThickness] (0,0) -- (\LocLength*\LocCosAngle,\LocLength*\LocSinAngle);
    \node [fill=white, draw=black, circle, minimum size=\NewNodeSize, inner sep=0, line width=\NodeBorderThickness] at (\LocLength*\LocCosAngle,\LocLength*\LocSinAngle) {};
    \node [fill=white, draw=black, circle, minimum size=\NewNodeSize, inner sep=0, line width=\NodeBorderThickness] at (0,0) {$\boldsymbol{j}$}; 
    \node [fill=white, draw=black, circle, minimum size=\NewNodeSize, inner sep=0, line width=\NodeBorderThickness] at (\TopNodeX,\TopNodeY) {$\boldsymbol{i}$}; 
%
\end{tikzpicture}
\caption{
\textbf{Types of causal arrows between dyads that share a node.}
\potentiallyImproveN{The black edges represent the dyads of the growing network and colors represent different types of causal arrows between them.}   
%
}
\label{Fig:SchemeWithLocalInfluences}
\end{figure}
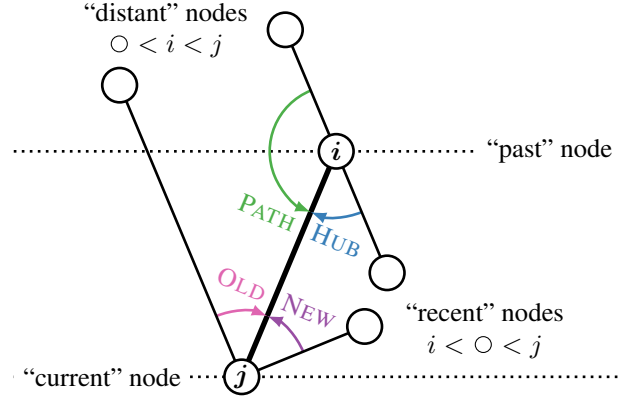

\spacebefsubsection
The first four types of causal arrows in {\refTypeCausalArrows} are 
from a parent dyad to a child dyad that have a node in common. 
Fig.~\ref{Fig:SchemeWithLocalInfluences} displays these relations in the growing network, and  Appendix~\ref{appendix:AppendixMetaDAGS} Fig.~\ref{Fig:ExampleGraphicalModelLocalDirect} shows their associated causal \mbox{meta-DAGs}. 

The {\hub} causal arrows 
mediate the tendency to reference a past concept 
due to how others have referenced it since then   
(\eg, citing a seminal paper).   

The {\transitive} causal arrows mediate the tendency to reference a concept 
due to the earlier concepts that it referenced  
(\eg, citing a paper due to its remarkable bibliography). 

The {\OldName} and {\NewName} causal arrows describe 
two ways in which one could sequentially decide which previous concepts to reference.  
A publication deciding what to cite using {\OldName} would do so chronologically,
making decisions about the oldest publications first, 
and allowing those choices to modulate its decisions about which recent publications to cite as well.  
Using {\NewName} is exactly the reverse; 
a publication begins by deciding which recent references to cite, 
and allowing these choices to modulate its decisions about older publications.  

Realistically, bibliographies are assembled more holistically, 
requiring both {\OldName} and {\NewName} causal influences.  
But since these causal arrows point in opposite directions, 
they cannot both be included in a causal meta-DAG  
as they would introduce cycles.  
However, in \Cref{sec:DorPAModel}, 
we describe a way to include both {\OldName} and {\NewName} causal arrows 
by using structural equations 
that can be unrolled into an asynchronous generative process.  
\spaceendsubsection
\subsubsection{Arrows between Dyads Not Sharing a Node} 
\label{sec:nonlocalcausalarrows}
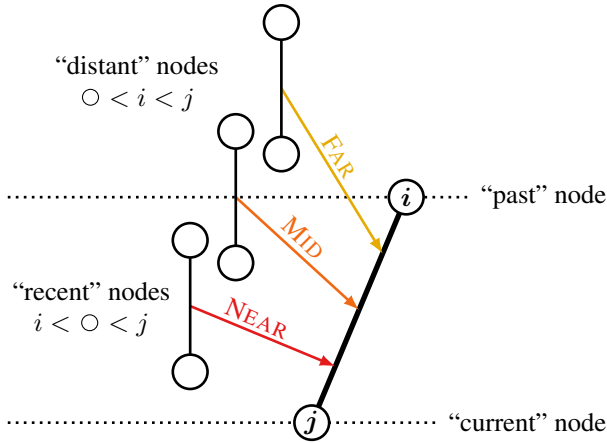
\begin{figure}[h]
\begin{tikzpicture}
    \newcommand{\LocLength}{1.75}
    \newcommand{\LocCosAngle}{0.92387953251} 
    \newcommand{\LocSinAngle}{0.38268343236} 
    \newcommand{\LocCosAngleTwo}{0.70710678118} 
    \newcommand{\TopNodeY}{3.0}
    \newcommand{\TopNodeX}{2.2426406871}
    \newcommand{\LeftDashedLength}{3.0}
    \newcommand{\RightDashedLength}{5.0}
    \newcommand{\NewEdgeThickness}{2pt}
    \newcommand{\NewNodeSize}{13pt}
    \newcommand{\OtherEdgeThickness}{1pt}
    \newcommand{\DottedThickness}{1pt}
    \newcommand{\CausalThickness}{1pt}
    \newcommand{\OtherNodeSize}{12pt}
    \newcommand{\NodeBorderThickness}{1pt}
    
    \newcommand{\BottomNodeX}{1.0}

    \newcommand{\NonLocSpacing}{0.6}
    \newcommand{\NonLocMidX}{-0.0}
    \newcommand{\NonLocMidY}{3.0}
    \newcommand{\NonLocLengthY}{\LocLength}

    \draw[line width=\NewEdgeThickness] (\BottomNodeX,0) -- (\TopNodeX,\TopNodeY);
    \draw[line width=\OtherEdgeThickness,dotted] (-\LeftDashedLength,0) -- (\RightDashedLength,0); 
    \draw[line width=\OtherEdgeThickness,dotted] (-\LeftDashedLength,\TopNodeY) -- (\RightDashedLength,\TopNodeY); 
    \node[right, anchor=east,fill=white,draw=none] at (\RightDashedLength,0) {``current'' node}; 
    \node[right, anchor=east,fill=white,draw=none] at (\RightDashedLength,\TopNodeY) {``past'' node}; 
    \newcommand{\TiltedOffsetX}{-0.5}
    \newcommand{\TiltedOffsetY}{-0.5}
    \node [label={[rotate=0,align=center]``recent'' nodes\\ \mbox{$i<\BoxNodeS{}<j$}}] at (-\LeftDashedLength+1.1,\TopNodeY/2-0.6) {};
    \node [label={[rotate=0,align=center]``distant'' nodes\\ \mbox{$\BoxNodeS{}<i<j$}}] at (-\LeftDashedLength+1.1+\NonLocSpacing,\TopNodeY*3/2-0.6) {};

    \draw[color=NonLocalInteriorColor,line width=\OtherEdgeThickness, -latex] (\NonLocMidX-\NonLocSpacing,\NonLocMidY-\NonLocSpacing*\LocCosAngle/\LocSinAngle) -- (\BottomNodeX*3/4+\TopNodeX/4,\NonLocMidY/4);
    \node [label={[rotate=-23.5,align=center,anchor=south,inner sep=0]\NearName}] at (\NonLocMidX/2-\NonLocSpacing/2+\TopNodeX*1/4/2+\BottomNodeX*2/4/2,\NonLocMidY/2-\NonLocSpacing*\LocCosAngle/\LocSinAngle/2 + \NonLocMidY/8) {};
    \draw[line width=\OtherEdgeThickness] (\NonLocMidX-\NonLocSpacing,\NonLocMidY-\NonLocSpacing*\LocCosAngle/\LocSinAngle - \NonLocLengthY/2) -- (\NonLocMidX-\NonLocSpacing,\NonLocMidY-\NonLocSpacing*\LocCosAngle/\LocSinAngle + \NonLocLengthY/2);
    \node [fill=white, draw=black, circle, minimum size=\NewNodeSize, inner sep=0, line width=\NodeBorderThickness] at (\NonLocMidX-\NonLocSpacing,\NonLocMidY-\NonLocSpacing*\LocCosAngle/\LocSinAngle - \NonLocLengthY/2) { };
    \node [fill=white, draw=black, circle, minimum size=\NewNodeSize, inner sep=0, line width=\NodeBorderThickness] at (\NonLocMidX-\NonLocSpacing,\NonLocMidY-\NonLocSpacing*\LocCosAngle/\LocSinAngle + \NonLocLengthY/2) { };

    \draw[color=NonLocalInterfaceColor,line width=\OtherEdgeThickness, -latex] (\NonLocMidX,\NonLocMidY) -- (\BottomNodeX*1/2+\TopNodeX/2,\NonLocMidY/2);
    \node [label={[rotate=-44.5,align=center,anchor=south,inner sep=0]\MidName}] at (\NonLocMidX/2+\TopNodeX*2/4/2+\BottomNodeX*2/4/2,\NonLocMidY/2+\NonLocMidY/4) {};
    \draw[line width=\OtherEdgeThickness] (\NonLocMidX,\NonLocMidY-\NonLocLengthY/2) -- (\NonLocMidX,\NonLocMidY+\NonLocLengthY/2); 
    \node [fill=white, draw=black, circle, minimum size=\NewNodeSize, inner sep=0, line width=\NodeBorderThickness] at (\NonLocMidX,\NonLocMidY-\NonLocLengthY/2) { };
    \node [fill=white, draw=black, circle, minimum size=\NewNodeSize, inner sep=0, line width=\NodeBorderThickness] at (\NonLocMidX,\NonLocMidY+\NonLocLengthY/2) { };

    \draw[color=NonLocalExteriorColor,line width=\OtherEdgeThickness, -latex] (\NonLocMidX+\NonLocSpacing,\NonLocMidY+\NonLocSpacing*\LocCosAngle/\LocSinAngle) -- (\BottomNodeX*1/4+\TopNodeX*3/4,\NonLocMidY*3/4);
    \node [label={[rotate=-58,align=center,anchor=south,inner sep=0]\FarName}] at (\NonLocMidX/2+\NonLocSpacing/2+\TopNodeX*3/4/2+\BottomNodeX*1/4/2,\NonLocMidY/2+\NonLocSpacing*\LocCosAngle/\LocSinAngle/2+\NonLocMidY*3/4/2) {};
    \draw[line width=\OtherEdgeThickness] (\NonLocMidX+\NonLocSpacing,\NonLocMidY+\NonLocSpacing*\LocCosAngle/\LocSinAngle - \NonLocLengthY/2) -- (\NonLocMidX+\NonLocSpacing,\NonLocMidY+\NonLocSpacing*\LocCosAngle/\LocSinAngle + \NonLocLengthY/2);
    \node [fill=white, draw=black, circle, minimum size=\NewNodeSize, inner sep=0, line width=\NodeBorderThickness] at (\NonLocMidX+\NonLocSpacing,\NonLocMidY+\NonLocSpacing*\LocCosAngle/\LocSinAngle - \NonLocLengthY/2) { };
    \node [fill=white, draw=black, circle, minimum size=\NewNodeSize, inner sep=0, line width=\NodeBorderThickness] at (\NonLocMidX+\NonLocSpacing,\NonLocMidY+\NonLocSpacing*\LocCosAngle/\LocSinAngle + \NonLocLengthY/2) { };

    \node [fill=white, draw=black, circle, minimum size=\NewNodeSize, inner sep=0, line width=\NodeBorderThickness] at (\BottomNodeX,0) {$\boldsymbol{j}$};
    \node [fill=white, draw=black, circle, minimum size=\NewNodeSize, inner sep=0, line width=\NodeBorderThickness] at (\TopNodeX,\TopNodeY) {$\boldsymbol{i}$};

\end{tikzpicture}
\caption{\textbf{Types of causal arrows between dyads that do not share a node.}  Horizontal dashed lines indicate the relative ordering of the nodes in the parent dyad ${\protect\DyadPairCircleNodes}$ with respect to the nodes in the child dyad $\dyadpair{i}{j}$.
}
\label{Fig:SchemeWithNonInfluences}
\end{figure}

\spacebefsubsection
The next three types of causal arrows in {\refTypeCausalArrows} 
are from a parent dyad to a child dyad with \textit{no} nodes in common. 
Fig.~\ref{Fig:SchemeWithNonInfluences} displays these relations in the growing network, and  Appendix~\ref{appendix:AppendixMetaDAGS} Fig.~\ref{Fig:ExampleGraphicalModelNonLocalDirect} shows their associated meta-DAGs. 

These types of causal arrows can be interpreted as a sort of ``context''. 
The {\nonLocalExterior} causal arrows allow current citation decisions to be influenced by the way historical documents referenced each other.  
The {\nonLocalInterface} causal arrows allow for influence from 
how more recent publications have been referencing historical documents.  
And the {\nonLocalInterior} causal arrows allow for influence from 
how recent publications have been referencing each other.  
\spaceendsubsection
\subsubsection{The Arrow from a Dyad to Itself} 
\label{sec:selfarrows}
\spacebefsubsection
Technically, the {\selfArrow} arrow in {\refTypeCausalArrows} is a directed cycle, 
so we do not include it on our meta-DAGs. 
Nevertheless, 
it can be reasonably interpreted in a number of ways,   
such as the intrinsic stochasticity of the dyad variable 
or as a mechanism for interventions on it.  
\potentiallyImproveN{For the enumeration} of \Cref{thm:21InvariantCausalModels}, 
it plays a useful role as the identity operator. 
\spaceendsubsection
\subsection{Invariance to Node Deletion}
\label{sec:deletion}
\spacebefsubsection
To enumerate all the possible deletion-invariant causal meta-DAGs with finite ancestral sets, 
we consider all combinations of these 7 causal arrow types that do not result in directed cycles.   
%
\potentiallyImproveN{For all 
 types of 
causal arrows except $\OldName$ and $\NewName$, 
the largest node index necessarily increases from parent dyad to child dyad, 
so any combination of those 5 causal arrow types cannot contain a cycle.}  
To any of these $2^5=32$ subsets, we may add either $\OldName$ or $\NewName$ 
(but not both) to create a valid deletion-invariant causal DAG.  

\begin{theorem}[\thmNameNodeDel]
\label{thm:NodeDeletionInvariantCausalModels}
There are \mbox{$2^5 3 = 96$} deletion-invariant causal meta-DAGs with finite ancestral sets, given by the subsets of $\{\FarName,\PathName,\MidName,\HubName,\NearName,\OldName,\NewName\}$ that do not contain both $\OldName$ and $\NewName$. 
\end{theorem}
\vspace{-5pt}
By requiring invariance to node deletion, 
we have described the set of causal DAGs over dyad variables that are ``the same'' 
for any subset of nodes that are actually included in the growing network.  
\potentiallyImproveN{To illustrate this idea, consider indexing publication nodes by the precise time that the authors made their final edit, 
in terms of milliseconds since 1 Jan 1970.} 
One can think of the resulting citation network as being initialized with over a trillion potential nodes, 
then deleting the vast majority of those that did not contain a publication.  
In a sense, invariance to node deletion is invariance to ``that which never existed''.  
\spaceendsubsection
\subsection{Invariance to Node Marginalization}
\label{sec:poset}
\spacebefsubsection
%
Similarly, one might be interested in invariance to ``that which was not observed''.  
Variables that exist but are not observed cannot simply be deleted from a causal model; 
in order to preserve the causal 
 and probabilistic 
relationships between the remaining variables, 
they must be \textit{marginalized}.  
For a causal DAG, marginalizing a variable is a two-step process: 
first add arrows from all of its parents to all of its children, 
then delete that variable \citep{richardson2002ancestral}. 
In general, one must also account for the stochastic component of this now-unobserved latent variable, 
often represented by introducing bidirected arrows between the children \citep{lauritzen1996graphical}.  
We will ignore such bidirected arrows for now, 
and address them momentarily.  
%

In our setting, the random variables are the dyads between the nodes of the growing network.  
So marginalization of a node in the growing network means 
marginalizing all dyad variables containing that node.  
Of the 96 deletion-invariant causal meta-DAGs, 
21 are also invariant to node marginalization, 
shown in Appendix~\ref{appendixFigHasse} Fig.~\ref{fig:PosetMetaDAG}.  
They are precisely the meta-DAGs that are \textit{transitively closed}, 
\potentiallyImproveN{that is, all dyad variables have the property that 
all of its parent dyads have arrows pointing to all of its child dyads.}  

To obtain the transitive closure \potentiallyImproveN{of a causal meta-DAG,  
one can} repeatedly perform the first step of marginalizing a variable, 
adding arrows from parents to children \textit{without} deleting the variable, 
until no more arrows can be added.  
For example, if the \potentiallyImproveN{meta-DAG} contains the arrows $\dyadpair{a}{b} \CausalArrow \dyadpair{i}{j}$ and $\dyadpair{i}{j} \CausalArrow \dyadpair{r}{s}$, then it must also include the arrow $\dyadpair{a}{b} \CausalArrow \dyadpair{r}{s}$.  
\potentiallyImproveN{As this first step is the only difference between deletion and marginalization, 
taking the transitive closure renders them equivalent, 
ensuring invariance to marginalization as well as deletion.}   

With the directed causal arrows accounted for, 
we return to the bidirected arrows mentioned earlier.  
Indeed, these must be included if one wishes to 
preserve conditional independence statements between the remaining variables.  
But for our setting, these statements do not distinguish 
between many of the resulting causal structures.  
For example, any causal meta-DAG with the $\FarName$ causal arrow
would require bidirected arrows between every dyad 
(marginalizing the first two nodes will suffice),  
resulting in zero conditional independence statements
for 13 of the 21 meta-DAGs in Appendix~\ref{appendixFigHasse} Fig.~\ref{fig:PosetMetaDAG}.  


However, causal models encode more than just conditional independence; 
they also encode how the distribution might change as a result of performing interventions.  
Consider the question of which dyad variables $\DyadPairCircleNodes$ \textit{cannot} be affected by an intervention on the outcome of $\dyadpair{i}{j}$. 
For causal DAGs, the answer is unaffected by the presence of bidirected arrows; 
\mbox{$\DyadPairCircleNodes \perp \text{do}\big(\dyadpair{i}{j}\big)$} if there is \textit{no} arrow \mbox{$\dyadpair{i}{j} \CausalArrow \DyadPairCircleNodes$} in its transitive closure.  
We therefore consider invariance (to node deletion and contraction) of this interventional structure 
to define and distinguish between the 21 meta-DAGs in Theorem~\ref{thm:21InvariantCausalModels}. 
\begin{theorem}[\thmNameHasse]
\label{thm:21InvariantCausalModels}
There are \mbox{$21$} causal meta-DAGs with finite ancestral sets whose interventional structure is 
invariant to both node deletion and marginalization. 
Their partial ordering is shown in Appendix~\ref{appendixFigHasse} Fig.~\ref{fig:PosetMetaDAG}.  
\end{theorem}
\vspace{-5pt}
\def\JoinArrowSpacing{\kern-3pt}
\def\ListArrowSpacing{\kern-2pt}
\spaceendsubsection
\subsubsection{Computing the Transitive Closure}
\spacebefsubsubsection
To take the transitive closure of a set of causal arrows, 
consider all the ways those arrows can compose.  
That is, for all cases in which the child dyad of the first causal arrow is the parent dyad of the next causal arrow, 
which types of causal arrows could point from the parent dyad of the first arrow to the child dyad of the second?  

For example: \mbox{$\CausalArrowColorNamePath\JoinArrowSpacing\CausalArrowColorNamePath=\{\CausalArrowColorNameFar\JoinArrowSpacing\}$}, 
since the effect of traversing two $\PathName$ causal arrows is always equivalent to
traversing a single $\FarName$ causal arrow.  
Some compositions have the potential to be equivalent to multiple arrow types: 
\mbox{$\CausalArrowColorNameHub\JoinArrowSpacing\CausalArrowColorNameOld=\{\CausalArrowColorNameMid\ListArrowSpacing,\CausalArrowColorNamePath\ListArrowSpacing,\CausalArrowColorNameFar\JoinArrowSpacing\}$}.  
Technically, this composition depends on the order: \mbox{$\CausalArrowColorNameOld\JoinArrowSpacing\CausalArrowColorNameHub=\{\CausalArrowColorNameMid\JoinArrowSpacing\}$}, 
but this noncommutativity will not be important for our purposes.    

Define the composition of two \textit{sets} of arrow types as the union over all combinations of arrows from the two sets.  
Then, to take the transitive closure, 
begin with an initial set of causal arrows 
(along with the {$\SelfName$} arrow, which plays the role of the identity).  
Compose this set of arrows with itself to obtain a (possibly) larger set of causal arrow types.  
Continue this process until the set is no longer increasing to obtain the transitive closure.  

For example, \mbox{$\{\CausalArrowColorNameSelf\ListArrowSpacing,\CausalArrowColorNameHub\ListArrowSpacing,\CausalArrowColorNamePath\JoinArrowSpacing\}$} composed with itself results in \mbox{$\{\CausalArrowColorNameSelf\ListArrowSpacing,\CausalArrowColorNameHub\ListArrowSpacing,\CausalArrowColorNamePath\ListArrowSpacing,\CausalArrowColorNameFar\JoinArrowSpacing\}$}.  
Composing \mbox{$\{\CausalArrowColorNameSelf\ListArrowSpacing,\CausalArrowColorNameHub\ListArrowSpacing,\CausalArrowColorNamePath\ListArrowSpacing,\CausalArrowColorNameFar\JoinArrowSpacing\}$} with itself results in the same set, so it is transitively closed.  
Starting with the 96 deletion-invariant subsets of arrows from Theorem~\ref{thm:NodeDeletionInvariantCausalModels}, 
one obtains the 21 fixed points of this process, 
corresponding to the meta-DAGs in Theorem~\ref{thm:21InvariantCausalModels}.  
%
%

Computationally, we implemented this by representing the arrows and their composition as matrices and matrix multiplication.  
While these 21 equivalence classes can easily be found by hand, 
this 
quickly become unwieldy for generalizations of the procedure.\footnote{
For example, 
for simple hypergraphs with cardinality 3, 
there are already 37 type of causal arrows between the triads 
(subsets of 3 nodes) with finite ancestral sets. 
}

\section{Streamlining Network Models}
\label{sec:parttwoppa}
\spacebefsection
%
Our classification of invariant causal meta-DAGs 
imposes 
requirements for where causal arrows can appear between the dyad variables, 
but so far we have said nothing of what the structural equations of a growing network might be. 
\potentiallyImproveN{
Indeed, one \textit{could} use a different function for each dyad
with a complex dependency on all of its parents, 
and such a model would still generate network distributions that are faithful to the causal structure of our meta-DAGs.}  

However, in the spirit of our framework of ``invariance of causal mechanisms'', 
we propose using the \textit{same} structural equation for all dyad variables in the model.  
Since the number of parents depends on the position of the child dyad variable, 
such a function must allow for an arbitrary number of inputs.  
\potentiallyImproveN{A natural choice is to define the function in terms of 
summary statistics of the different types of causal parents.} 
%

In Section~\ref{sec:ourppamodel}, 
we present a simple model for binary (edge or no-edge) dyad variables  
that implements preferential attachment using the $\HubName$ and $\PathName$ causal arrows.  
More important than the model itself is the rather surprising insight it provides:
\textit{reducing} dependencies between the dyad variables leads to \textit{increased} diversity in 
the 
asymptotic behavior of the growing network.  
\spaceendsection
\subsection{Basic Preferential Attachment}
\label{sec:prefattachmentcontext}
\spacebefsubsection
While the relationship between preferential growth and scale-free distributions had 
already been described by several authors \citep{eggenberger1923statistik,simon1955class,price1965networks},  
the effect itself is perhaps best exemplified by the overwhelming number of citations garnered by 
\citet{barabasi1999emergence}. 

The simplest statement of their model 
\citep{posfai2016network}
has a single parameter, $m$.  
Initialize the network with clique of $m$ nodes. 
At each iteration, select $m$ nodes proportional to their current degree, 
and add a new node connected to each of these selected nodes.  

Many extensions to this model have been described 
\citep{ray2024stochasticthesis}: 
introducing parameters to control correlations between neighboring degrees \citep{avin2020mixed} or to promote clustering \citep{eikmeier2019triangle}; 
adding node covariates \citep{bianconi2001competition, lee2015preferential}; 
and considering alternative attachment functions  \citep{krapivsky2000connectivity}.   
Here, we draw attention to a particular line of modifications that reduce the \potentiallyImproveN{statistical} correlations between pairs of edges \citep{bollobas2007phase,wang2020directed}.  
\spaceendsubsubsection
\subsubsection{Poissonified Preferential Attachment} 
\label{sec:theirsppamodel}
\spacebefsubsubsection
\potentiallyImproveN{Most models of preferential attachment have a  
parameter $m$, 
specifying the precise number of edges that each new node $j$ makes with the previous nodes.  
This results in a small anticorrelation between the dyad variables \dyadpair{\BoxNodeS{}}{j} within each iteration.} 
%
Alternatively, one could compute probabilities for each edge that are proportional to their degrees, such that $m$ edges \potentiallyImproveN{will be} 
added in expectation \citep{van2024random}.\footnote{Setting aside cases with probabilities greater than $1$. }
This does not qualitatively change the asymptotic behavior; 
the degree distribution has the same power-law tail \mbox{$p(d)\propto d^{-3}$} 
and average degree \mbox{$\langle d\rangle = 2m$} as before.  
%

By allowing the number of edges to be an implicit random variable, 
this modification renders the dyad variables \dyadpair{\BoxNodeS{}}{j} conditionally independent 
given the current degrees of the previous nodes.  
However, their outcomes still depend on the entire network up to the previous iteration.  
This is due to the fact that the sum of their edge probabilities has been scaled to be equal to $m$.  
Is this dependence on all previous dyad variables necessary? 
How much dependence can we remove  
while still retaining the hallmark features of preferential attachment models? 
\spaceendsubsection
\subsection{{\ppaName}}
\label{sec:ourppamodel}
\spacebefsubsection
Motivated by this question, 
we note that only two of the seven 
types of causal arrows are essentially being asked for by preferential attachment: 
$\HubName$ and $\PathName$, 
and that their transitive closure implies 
significantly less dependence on previous dyad variables (see~\Cref{fig:PosetMetaDAG}). 
As summary statistics for the structural equation, 
we use node degrees of the older node: 
$d_i^{\text{in}}$ corresponds to $\HubName$, 
and $d_i^{\text{out}}$ corresponds to $\PathName$.  
We take the edge probabilities to be an affine function of these statistics, 
leading to the model we call 
\emphWord{{\mbox{\ppaName}}}, or \emphWord{\mbox{\ppaAcron}}, 
as sampling from it is highly parallelizable (see discussion in Section~\ref{sec:discussion}).  
%
Explicitly:
\begin{align}
    x_{ij}^{ } &\sim \text{Bernoulli}\big(p_{ij}^{ }\big) \label{eq:PPAbernoulli} \\
    p_{ij}^{ } &= \frac{\alphaP + \thetain d_i^{\text{in}} + \thetaout d_i^{\text{out}}}{j-2+\alphaP +\betaP} \label{eq:PPAprob}\\
    d_i^{\text{in}} &= \sum_{\BoxNodeSS{}=i+1}^{j-1} x_{i\BoxNodeSS{}}^{ }  \quad\quad
    d_i^{\text{out}} = \sum_{\BoxNodeSS{}=1}^{i-1} x_{\BoxNodeSS{} i}^{ } \label{eq:PPAdout}    
\end{align}
where \mbox{$x_{ij} = 1$} indicates an edge between nodes $i$ and $j$ (and $\mbox{$x_{ij} = 0$}$ indicates no edge). 


Surprisingly, while drastically \textit{decreasing the dependence} between the dyads, 
this model exhibits \textit{increased diversity} in its asymptotic behavior 
(see Fig.~\ref{fig:PPARegimes}, Theorems~\ref{thm:PPAPhaseTransition} and \ref{thm:PPAPhaseTransitiondegdist}, proofs in Appendix~\ref{appendixDAPAProofs}).
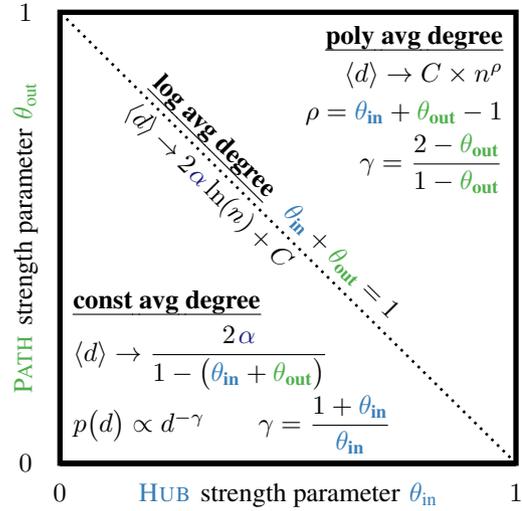
\begin{figure}[h]
    \centering
    \begin{tikzpicture}[remember picture]
        \tikzmath{\SquareSize=6.0;}
        \tikzmath{\BottomRightX=0.0;}
        \tikzmath{\BottomRightY={-\SquareSize};}
        \tikzmath{\UpperLeftX={\BottomRightX+\SquareSize};}
        \tikzmath{\UpperLeftY={\BottomRightY+\SquareSize};}
        \tikzmath{\BoxThickness=2;}
        \tikzmath{\DiagThickness=1;}

        \draw [line width=\BoxThickness] (\BottomRightX,\BottomRightY) -- (\UpperLeftX,\BottomRightY) -- (\UpperLeftX,\UpperLeftY) -- (\BottomRightX,\UpperLeftY) -- (\BottomRightX,\BottomRightY) -- (\UpperLeftX,\BottomRightY);
    
        \draw [line width=\DiagThickness,dotted] (\UpperLeftX,\BottomRightY) -- (\BottomRightX,\UpperLeftY);
    
        \tikzmath{\offset=0.5;}
        \tikzmath{\X=\BottomRightX+\SquareSize/2;}
        \tikzmath{\Y=\BottomRightY-\offset;}
        \node[text=black,rotate=0,anchor=base] at (\X,\Y) {$\HubName$ \kern1pt strength parameter \kern1pt $\thetain$};
        \node[text=black,rotate=0,anchor=base] at (\BottomRightX,\Y) {$0$};
        \node[text=black,rotate=0,anchor=base] at (\UpperLeftX,\Y) {$1$};
    
        \tikzmath{\offset=0.45;}
        \tikzmath{\X=\BottomRightX-\offset;}
        \tikzmath{\Y=\BottomRightY+\SquareSize/2;}
        \node[text=black,rotate=90] at (\X,\Y) {$\PathName$ \kern1pt strength parameter \kern1pt $\thetaout$};
        \node[text=black,rotate=0] at (\X,\BottomRightY) {$0$};
        \node[text=black,rotate=0] at (\X,\UpperLeftY) {$1$};

        \def\SpacingObject{\vphantom{\big(}}
        \tikzmath{\x=1.0;\y=-1.0;}
        \node[text=\logColor,rotate=-45,anchor=south west] at (\x-0.05,\y-0.05) {\SpacingObject\textbf{\uline{log avg degree}} \kern6pt \textbf{\mbox{$\thetain+\thetaout=1$}}};
        \node[text=\logColor,rotate=-45,anchor=north west] at (\x+0.05,\y+0.05) {\SpacingObject\textbf{$\langle d \rangle$ $\rightarrow$ $2\kern1pt \alphaP\ln(n) + C$}};
        \tikzmath{\CenterX=\BottomRightX+\SquareSize/2;}
        \tikzmath{\CenterY=\BottomRightY+\SquareSize/2;}

        \tikzmath{\x=\SquareSize-0.05;\y=0.0;}
        \node[text=\polyColor,rotate=0,anchor=north east] at (\x,\y) 
            {\SpacingObject\textbf{\uline{poly avg degree}} };
        \tikzmath{\x=\x;\y=\y-0.5;}
        \node[text=\polyColor,rotate=0,anchor=north east] at (\x,\y) {\SpacingObject\textbf{\mbox{$\langle d\rangle\rightarrow C\times n^{\rho}_{ }$}}};
        \tikzmath{\x=\x;\y=\y-0.5;}
        \node[text=\polyColor,rotate=0,anchor=north east] at (\x,\y) 
            {\SpacingObject\textbf{\mbox{$\rho=\thetain+\thetaout-1$}}};
        \tikzmath{\x=\x;\y=\y-0.5;}
        \node[text=\constColor,rotate=0,anchor=north east] at (\x,\y) {\textbf{$\gamma = \displaystyle\frac{2-\thetaout}{1-\thetaout}$}};

        \tikzmath{\x=0.05;\y=-3.65;}
        \renewcommand{\eqspace}{\kern2pt}
        \tikzmath{\x=\x;\y=\y-0.5;}
        \node[text=\constColor,rotate=0,anchor=south west] at (\x,\y) {\textbf{\uline{const avg degree}}};
        \tikzmath{\x=\x;\y=\y-1.0;}
        \node[text=\constColor,rotate=0,anchor=south west] at (\x,\y) {\textbf{$\langle d \rangle \rightarrow \displaystyle\frac{2\kern1pt\alphaP}{1-\big(\thetain+\thetaout\big)}$}};
        \tikzmath{\x=\x;\y=\y-0.85;}
        \node[text=\constColor,rotate=0,anchor=south west] at (\x,\y) {\textbf{$p\big(d\big)  \propto  d^{-\gamma} \qquad \gamma  =  \displaystyle\frac{1+\thetain}{\thetain}$}};


    
    \end{tikzpicture}
    \vspace{-5pt}
    \caption{\textbf{Sparsity and power-laws in the {\ppaAcron} model.} 
    }
    \label{fig:PPARegimes}
\end{figure}
\spaceendsubsubsection
\subsubsection{Three Sparsity Regimes}
\spacebefsubsubsection
Sparsity can be characterized in terms of the average degree as a function of the number of nodes ${\avgdegree = \frac{2E(n)}{n}}$.  
For dense networks, the average degree grows linearly in the number of nodes: ${\avgdegree = \bigO{n}}$, 
while for sparse networks, the average degree grows sublinearly, ${\avgdegree = \littleO{n}}$ \citep{van2024random}. 
%
Networks with power-law degree distributions are naturally sparse;  
in order for the degrees to span arbitrarily many orders of magnitude, 
``most'' nodes have a vanishingly small degree compared to the maximum.  

Despite 
the apparent similarity of our {\ppaAcron} model to preferential attachment models that ``hard-code'' the average degree, 
our model exhibits an emergent phase transition to sparse scalings with growing average degree. 
\begin{theorem}[\thmNamePPA]
\label{thm:PPAPhaseTransition}
    Our {\ppaAcron} model exhibits three qualitatively different asymptotic behaviors for the average degree $\avgdegree$. \\
    \def\TempKern{\kern4pt}
\begin{tabular}{llr}
    \vphantom{\Bigg(}constant:      &\TempKern$\displaystyle\frac{2\kern1pt\alphaP}{1-\big(\thetain+\thetaout\big)}$     &\TempKern$0<\thetain+\thetaout<1$ \\
    \vphantom{\Bigg(}logarithmic:   &\TempKern$2\kern1pt\alphaP \log\big(n\big) + C$                            &\TempKern$\thetain+\thetaout=1$ \\
    \vphantom{\Bigg(}polynomial:    &\TempKern$C \times n^{\thetain+\thetaout-1}$                     &\TempKern$1<\thetain+\thetaout<2$
\end{tabular}
\end{theorem}
\spaceendsubsubsection
\subsubsection{And a Flexible Power-law Degree Distribution}
\label{sec:powerlawflexible}
\spacebefsubsubsection
For all three sparsity regimes, 
the degree distributions of the growing networks have a range of power-law scalings.
\begin{theorem}[\thmNamePPAdegdist]
\label{thm:PPAPhaseTransitiondegdist}
   The asymptotic probability that a random node has degree $d$ has a tail of the form \mbox{$p(d) \propto d^{-\gamma}$}, where the scaling exponent depends on either $\thetain$ or $\thetaout$. \\
   \def\TempKern{\kern4pt}
    \begin{tabular}{llr}
    \vphantom{\Bigg(}\text{constant:}      &\TempKern$\displaystyle\gamma = \frac{1+\thetain}{\thetain}$     &\TempKern$0<\thetain+\thetaout\leq1$ \\
    \vphantom{\Bigg(}polynomial:    &\TempKern$\displaystyle\gamma = \frac{2-\thetaout}{1-\thetaout}$                     &\TempKern$1\leq\thetain+\thetaout<2$
\end{tabular}
\end{theorem}
Notice that these two expressions result in the same scaling exponent precisely when \mbox{$0<\thetain+\thetaout\leq1$}.  


\spaceendsubsubsection
\section{Some Applications} 
\label{sec:extensions}
\spacebefsection
\spaceendsubsubsection
\subsection{Inference and Generalization}
\label{sec:inference}
\spacebefsection
This flexibility of asymptotic behaviors from a simple model is a useful property for extrapolating from limited data.  
For instance, 
consider observing a growing network that is still in its relative infancy.  
\potentiallyImproveN{The average degree is increasing as a function of the number of nodes, but it is slowing down; 
will it converge to some constant value, or if not, at what rate might it increase?}  
The degree distribution is currently more spread than a network with independent edges, 
 but there is not yet a region that looks linear on a log-log plot;\footnote{\potentiallyImproveN{Estimating the power-law exponent of a degree distribution is notoriously tricky \citep{clauset2009power}.  
No finite network is truly scale-free; 
even if there is an obvious power-law that fits the majority of the degree distribution, 
there are necessarily deviations at the extremities.}}  
what might its scaling exponent be once many more nodes are added?  

By fitting the parameters of a simple structural equation to initial observations, 
one might still be able to predict the 
the asymptotic behavior of the growing network, 
despite those features not yet being present.  
%
%
%
\spaceendsubsection 
\subsection{Interventions and Counterfactuals}
\label{sec:interventionandcounterfactuals}
\spacebefsubsection
\potentiallyImprove{
In this section, 
we 
illustrate  
how 
to use our framework 
to answer interventional and counterfactual causal queries 
using our beloved running examples.} 

Suppose you are about to submit a publication, 
and you want to add a few more citations to your bibliography to help it reach a larger audience.  
To estimate the net effect of such strategic citations, 
one could fit the parameters of 
a causal model 
(such as the {\ppaAcron} model or its extensions)
to the current citation network.  

By approximating the strength of various causal mechanisms, 
one can run the model forward to estimate the number of 
additional citations one might receive as a result.  
This is an example of an \textit{interventional} question, 
as the answer involves quantifying (the result of performing an action) 
over a \textit{distribution} of possible futures.  

Now suppose you have a older publication that you really feel should have more citations, 
and you are deciding how much to regret not promoting it more at the time.  
This is an example of a \textit{counterfactual} question,  
as now the answer involves quantifying the difference between 
one \textit{particular} outcome (that was actually observed), 
and another (that \textit{could} have occurred, but did not).  

To estimate net effect of such fictional actions, 
one can use the structural equations of \potentiallyImproveN{a causal model.}   
For example, the randomness in Eq.~\eqref{eq:PPAbernoulli} can be represented explicitly by introducing an (unobserved) random variable: 
\begin{align}
       \epsilon_{ij}^{ } &\sim \mathcal{U}\textit{niform}\big(0,1\big) \\
    x_{ij}^{ } &= \text{sign}\big(p_{ij}^{ } - \epsilon_{ij}^{ }\big) \label{eq:PPASEM}
\end{align}
\potentiallyImproveN{From the estimated parameters of the model, 
and the actual observed data, 
one can use this form to estimate the likelihood of such counterfactual changes.}  
Essentially, this involves performing bayesian updates 
to the $p_{ij}^{ }$ and $x_{ij}^{ }$, while treating the $\epsilon_{ij}^{ }$ as fixed \citep{pearl2009causality}.

\section{Distributed Discussion}
\label{sec:discussion}
\spacebefsection
Initially, we set out to classify similar causal models for networks that grow one node at a time.  
This requirement turned out to be overly restrictive, 
and we were 
pleasantly surprised 
to find causal structures that were less rigid in the order of their generation.  

In particular, causal models in which the dyads only depend on one of the two ``quadrants'' of past dyads---such as the ``{\ppaAcron} w/ clustering'' or ``bottom-up causality'' in Fig.~\ref{fig:PosetMetaDAG}, or any subset of their arrows---can be evaluated in a distributed manner (as illustrated in Fig.~\ref{fig:FirstNatureFigure}).  
For these models, coarse-graining the rows and columns of the grid of dyad variables 
results in blocks of dyads with a similar causal structure.  
Thus, one can assign workers to different blocks of dyads to evaluate them in parallel 
requiring communication only when workers move to the next block.  
For example, in the model below, $w$ workers can alternate between evaluating blocks of size \mbox{$\frac{n}{w}$-by-$\frac{n}{w}$} and a total of $2w$ rounds of communication.  

\paragraph{{\ppaAcron} model with clustering  --- $\HubName+\PathName+\OldName$.} 
As an extension of the {\ppaAcron} model, 
one could include $\OldName$ causal arrows 
(see bottom-right of Fig.~\ref{Fig:ExampleGraphicalModelLocalDirectSeveral} for its causal meta-DAG.).  
This addition would, for example, 
allow for the in-degrees to also exhibit a power-law.  
Moreover, when both $\OldName$ and $\PathName$ arrows are present, 
it is possible to promote clustering via triadic closure, 
as the similarity between the connections that nodes $i$ and $j$ make with the ``distant'' nodes \mbox{$\BoxNodeS{}<i<j$} can influence the likelihood that $i$ and $j$ themselves form a connection.  

\paragraph{A ``bottom-up'' causality  --- $\HubName+\NewName$.} 
The {\ppaAcron} model and its extension including $\OldName$ have a ``top-down'' sort of causal structure, 
with dyads containing older nodes influencing dyads containing newer nodes. 
Conversely, the causal model with $\HubName$ and $\NewName$ depends on the other ``quadrant'' of dyads, 
and instead has a sort of ``bottom-up'' sort of causal structure 
(see its causal meta-DAG at the top-right of Fig.~\ref{Fig:ExampleGraphicalModelLocalDirectSeveral}).  
That is, the dyads containing nodes that are \textit{closer together} in the ordering 
influence the outcomes of dyads containing nodes that are \textit{further apart}.  
This causal meta-DAG could be useful for modeling ``local'' clustering  
between nodes that occur at similar times.  


\spaceendsubsection
\subsection{Sparse composable structural equations}
\label{sec:DorPAModel}
\spacebefsubsection
As previously mentioned, the {$\OldName$} and {$\NewName$} causal arrows cannot be included in the same meta-DAG.  
Here we propose a way to do essentially that.  

The choice of an affine structural equation in the {\ppaAcron} model was motivated in part by its similarity to other classical growing models, such as Pólya's urn \citep{eggenberger1923statistik,mahmoud2008polya}, the process of Pitman-Yor \citep{pitman1997two}, and various Canonical Restaurant processes \citep{aldous2006ecole,ghahramani2005infinite}.  

Here is another option for \Cref{eq:PPAprob}, with ellipses to suggest its straightforward generalization: 
\begin{align}
    p_{ij}^{ } &= 1 - \exp\!\bigg( - \frac{\alphaP + \thetain d_i^{\text{in}} + \thetaout d_i^{\text{out}} + \cdots}{j+\betaP}\bigg) \label{eq:DORPAprob}
\end{align}
This simple transformation of the affine model retains all the asymptotics of the original {\ppaAcron} model  
(since \mbox{\smash{$1-\exp(-p_{ij})\approx p_{ij}$}} for \mbox{$p_{ij}\ll1$}).  
But now the inclusion of additional terms such as \mbox{\smash{$\thetaold d^{\text{old}}$}} predictably increase the probability of an edge 
(without becoming greater than $1$).  
Also, different choices for the denominators might allow for better modeling of growing networks over a wide range of scales.  

However, the most compelling property of \Cref{eq:DORPAprob} is that it allows one to effectively sample from causal models that apparently have cycles!  
Moreover, the algorithm is naturally asynchronous, and exploits the sparsity of the resulting network.  

Here is a sketch of the algorithm.  
Initialize all dyads in the graph as ``empty''.  
Add an edge independently for each dyad variable with probability \mbox{$\exp\!\big(-\frac{\alphaP}{j+\betaP}\big)$}, and set those dyads as ``active''.  
Iteratively take an active dyad, set it as ``completed'', 
and sample its {$\HubName$} children independently with probability \mbox{$\exp(-\frac{\thetain}{j+\betaP})$}, 
and likewise for {$\PathName$} children.  
For any sampled children dyads that are currently ``empty'', 
add that edge to the graph and set those dyads as ``active''.  
Continue until there are no ``active'' dyads.  

This perspective allows for directed cycles between the dyads, 
essentially ``fine-graining'' them into a series of back-and-forth communication, 
while the acyclic transitions \mbox{$\text{``empty''}\rightarrow\text{``active''}\rightarrow\text{``completed''}$} ensure the algorithm will terminate for a finite graph.  
Decomposing the behavior of complex interconnected networks into a elemental ``event-based'' activity appears to be a promising direction for future study.

\bibliography{refs}

\begin{thebibliography}{46}
\providecommand{\natexlab}[1]{#1}
\providecommand{\url}[1]{\texttt{#1}}
\expandafter\ifx\csname urlstyle\endcsname\relax
  \providecommand{\doi}[1]{doi: #1}\else
  \providecommand{\doi}{doi: \begingroup \urlstyle{rm}\Url}\fi

\bibitem[Adamic et~al.(2017)Adamic, Brunetti, Harris, and Kirilenko]{adamic2017trading}
Lada Adamic, Celso Brunetti, Jeffrey~H Harris, and Andrei Kirilenko.
\newblock Trading networks.
\newblock \emph{The Econometrics Journal}, 20\penalty0 (3):\penalty0 S126--S149, 2017.

\bibitem[Aldous et~al.(2006)Aldous, Ibragimov, and Jacod]{aldous2006ecole}
David~J Aldous, Illdar~A Ibragimov, and Jean Jacod.
\newblock \emph{Ecole d'Ete de Probabilites de Saint-Flour XIII, 1983}, volume 1117.
\newblock Springer, 2006.

\bibitem[Arnold et~al.(2024)Arnold, Zhong, Ba, Steer, Mondragon, Cuadrado, Lambiotte, and Clegg]{arnold2024insights}
Naomi~A Arnold, Peijie Zhong, Cheick~Tidiane Ba, Ben Steer, Raul Mondragon, Felix Cuadrado, Renaud Lambiotte, and Richard~G Clegg.
\newblock Insights and caveats from mining local and global temporal motifs in cryptocurrency transaction networks.
\newblock \emph{Scientific Reports}, 14\penalty0 (1):\penalty0 26569, 2024.

\bibitem[Avin et~al.(2020)Avin, Daltrophe, Keller, Lotker, Mathieu, Peleg, and Pignolet]{avin2020mixed}
Chen Avin, Hadassa Daltrophe, Barbara Keller, Zvi Lotker, Claire Mathieu, David Peleg, and Yvonne-Anne Pignolet.
\newblock Mixed preferential attachment model: Homophily and minorities in social networks.
\newblock \emph{Physica A: Statistical Mechanics and its Applications}, 555:\penalty0 124723, 2020.

\bibitem[Barab{\'a}si and Albert(1999)]{barabasi1999emergence}
Albert-L{\'a}szl{\'o} Barab{\'a}si and R{\'e}ka Albert.
\newblock Emergence of scaling in random networks.
\newblock \emph{Science}, 286\penalty0 (5439):\penalty0 509--512, 1999.

\bibitem[Bianconi and Barab{\'a}si(2001)]{bianconi2001competition}
Ginestra Bianconi and A-L Barab{\'a}si.
\newblock Competition and multiscaling in evolving networks.
\newblock \emph{Europhysics letters}, 54\penalty0 (4):\penalty0 436, 2001.

\bibitem[Bollob{\'a}s et~al.(2007)Bollob{\'a}s, Janson, and Riordan]{bollobas2007phase}
B{\'e}la Bollob{\'a}s, Svante Janson, and Oliver Riordan.
\newblock The phase transition in inhomogeneous random graphs.
\newblock \emph{Random Structures \& Algorithms}, 31\penalty0 (1):\penalty0 3--122, 2007.

\bibitem[Bravo-Hermsdorff(2023)]{bravo2023quantifyinghuman}
Gecia Bravo-Hermsdorff.
\newblock Quantifying human priors over social and navigation networks.
\newblock In \emph{International Conference on Machine Learning}, pages 3063--3105. PMLR, 2023.

\bibitem[Bravo-Hermsdorff et~al.(2024)Bravo-Hermsdorff, Watson, Yu, Zeitler, and Silva]{bravo2024intervention}
Gecia Bravo-Hermsdorff, David Watson, Jialin Yu, Jakob Zeitler, and Ricardo Silva.
\newblock Intervention generalization: A view from factor graph models.
\newblock \emph{Advances in Neural Information Processing Systems}, 36, 2024.

\bibitem[Chazelle(2012)]{chazelle2012dynamics}
Bernard Chazelle.
\newblock The dynamics of influence systems.
\newblock In \emph{2012 IEEE 53rd Annual Symposium on Foundations of Computer Science}, pages 311--320. IEEE, 2012.

\bibitem[Chazelle and Wang(2019)]{chazelle2019iterated}
Bernard Chazelle and Chu Wang.
\newblock Iterated learning in dynamic social networks.
\newblock \emph{Journal of Machine Learning Research}, 20\penalty0 (29):\penalty0 1--28, 2019.

\bibitem[Clauset et~al.(2009)Clauset, Shalizi, and Newman]{clauset2009power}
Aaron Clauset, Cosma~Rohilla Shalizi, and Mark~EJ Newman.
\newblock Power-law distributions in empirical data.
\newblock \emph{SIAM review}, 51\penalty0 (4):\penalty0 661--703, 2009.

\bibitem[Coecke(2023)]{coecke2023compositionality}
Bob Coecke.
\newblock Compositionality as we see it, everywhere around us.
\newblock In \emph{The Quantum-Like Revolution: A Festschrift for Andrei Khrennikov}, pages 247--267. Springer, 2023.

\bibitem[Deutsch(2011)]{deutsch2011beginning}
David Deutsch.
\newblock \emph{The beginning of infinity: Explanations that transform the world}.
\newblock penguin uK, 2011.

\bibitem[Eggenberger and P{\'o}lya(1923)]{eggenberger1923statistik}
Florian Eggenberger and George P{\'o}lya.
\newblock {\"U}ber die statistik verketteter vorg{\"a}nge.
\newblock \emph{ZAMM-Journal of Applied Mathematics and Mechanics/Zeitschrift f{\"u}r Angewandte Mathematik und Mechanik}, 3\penalty0 (4):\penalty0 279--289, 1923.

\bibitem[Eikmeier and Gleich(2019)]{eikmeier2019triangle}
Nicole Eikmeier and David~F Gleich.
\newblock Triangle preferential attachment has power-law degrees and eigenvalues; eigenvalues are more stable to network sampling.
\newblock \emph{arXiv preprint arXiv:1904.12989}, 2019.

\bibitem[Ghahramani and Griffiths(2005)]{ghahramani2005infinite}
Zoubin Ghahramani and Thomas Griffiths.
\newblock Infinite latent feature models and the indian buffet process.
\newblock \emph{Advances in neural information processing systems}, 18, 2005.

\bibitem[Goglia and Vega(2024)]{goglia2024structure}
Diletta Goglia and Davide Vega.
\newblock Structure and dynamics of growing networks of reddit threads.
\newblock \emph{Applied Network Science}, 9\penalty0 (1):\penalty0 48, 2024.

\bibitem[Gunderson et~al.(2024)Gunderson, Bravo-Hermsdorff, and Orbanz]{gunderson2024graph}
Lee Gunderson, Gecia Bravo-Hermsdorff, and Peter Orbanz.
\newblock The graph pencil method: mapping subgraph densities to stochastic block models.
\newblock \emph{Advances in Neural Information Processing Systems}, 36, 2024.

\bibitem[Harris(2013)]{harris2013introduction}
Jenine~K Harris.
\newblock \emph{An introduction to exponential random graph modeling}, volume 173.
\newblock Sage Publications, 2013.

\bibitem[Hofstadter(1999)]{hofstadter1999godel}
Douglas~R Hofstadter.
\newblock \emph{G{\"o}del, Escher, Bach: an eternal golden braid}.
\newblock Basic books, 1999.

\bibitem[Krapivsky et~al.(2000)Krapivsky, Redner, and Leyvraz]{krapivsky2000connectivity}
Paul~L Krapivsky, Sidney Redner, and Francois Leyvraz.
\newblock Connectivity of growing random networks.
\newblock \emph{Physical Review Letters}, 85\penalty0 (21):\penalty0 4629, 2000.

\bibitem[Lauritzen et~al.(2018)Lauritzen, Rinaldo, and Sadeghi]{lauritzen2018random}
Steffen Lauritzen, Alessandro Rinaldo, and Kayvan Sadeghi.
\newblock Random networks, graphical models and exchangeability.
\newblock \emph{Journal of the Royal Statistical Society Series B: Statistical Methodology}, 80\penalty0 (3):\penalty0 481--508, 2018.

\bibitem[Lauritzen(1996)]{lauritzen1996graphical}
Steffen~L Lauritzen.
\newblock \emph{Graphical models}, volume~17.
\newblock Clarendon Press, 1996.

\bibitem[Lee et~al.(2015)Lee, Zaheer, G{\"u}nnemann, and Smola]{lee2015preferential}
Jay Lee, Manzil Zaheer, Stephan G{\"u}nnemann, and Alex Smola.
\newblock Preferential attachment in graphs with affinities.
\newblock In \emph{Artificial Intelligence and Statistics}, pages 571--580. PMLR, 2015.

\bibitem[Lorenz and Tull(2023)]{lorenz2023causal}
Robin Lorenz and Sean Tull.
\newblock Causal models in string diagrams.
\newblock \emph{arXiv preprint arXiv:2304.07638}, 2023.

\bibitem[Lov{\'a}sz(2012)]{lovasz2012large}
L{\'a}szl{\'o} Lov{\'a}sz.
\newblock \emph{Large networks and graph limits}, volume~60.
\newblock American Mathematical Soc., 2012.

\bibitem[Mahmoud(2008)]{mahmoud2008polya}
Hosam Mahmoud.
\newblock \emph{P{\'o}lya urn models}.
\newblock Chapman and Hall/CRC, 2008.

\bibitem[Mazur(2008)]{mazur2008one}
Barry Mazur.
\newblock When is one thing equal to some other thing.
\newblock \emph{Proof and other dilemmas: Mathematics and philosophy}, pages 221--241, 2008.

\bibitem[Orbanz(2017)]{orbanz2017subsampling}
Peter Orbanz.
\newblock Subsampling large graphs and invariance in networks.
\newblock \emph{arXiv:1710.04217}, 2017.

\bibitem[Orbanz and Roy(2014)]{orbanz2014bayesian}
Peter Orbanz and Daniel~M Roy.
\newblock Bayesian models of graphs, arrays and other exchangeable random structures.
\newblock \emph{IEEE Transactions on Pattern Analysis and Machine Intelligence}, 37\penalty0 (2):\penalty0 437--461, 2014.

\bibitem[Pearl(1994)]{pearl1994probabilistic}
Judea Pearl.
\newblock A probabilistic calculus of actions.
\newblock In \emph{Uncertainty in Artificial Intelligence}, pages 454--462. Elsevier, 1994.

\bibitem[Pearl(2009)]{pearl2009causality}
Judea Pearl.
\newblock \emph{Causality}.
\newblock Cambridge university press, 2009.

\bibitem[Pek{\"o}z et~al.(2029)Pek{\"o}z, R{\"o}llin, and Ross]{pekoz2019polya}
Erol Pek{\"o}z, Adrian R{\"o}llin, and Nathan Ross.
\newblock P{\'o}lya urns with immigration at random times.
\newblock \emph{Bernoulli}, 25\penalty0 (1):\penalty0 189--220, 2029.

\bibitem[Peters and Halpern(2021)]{peters2021causal}
Spencer Peters and Joseph~Y Halpern.
\newblock Causal modeling with infinitely many variables.
\newblock \emph{arXiv preprint arXiv:2112.09171}, 2021.

\bibitem[Pitman and Yor(1997)]{pitman1997two}
Jim Pitman and Marc Yor.
\newblock The two-parameter poisson-dirichlet distribution derived from a stable subordinator.
\newblock \emph{The Annals of Probability}, pages 855--900, 1997.

\bibitem[P{\'o}sfai and Barab{\'a}si(2016)]{posfai2016network}
M{\'a}rton P{\'o}sfai and Albert-L{\'a}szl{\'o} Barab{\'a}si.
\newblock \emph{Network science}.
\newblock Citeseer, 2016.

\bibitem[Price(1965)]{price1965networks}
Derek J De~Solla Price.
\newblock Networks of scientific papers: The pattern of bibliographic references indicates the nature of the scientific research front.
\newblock \emph{Science}, 149\penalty0 (3683):\penalty0 510--515, 1965.

\bibitem[Radicchi et~al.(2011)Radicchi, Fortunato, and Vespignani]{radicchi2011citation}
Filippo Radicchi, Santo Fortunato, and Alessandro Vespignani.
\newblock Citation networks.
\newblock \emph{Models of science dynamics: Encounters between complexity theory and information sciences}, pages 233--257, 2011.

\bibitem[Ray(2024)]{ray2024stochasticthesis}
Rounak Ray.
\newblock \emph{Stochastic processes on preferential attachment models}.
\newblock Phd thesis, Eindhoven University of Technology, 2024.
\newblock Available at \url{https://arxiv.org/abs/2411.14111}.

\bibitem[Richardson and Spirtes(2002)]{richardson2002ancestral}
Thomas Richardson and Peter Spirtes.
\newblock Ancestral graph markov models.
\newblock \emph{The Annals of Statistics}, 30\penalty0 (4):\penalty0 962--1030, 2002.

\bibitem[Rovelli(2021)]{Rovelli2021-ROVHMS}
Carlo Rovelli.
\newblock \emph{Helgoland: Making Sense of the Quantum Revolution}.
\newblock Riverhead Books, New York, 2021.

\bibitem[Simon(1955)]{simon1955class}
Herbert~A Simon.
\newblock On a class of skew distribution functions.
\newblock \emph{Biometrika}, 42\penalty0 (3/4):\penalty0 425--440, 1955.

\bibitem[Van Der~Hofstad(2024)]{van2024random}
Remco Van Der~Hofstad.
\newblock \emph{Random graphs and complex networks}, volume~54.
\newblock Cambridge university press, 2024.

\bibitem[Villar et~al.(2023)Villar, Hogg, Yao, Kevrekidis, and Sch{\"o}lkopf]{villar2023towards}
Soledad Villar, David~W Hogg, Weichi Yao, George~A Kevrekidis, and Bernhard Sch{\"o}lkopf.
\newblock Towards fully covariant machine learning.
\newblock \emph{arXiv preprint arXiv:2301.13724}, 2023.

\bibitem[Wang and Resnick(2020)]{wang2020directed}
Tiandong Wang and Sidney~I Resnick.
\newblock A directed preferential attachment model with poisson measurement.
\newblock \emph{arXiv preprint arXiv:2008.07005}, 2020.

\end{thebibliography}

\onecolumn

\title{
\paperTitleAppendix 
}

\appendix


\spacestartappendixsection
\section{Examples of Causal Meta-DAGs}
\label{appendix:AppendixMetaDAGS}


\potentiallyImprove{In this section, 
we display the causal meta-DAGs associated with the various types of causal arrows (defined in \refTypeCausalArrows) for a growing network with five nodes, where the $X_{ij}$ represent the dyad variables.  
Bear in mind that these structures continue indefinitely for networks with any number of nodes.}  

\newcommand{\spaceDAGfivefig}{\vspace{20pt}}
\newcommand{\hspaceDAGfivefig}{\hspace{120pt}}
\newcommand{\spaceDAGfivefigcaption}{\vspace{5pt}}

 \newcommand{\spaceaftercaptionthree}{\vspace{4pt}}
\newcommand{\spaceaftercaptiontwo}{\vspace{0pt}}
 \newcommand{\spaceaftercaptionone}{\vspace{4pt}}

\newcommand{\arrowlength}{2mm}
\newcommand{\arrowwidth}{2mm}
\newcommand{\nodesDistanceDAGs}{14.5mm}
\newcommand{\verticalDistanceDAGFig}{14.5mm}
\newcommand{\nodeSizeDAGs}{4.0mm}
\newcommand{\nodetextSeparationDAGs}{-0.5mm}

\newcommand{\dyadname}[1]{\small{$X_{#1}$}}
\tikzset{square/.style={regular polygon,regular polygon sides=4,font=\sffamily}}
\newcommand{\nodeshape}{square} 


\pgfkeys{
  /nodeListX/.initial={ }, 
  /nodeListX/append/.code args={#1}{%
    \pgfkeysgetvalue{/nodeListX}\currentlist%
    \ifx\currentlist\empty
      \pgfkeyssetvalue{/nodeListX}{#1}%
    \else
      \pgfkeyssetvalue{/nodeListX}{\currentlist,#1}%
    \fi
  }
}
\pgfkeys{
  /nodeListY/.initial={ }, 
  /nodeListY/append/.code args={#1}{%
    \pgfkeysgetvalue{/nodeListY}\currentlist%
    \ifx\currentlist\empty
      \pgfkeyssetvalue{/nodeListY}{#1}%
    \else
      \pgfkeyssetvalue{/nodeListY}{\currentlist,#1}%
    \fi
  }
}

\newcommand{\verticalspacingnewdags}{\vspace{40pt}}
\newcommand{\horizontalspacingnewdags}{\hspace{42pt}}
      \def\NewNodeSize{12pt}
    \tikzmath{\NodeRowDelta=2.0;}
    \tikzmath{\NodeColDelta=2.0;}

    \def\DyadTextSize{12}
    \def\DyadBoxSize{0.9cm}
    \def\DyadBoxBorderThickness{1pt} 
    \newcommand{\arrowlengthh}{3mm}
    \newcommand{\arrowwidthh}{3mm}
    \def\ArrowLineWidth{1.5pt}
    
\def\retangleCoverPosX{-0.1}
\def\retangleCoverPosY{1.9}
\def\retangleCoverDimX{7.5}
\def\retangleCoverDimY{9.5}
\def\retangleBoxColor{white}

\vspace{30pt}

\begin{figure}[H]
  \centering
  \begin{tikzpicture}[remember picture]
    \draw[draw=\retangleBoxColor, fill=none] (\retangleCoverPosX,\retangleCoverPosY) rectangle (\retangleCoverDimX,\retangleCoverDimY);

    \tikzmath{\xStart=1.0;\yStart=11.0;}
    \foreach [count=\j from 2] \dyadY in {2,...,5} {
        \tikzmath{\iMax=\j-1;}
        \foreach [count=\i from 1] \dyadX in {1,...,\iMax} {
            \node (dyad\i\j) [draw=white,fill=none,inner sep=0cm, regular polygon sides=4, minimum size=\DyadBoxSize, line width=\DyadBoxBorderThickness] at (\xStart+\NodeColDelta*\i-\NodeColDelta,\yStart-\NodeRowDelta*\j+\NodeRowDelta) { };
        };
    };

    \tikzmath{\AngleDiagOffset=45;}
    \tikzmath{\NearOutAngle=180+\AngleDiagOffset;\NearInAngle=\AngleDiagOffset;}
    \foreach \i in {2,...,3} {
        \tikzmath{\jMin=int(\i+1);\jMax=int(4);};
        \foreach \j in {\jMin,...,\jMax} {
            \tikzmath{\kMin=int(1);\kMax=int(\i-1);};
            \foreach \k in {\kMin,...,\kMax} {
                \tikzmath{\lMin=int(\j+1);\lMax=int(5);};
                \foreach \l in {\lMin,...,\lMax} {
                    \draw[->,line width=\ArrowLineWidth, >={Stealth[length=\arrowlengthh, width=\arrowwidthh]}, color=NonLocalInteriorColor,opacity=0,  shorten <= -0.25mm] (dyad\i\j)  to[out=\NearOutAngle, in=\NearInAngle] (dyad\k\l); 
                };
            };
        };
    };

    \tikzmath{\AngleDiagOffset=0;}
    \tikzmath{\FarOutAngle=0-\AngleDiagOffset;\FarInAngle=90+\AngleDiagOffset;}
    \foreach \i in {1,...,2} {
        \tikzmath{\jMin=int(\i+1);\jMax=int(3);};
        \foreach \j in {\jMin,...,\jMax} {
            \tikzmath{\kMin=int(\j+1);\kMax=int(5-1);};
            \foreach \k in {\kMin,...,\kMax} {
                \tikzmath{\lMin=int(\k+1);\lMax=int(5);};
                \foreach \l in {\lMin,...,\lMax} {
                    \draw[->,line width=\ArrowLineWidth, >={Stealth[length=\arrowlengthh, width=\arrowwidthh]}, color=NonLocalExteriorColor, opacity=0,  shorten <= -0.25mm] (dyad\i\j)  to[out=\FarOutAngle, in=\FarInAngle] (dyad\k\l); 
                };
            };
        };
    };
    \tikzmath{\AngleDiagOffset=30;}
    \tikzmath{\HubOutAngle=270-\AngleDiagOffset;\HubInAngle=90+\AngleDiagOffset;}
    \foreach \i in {1,...,3} {
        \tikzmath{\jMin=int(\i+1);\jMax=4;};
        \foreach \j in {\jMin,...,\jMax} {
            \tikzmath{\kMin=int(\j+1);\kMax=5;};
            \foreach \k in {\kMin,...,\kMax} {
                \draw[->,line width=\ArrowLineWidth, >={Stealth[length=\arrowlengthh, width=\arrowwidthh]}, color=HubColor, shorten <= -0.25mm] (dyad\i\j)  to[out=\HubOutAngle, in=\HubInAngle] (dyad\i\k); 
            };
        };
    };
    \tikzmath{\AngleDiagOffset=30;}
    \tikzmath{\ForwardOutAngle=-\AngleDiagOffset;\ForwardInAngle=180+\AngleDiagOffset;}
    \foreach \i in {3,...,5} {
        \tikzmath{\jMin=int(1);\jMax=int(\i-2);};
        \foreach \j in {\jMin,...,\jMax} {
            \tikzmath{\kMin=int(\j+1);\kMax=int(\i-1);};
            \foreach \k in {\kMin,...,\kMax} {
                \draw[->,line width=\ArrowLineWidth, >={Stealth[length=\arrowlengthh, width=\arrowwidthh]}, color=ForwardColor, opacity=0, shorten <= -0.25mm] (dyad\j\i)  to[out=\ForwardOutAngle, in=\ForwardInAngle] (dyad\k\i); 
            };
        };
    };
    \tikzmath{\AngleDiagOffset=30;}
    \tikzmath{\PathOutAngle=270+\AngleDiagOffset;\PathInAngle=180-\AngleDiagOffset;}
    \foreach \i in {1,...,3} {
        \tikzmath{\jMin=int(\i+1);\jMax=int(4);};
        \foreach \j in {\jMin,...,\jMax} {
            \tikzmath{\kMin=int(\j+1);\kMax=int(5);};
            \foreach \k in {\kMin,...,\kMax} {
                \draw[->,line width=\ArrowLineWidth, >={Stealth[length=\arrowlengthh, width=\arrowwidthh]}, color=TransitiveColor, opacity=0, shorten <= -0.25mm] (dyad\i\j)  to[out=\PathOutAngle, in=\PathInAngle] (dyad\j\k); 
            };
        };
    };

 \tikzmath{\AngleDiagOffset=60;}
    \tikzmath{\MidOutAngle=0-\AngleDiagOffset;\MidInAngle=90+\AngleDiagOffset;}
        \tikzmath{\AngleDiagOffset=60;}
    \tikzmath{\MidOutAngle=0-\AngleDiagOffset;\MidInAngle=90+\AngleDiagOffset;}
    \foreach \i in {1,...,2} {
        \tikzmath{\jMin=int(\i+2);\jMax=int(4);};
        \foreach \j in {\jMin,...,\jMax} {
            \tikzmath{\kMin=int(\i+1);\kMax=int(\j-1);};
            \foreach \k in {\kMin,...,\kMax} {
                \tikzmath{\lMin=int(\j+1);\lMax=int(5);};
                \foreach \l in {\lMin,...,\lMax} {
                    \draw[->,line width=\ArrowLineWidth, >={Stealth[length=\arrowlengthh, width=\arrowwidthh]}, color=NonLocalInterfaceColor, opacity=0, shorten <= -0.25mm] (dyad\i\j)  to[out=\MidOutAngle, in=\MidInAngle] (dyad\k\l); 
                };
            };
        };
    };
    \foreach [count=\j from 2] \dyadY in {2,...,5} {
        \tikzmath{\iMax=\j-1;}
        \foreach [count=\i from 1] \dyadX in {1,...,\iMax} {
            \node (dyad\i\j) [draw=black,fill=white,inner sep=0cm, regular polygon sides=4, minimum size=\DyadBoxSize, line width=\DyadBoxBorderThickness,font=\fontsize{\DyadTextSize}{\DyadTextSize}\selectfont] at (\xStart+\NodeColDelta*\i-\NodeColDelta,\yStart-\NodeRowDelta*\j+\NodeRowDelta) {$X_{\i\j}^{ }$};
        };
    };
  \end{tikzpicture}
  \horizontalspacingnewdags
    \begin{tikzpicture}[remember picture]
        \draw[draw=\retangleBoxColor, fill=none] (\retangleCoverPosX,\retangleCoverPosY) rectangle (\retangleCoverDimX,\retangleCoverDimY);

    \tikzmath{\xStart=1.0;\yStart=11.0;}
    \foreach [count=\j from 2] \dyadY in {2,...,5} {
        \tikzmath{\iMax=\j-1;}
        \foreach [count=\i from 1] \dyadX in {1,...,\iMax} {
            \node (dyad\i\j) [draw=white,fill=none,inner sep=0cm, regular polygon sides=4, minimum size=\DyadBoxSize, line width=\DyadBoxBorderThickness] at (\xStart+\NodeColDelta*\i-\NodeColDelta,\yStart-\NodeRowDelta*\j+\NodeRowDelta) { };
        };
    };

    \tikzmath{\AngleDiagOffset=45;}
    \tikzmath{\NearOutAngle=180+\AngleDiagOffset;\NearInAngle=\AngleDiagOffset;}
    \foreach \i in {2,...,3} {
        \tikzmath{\jMin=int(\i+1);\jMax=int(4);};
        \foreach \j in {\jMin,...,\jMax} {
            \tikzmath{\kMin=int(1);\kMax=int(\i-1);};
            \foreach \k in {\kMin,...,\kMax} {
                \tikzmath{\lMin=int(\j+1);\lMax=int(5);};
                \foreach \l in {\lMin,...,\lMax} {
                    \draw[->,line width=\ArrowLineWidth, >={Stealth[length=\arrowlengthh, width=\arrowwidthh]}, color=NonLocalInteriorColor, opacity=0, shorten <= -0.25mm] (dyad\i\j)  to[out=\NearOutAngle, in=\NearInAngle] (dyad\k\l); 
                };
            };
        };
    };

    \tikzmath{\AngleDiagOffset=0;}
    \tikzmath{\FarOutAngle=0-\AngleDiagOffset;\FarInAngle=90+\AngleDiagOffset;}
    \foreach \i in {1,...,2} {
        \tikzmath{\jMin=int(\i+1);\jMax=int(3);};
        \foreach \j in {\jMin,...,\jMax} {
            \tikzmath{\kMin=int(\j+1);\kMax=int(5-1);};
            \foreach \k in {\kMin,...,\kMax} {
                \tikzmath{\lMin=int(\k+1);\lMax=int(5);};
                \foreach \l in {\lMin,...,\lMax} {
                    \draw[->,line width=\ArrowLineWidth, >={Stealth[length=\arrowlengthh, width=\arrowwidthh]}, color=NonLocalExteriorColor, opacity=0, shorten <= -0.25mm] (dyad\i\j)  to[out=\FarOutAngle, in=\FarInAngle] (dyad\k\l); 
                };
            };
        };
    };
    \tikzmath{\AngleDiagOffset=30;}
    \tikzmath{\HubOutAngle=270-\AngleDiagOffset;\HubInAngle=90+\AngleDiagOffset;}
    \foreach \i in {1,...,3} {
        \tikzmath{\jMin=int(\i+1);\jMax=4;};
        \foreach \j in {\jMin,...,\jMax} {
            \tikzmath{\kMin=int(\j+1);\kMax=5;};
            \foreach \k in {\kMin,...,\kMax} {
                \draw[->,line width=\ArrowLineWidth, >={Stealth[length=\arrowlengthh, width=\arrowwidthh]}, color=HubColor, opacity=0, shorten <= -0.25mm] (dyad\i\j)  to[out=\HubOutAngle, in=\HubInAngle] (dyad\i\k); 
            };
        };
    };
    \tikzmath{\AngleDiagOffset=30;}
    \tikzmath{\ForwardOutAngle=-\AngleDiagOffset;\ForwardInAngle=180+\AngleDiagOffset;}
    \foreach \i in {3,...,5} {
        \tikzmath{\jMin=int(1);\jMax=int(\i-2);};
        \foreach \j in {\jMin,...,\jMax} {
            \tikzmath{\kMin=int(\j+1);\kMax=int(\i-1);};
            \foreach \k in {\kMin,...,\kMax} {
                \draw[->,line width=\ArrowLineWidth, >={Stealth[length=\arrowlengthh, width=\arrowwidthh]}, color=ForwardColor,opacity=0,  shorten <= -0.25mm] (dyad\j\i)  to[out=\ForwardOutAngle, in=\ForwardInAngle] (dyad\k\i); 
            };
        };
    };

    \tikzmath{\AngleDiagOffset=30;}
    \tikzmath{\PathOutAngle=270+\AngleDiagOffset;\PathInAngle=180-\AngleDiagOffset;}
    \foreach \i in {1,...,3} {
        \tikzmath{\jMin=int(\i+1);\jMax=int(4);};
        \foreach \j in {\jMin,...,\jMax} {
            \tikzmath{\kMin=int(\j+1);\kMax=int(5);};
            \foreach \k in {\kMin,...,\kMax} {
                \draw[->,line width=\ArrowLineWidth, >={Stealth[length=\arrowlengthh, width=\arrowwidthh]}, color=TransitiveColor, shorten <= -0.25mm] (dyad\i\j)  to[out=\PathOutAngle, in=\PathInAngle] (dyad\j\k); 
            };
        };
    };

 \tikzmath{\AngleDiagOffset=60;}
    \tikzmath{\MidOutAngle=0-\AngleDiagOffset;\MidInAngle=90+\AngleDiagOffset;}
        \tikzmath{\AngleDiagOffset=60;}
    \tikzmath{\MidOutAngle=0-\AngleDiagOffset;\MidInAngle=90+\AngleDiagOffset;}
    \foreach \i in {1,...,2} {
        \tikzmath{\jMin=int(\i+2);\jMax=int(4);};
        \foreach \j in {\jMin,...,\jMax} {
            \tikzmath{\kMin=int(\i+1);\kMax=int(\j-1);};
            \foreach \k in {\kMin,...,\kMax} {
                \tikzmath{\lMin=int(\j+1);\lMax=int(5);};
                \foreach \l in {\lMin,...,\lMax} {
                    \draw[->,line width=\ArrowLineWidth, >={Stealth[length=\arrowlengthh, width=\arrowwidthh]}, color=NonLocalInterfaceColor,opacity=0,  shorten <= -0.25mm] (dyad\i\j)  to[out=\MidOutAngle, in=\MidInAngle] (dyad\k\l); 
                };
            };
        };
    };

    \foreach [count=\j from 2] \dyadY in {2,...,5} {
        \tikzmath{\iMax=\j-1;}
        \foreach [count=\i from 1] \dyadX in {1,...,\iMax} {
            \node (dyad\i\j) [draw=black,fill=white,inner sep=0cm, regular polygon sides=4, minimum size=\DyadBoxSize, line width=\DyadBoxBorderThickness,font=\fontsize{\DyadTextSize}{\DyadTextSize}\selectfont] at (\xStart+\NodeColDelta*\i-\NodeColDelta,\yStart-\NodeRowDelta*\j+\NodeRowDelta) {$X_{\i\j}^{ }$};
        };
    };
  \end{tikzpicture}\\ \verticalspacingnewdags
  \begin{tikzpicture}[remember picture]
        \draw[draw=\retangleBoxColor, fill=none] (\retangleCoverPosX,\retangleCoverPosY) rectangle (\retangleCoverDimX,\retangleCoverDimY);

    \tikzmath{\xStart=1.0;\yStart=11.0;}
    \foreach [count=\j from 2] \dyadY in {2,...,5} {
        \tikzmath{\iMax=\j-1;}
        \foreach [count=\i from 1] \dyadX in {1,...,\iMax} {
            \node (dyad\i\j) [draw=white,fill=none,inner sep=0cm, regular polygon sides=4, minimum size=\DyadBoxSize, line width=\DyadBoxBorderThickness] at (\xStart+\NodeColDelta*\i-\NodeColDelta,\yStart-\NodeRowDelta*\j+\NodeRowDelta) { };
        };
    };

    \tikzmath{\AngleDiagOffset=45;}
    \tikzmath{\NearOutAngle=180+\AngleDiagOffset;\NearInAngle=\AngleDiagOffset;}
    \foreach \i in {2,...,3} {
        \tikzmath{\jMin=int(\i+1);\jMax=int(4);};
        \foreach \j in {\jMin,...,\jMax} {
            \tikzmath{\kMin=int(1);\kMax=int(\i-1);};
            \foreach \k in {\kMin,...,\kMax} {
                \tikzmath{\lMin=int(\j+1);\lMax=int(5);};
                \foreach \l in {\lMin,...,\lMax} {
                    \draw[->,line width=\ArrowLineWidth, >={Stealth[length=\arrowlengthh, width=\arrowwidthh]}, color=NonLocalInteriorColor, opacity=0, shorten <= -0.25mm] (dyad\i\j)  to[out=\NearOutAngle, in=\NearInAngle] (dyad\k\l); 
                };
            };
        };
    };

    \tikzmath{\AngleDiagOffset=0;}
    \tikzmath{\FarOutAngle=0-\AngleDiagOffset;\FarInAngle=90+\AngleDiagOffset;}
    \foreach \i in {1,...,2} {
        \tikzmath{\jMin=int(\i+1);\jMax=int(3);};
        \foreach \j in {\jMin,...,\jMax} {
            \tikzmath{\kMin=int(\j+1);\kMax=int(5-1);};
            \foreach \k in {\kMin,...,\kMax} {
                \tikzmath{\lMin=int(\k+1);\lMax=int(5);};
                \foreach \l in {\lMin,...,\lMax} {
                    \draw[->,line width=\ArrowLineWidth, >={Stealth[length=\arrowlengthh, width=\arrowwidthh]}, color=NonLocalExteriorColor, opacity=0, shorten <= -0.25mm] (dyad\i\j)  to[out=\FarOutAngle, in=\FarInAngle] (dyad\k\l); 
                };
            };
        };
    };
    \tikzmath{\AngleDiagOffset=30;}
    \tikzmath{\HubOutAngle=270-\AngleDiagOffset;\HubInAngle=90+\AngleDiagOffset;}
    \foreach \i in {1,...,3} {
        \tikzmath{\jMin=int(\i+1);\jMax=4;};
        \foreach \j in {\jMin,...,\jMax} {
            \tikzmath{\kMin=int(\j+1);\kMax=5;};
            \foreach \k in {\kMin,...,\kMax} {
                \draw[->,line width=\ArrowLineWidth, >={Stealth[length=\arrowlengthh, width=\arrowwidthh]}, color=HubColor, opacity=0, shorten <= -0.25mm] (dyad\i\j)  to[out=\HubOutAngle, in=\HubInAngle] (dyad\i\k); 
            };
        };
    };
    \tikzmath{\AngleDiagOffset=30;}
    \tikzmath{\ForwardOutAngle=-\AngleDiagOffset;\ForwardInAngle=180+\AngleDiagOffset;}
    \foreach \i in {3,...,5} {
        \tikzmath{\jMin=int(1);\jMax=int(\i-2);};
        \foreach \j in {\jMin,...,\jMax} {
            \tikzmath{\kMin=int(\j+1);\kMax=int(\i-1);};
            \foreach \k in {\kMin,...,\kMax} {
                \draw[->,line width=\ArrowLineWidth, >={Stealth[length=\arrowlengthh, width=\arrowwidthh]}, color=ForwardColor, shorten <= -0.25mm] (dyad\j\i)  to[out=\ForwardOutAngle, in=\ForwardInAngle] (dyad\k\i); 
            };
        };
    };
    \tikzmath{\AngleDiagOffset=30;}
    \tikzmath{\PathOutAngle=270+\AngleDiagOffset;\PathInAngle=180-\AngleDiagOffset;}
    \foreach \i in {1,...,3} {
        \tikzmath{\jMin=int(\i+1);\jMax=int(4);};
        \foreach \j in {\jMin,...,\jMax} {
            \tikzmath{\kMin=int(\j+1);\kMax=int(5);};
            \foreach \k in {\kMin,...,\kMax} {
                \draw[->,line width=\ArrowLineWidth, >={Stealth[length=\arrowlengthh, width=\arrowwidthh]}, color=TransitiveColor, opacity=0, shorten <= -0.25mm] (dyad\i\j)  to[out=\PathOutAngle, in=\PathInAngle] (dyad\j\k); 
            };
        };
    };

 \tikzmath{\AngleDiagOffset=60;}
    \tikzmath{\MidOutAngle=0-\AngleDiagOffset;\MidInAngle=90+\AngleDiagOffset;}
        \tikzmath{\AngleDiagOffset=60;}
    \tikzmath{\MidOutAngle=0-\AngleDiagOffset;\MidInAngle=90+\AngleDiagOffset;}
    \foreach \i in {1,...,2} {
        \tikzmath{\jMin=int(\i+2);\jMax=int(4);};
        \foreach \j in {\jMin,...,\jMax} {
            \tikzmath{\kMin=int(\i+1);\kMax=int(\j-1);};
            \foreach \k in {\kMin,...,\kMax} {
                \tikzmath{\lMin=int(\j+1);\lMax=int(5);};
                \foreach \l in {\lMin,...,\lMax} {
                    \draw[->,line width=\ArrowLineWidth, >={Stealth[length=\arrowlengthh, width=\arrowwidthh]}, color=NonLocalInterfaceColor, opacity=0, shorten <= -0.25mm] (dyad\i\j)  to[out=\MidOutAngle, in=\MidInAngle] (dyad\k\l); 
                };
            };
        };
    };

    \foreach [count=\j from 2] \dyadY in {2,...,5} {
        \tikzmath{\iMax=\j-1;}
        \foreach [count=\i from 1] \dyadX in {1,...,\iMax} {
            \node (dyad\i\j) [draw=black,fill=white,inner sep=0cm, regular polygon sides=4, minimum size=\DyadBoxSize, line width=\DyadBoxBorderThickness,font=\fontsize{\DyadTextSize}{\DyadTextSize}\selectfont] at (\xStart+\NodeColDelta*\i-\NodeColDelta,\yStart-\NodeRowDelta*\j+\NodeRowDelta) {$X_{\i\j}^{ }$};
        };
    };
  \end{tikzpicture}
  \horizontalspacingnewdags
    \begin{tikzpicture}[remember picture]
        \draw[draw=\retangleBoxColor, fill=none] (\retangleCoverPosX,\retangleCoverPosY) rectangle (\retangleCoverDimX,\retangleCoverDimY);

    \tikzmath{\xStart=1.0;\yStart=11.0;}
    \foreach [count=\j from 2] \dyadY in {2,...,5} {
        \tikzmath{\iMax=\j-1;}
        \foreach [count=\i from 1] \dyadX in {1,...,\iMax} {
            \node (dyad\i\j) [draw=white,fill=none,inner sep=0cm, regular polygon sides=4, minimum size=\DyadBoxSize, line width=\DyadBoxBorderThickness] at (\xStart+\NodeColDelta*\i-\NodeColDelta,\yStart-\NodeRowDelta*\j+\NodeRowDelta) { };
        };
    };

    \tikzmath{\AngleDiagOffset=45;}
    \tikzmath{\NearOutAngle=180+\AngleDiagOffset;\NearInAngle=\AngleDiagOffset;}
    \foreach \i in {2,...,3} {
        \tikzmath{\jMin=int(\i+1);\jMax=int(4);};
        \foreach \j in {\jMin,...,\jMax} {
            \tikzmath{\kMin=int(1);\kMax=int(\i-1);};
            \foreach \k in {\kMin,...,\kMax} {
                \tikzmath{\lMin=int(\j+1);\lMax=int(5);};
                \foreach \l in {\lMin,...,\lMax} {
                    \draw[->,line width=\ArrowLineWidth, >={Stealth[length=\arrowlengthh, width=\arrowwidthh]}, color=NonLocalInteriorColor, opacity=0, shorten <= -0.25mm] (dyad\i\j)  to[out=\NearOutAngle, in=\NearInAngle] (dyad\k\l); 
                };
            };
        };
    };

    \tikzmath{\AngleDiagOffset=0;}
    \tikzmath{\FarOutAngle=0-\AngleDiagOffset;\FarInAngle=90+\AngleDiagOffset;}
    \foreach \i in {1,...,2} {
        \tikzmath{\jMin=int(\i+1);\jMax=int(3);};
        \foreach \j in {\jMin,...,\jMax} {
            \tikzmath{\kMin=int(\j+1);\kMax=int(5-1);};
            \foreach \k in {\kMin,...,\kMax} {
                \tikzmath{\lMin=int(\k+1);\lMax=int(5);};
                \foreach \l in {\lMin,...,\lMax} {
                    \draw[->,line width=\ArrowLineWidth, >={Stealth[length=\arrowlengthh, width=\arrowwidthh]}, color=NonLocalExteriorColor,opacity=0,  shorten <= -0.25mm] (dyad\i\j)  to[out=\FarOutAngle, in=\FarInAngle] (dyad\k\l); 
                };
            };
        };
    };
    \tikzmath{\AngleDiagOffset=30;}
    \tikzmath{\HubOutAngle=270-\AngleDiagOffset;\HubInAngle=90+\AngleDiagOffset;}
    \foreach \i in {1,...,3} {
        \tikzmath{\jMin=int(\i+1);\jMax=4;};
        \foreach \j in {\jMin,...,\jMax} {
            \tikzmath{\kMin=int(\j+1);\kMax=5;};
            \foreach \k in {\kMin,...,\kMax} {
                \draw[->,line width=\ArrowLineWidth, >={Stealth[length=\arrowlengthh, width=\arrowwidthh]}, color=HubColor, opacity=0, shorten <= -0.25mm] (dyad\i\j)  to[out=\HubOutAngle, in=\HubInAngle] (dyad\i\k); 
            };
        };
    };
        \tikzmath{\AngleDiagOffset=30;}
    \tikzmath{\ForwardOutAngle=-\AngleDiagOffset;\ForwardInAngle=180+\AngleDiagOffset;}
    \foreach \i in {3,...,5} {
        \tikzmath{\jMin=int(1);\jMax=int(\i-2);};
        \foreach \j in {\jMin,...,\jMax} {
            \tikzmath{\kMin=int(\j+1);\kMax=int(\i-1);};
            \foreach \k in {\kMin,...,\kMax} {
                \draw[->,line width=\ArrowLineWidth, >={Stealth[length=\arrowlengthh, width=\arrowwidthh]}, color=BackwardColor, shorten <= -0.25mm] (dyad\k\i)  to[out=\ForwardInAngle, in=\ForwardOutAngle] (dyad\j\i); 
            };
        };
    };

    \tikzmath{\AngleDiagOffset=30;}
    \tikzmath{\PathOutAngle=270+\AngleDiagOffset;\PathInAngle=180-\AngleDiagOffset;}
    \foreach \i in {1,...,3} {
        \tikzmath{\jMin=int(\i+1);\jMax=int(4);};
        \foreach \j in {\jMin,...,\jMax} {
            \tikzmath{\kMin=int(\j+1);\kMax=int(5);};
            \foreach \k in {\kMin,...,\kMax} {
                \draw[->,line width=\ArrowLineWidth, >={Stealth[length=\arrowlengthh, width=\arrowwidthh]}, color=TransitiveColor, opacity=0, shorten <= -0.25mm] (dyad\i\j)  to[out=\PathOutAngle, in=\PathInAngle] (dyad\j\k); 
            };
        };
    };

   \tikzmath{\AngleDiagOffset=60;}
    \tikzmath{\MidOutAngle=0-\AngleDiagOffset;\MidInAngle=90+\AngleDiagOffset;}
        \tikzmath{\AngleDiagOffset=60;}
    \tikzmath{\MidOutAngle=0-\AngleDiagOffset;\MidInAngle=90+\AngleDiagOffset;}
    \foreach \i in {1,...,2} {
        \tikzmath{\jMin=int(\i+2);\jMax=int(4);};
        \foreach \j in {\jMin,...,\jMax} {
            \tikzmath{\kMin=int(\i+1);\kMax=int(\j-1);};
            \foreach \k in {\kMin,...,\kMax} {
                \tikzmath{\lMin=int(\j+1);\lMax=int(5);};
                \foreach \l in {\lMin,...,\lMax} {
                    \draw[->,line width=\ArrowLineWidth, >={Stealth[length=\arrowlengthh, width=\arrowwidthh]}, color=NonLocalInterfaceColor, opacity=0, shorten <= -0.25mm] (dyad\i\j)  to[out=\MidOutAngle, in=\MidInAngle] (dyad\k\l); 
                };
            };
        };
    };

    \foreach [count=\j from 2] \dyadY in {2,...,5} {
        \tikzmath{\iMax=\j-1;}
        \foreach [count=\i from 1] \dyadX in {1,...,\iMax} {
            \node (dyad\i\j) [draw=black,fill=white,inner sep=0cm, regular polygon sides=4, minimum size=\DyadBoxSize, line width=\DyadBoxBorderThickness,font=\fontsize{\DyadTextSize}{\DyadTextSize}\selectfont] at (\xStart+\NodeColDelta*\i-\NodeColDelta,\yStart-\NodeRowDelta*\j+\NodeRowDelta) {$X_{\i\j}^{ }$};
        };
    };
  \end{tikzpicture}
   \spaceaftercaptionone
  \caption{\textbf{Causal graphs for each of the types of causal arrows between dyads that share a node. 
}\\ 
Causal \mbox{meta-DAGs} between dyads of a growing network with $5$ nodes \potentiallyImprove{that are compatible with network models having the following types of causal arrows:} 
{\hub} (\textit{top-left}); 
{\transitive} (\textit{top-right}); 
{\forward} (\textit{bottom-left}); 
and  {\backward} (\textit{bottom-right}). 
}
\label{Fig:ExampleGraphicalModelLocalDirect}
\end{figure}

\newpage 

\spaceafterfivedyadsfig

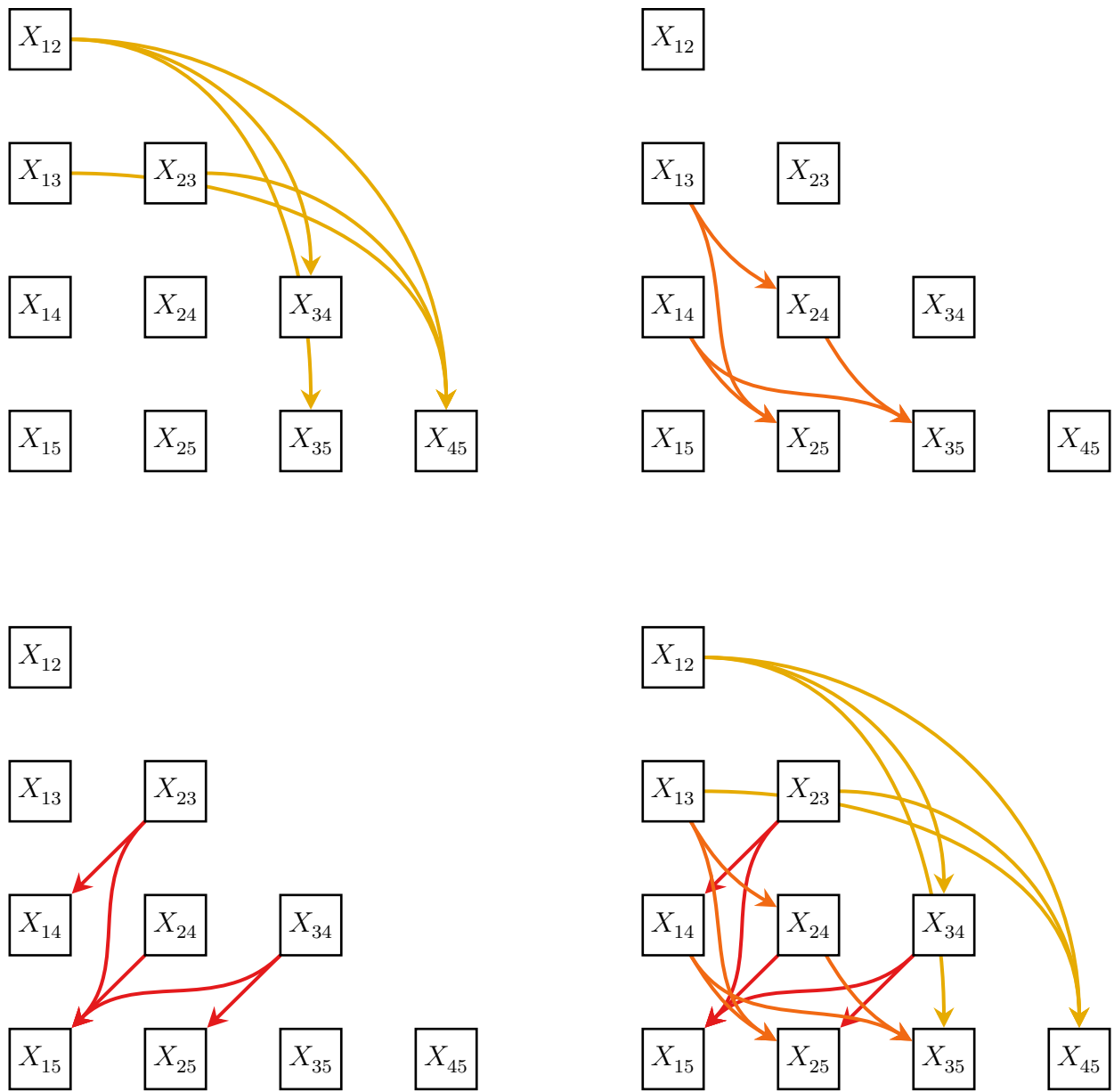
\begin{figure}[H]
  \centering
  \begin{tikzpicture}[remember picture]
    \draw[draw=\retangleBoxColor, fill=none] (\retangleCoverPosX,\retangleCoverPosY) rectangle (\retangleCoverDimX,\retangleCoverDimY);

    \tikzmath{\xStart=1.0;\yStart=11.0;}
    \foreach [count=\j from 2] \dyadY in {2,...,5} {
        \tikzmath{\iMax=\j-1;}
        \foreach [count=\i from 1] \dyadX in {1,...,\iMax} {
            \node (dyad\i\j) [draw=white,fill=none,inner sep=0cm, regular polygon sides=4, minimum size=\DyadBoxSize, line width=\DyadBoxBorderThickness] at (\xStart+\NodeColDelta*\i-\NodeColDelta,\yStart-\NodeRowDelta*\j+\NodeRowDelta) { };
        };
    };

    \tikzmath{\AngleDiagOffset=45;}
    \tikzmath{\NearOutAngle=180+\AngleDiagOffset;\NearInAngle=\AngleDiagOffset;}
    \foreach \i in {2,...,3} {
        \tikzmath{\jMin=int(\i+1);\jMax=int(4);};
        \foreach \j in {\jMin,...,\jMax} {
            \tikzmath{\kMin=int(1);\kMax=int(\i-1);};
            \foreach \k in {\kMin,...,\kMax} {
                \tikzmath{\lMin=int(\j+1);\lMax=int(5);};
                \foreach \l in {\lMin,...,\lMax} {
                    \draw[->,line width=\ArrowLineWidth, >={Stealth[length=\arrowlengthh, width=\arrowwidthh]}, color=white, shorten <= -0.25mm] (dyad\i\j)  to[out=\NearOutAngle, in=\NearInAngle] (dyad\k\l); 
                };
            };
        };
    };

    \tikzmath{\AngleDiagOffset=0;}
    \tikzmath{\FarOutAngle=0-\AngleDiagOffset;\FarInAngle=90+\AngleDiagOffset;}
    \foreach \i in {1,...,2} {
        \tikzmath{\jMin=int(\i+1);\jMax=int(3);};
        \foreach \j in {\jMin,...,\jMax} {
            \tikzmath{\kMin=int(\j+1);\kMax=int(5-1);};
            \foreach \k in {\kMin,...,\kMax} {
                \tikzmath{\lMin=int(\k+1);\lMax=int(5);};
                \foreach \l in {\lMin,...,\lMax} {
                    \draw[->,line width=\ArrowLineWidth, >={Stealth[length=\arrowlengthh, width=\arrowwidthh]}, color=NonLocalExteriorColor, shorten <= -0.25mm] (dyad\i\j)  to[out=\FarOutAngle, in=\FarInAngle] (dyad\k\l); 
                };
            };
        };
    };
    \tikzmath{\AngleDiagOffset=30;}
    \tikzmath{\HubOutAngle=270-\AngleDiagOffset;\HubInAngle=90+\AngleDiagOffset;}
    \foreach \i in {1,...,3} {
        \tikzmath{\jMin=int(\i+1);\jMax=4;};
        \foreach \j in {\jMin,...,\jMax} {
            \tikzmath{\kMin=int(\j+1);\kMax=5;};
            \foreach \k in {\kMin,...,\kMax} {
                \draw[->,line width=\ArrowLineWidth, >={Stealth[length=\arrowlengthh, width=\arrowwidthh]}, color=white, opacity=0, shorten <= -0.25mm] (dyad\i\j)  to[out=\HubOutAngle, in=\HubInAngle] (dyad\i\k); 
            };
        };
    };
    \tikzmath{\AngleDiagOffset=30;}
    \tikzmath{\ForwardOutAngle=-\AngleDiagOffset;\ForwardInAngle=180+\AngleDiagOffset;}
    \foreach \i in {3,...,5} {
        \tikzmath{\jMin=int(1);\jMax=int(\i-2);};
        \foreach \j in {\jMin,...,\jMax} {
            \tikzmath{\kMin=int(\j+1);\kMax=int(\i-1);};
            \foreach \k in {\kMin,...,\kMax} {
                \draw[->,line width=\ArrowLineWidth, >={Stealth[length=\arrowlengthh, width=\arrowwidthh]}, color=white, opacity=0, shorten <= -0.25mm] (dyad\j\i)  to[out=\ForwardOutAngle, in=\ForwardInAngle] (dyad\k\i); 
            };
        };
    };

    \tikzmath{\AngleDiagOffset=30;}
    \tikzmath{\PathOutAngle=270+\AngleDiagOffset;\PathInAngle=180-\AngleDiagOffset;}
    \foreach \i in {1,...,3} {
        \tikzmath{\jMin=int(\i+1);\jMax=int(4);};
        \foreach \j in {\jMin,...,\jMax} {
            \tikzmath{\kMin=int(\j+1);\kMax=int(5);};
            \foreach \k in {\kMin,...,\kMax} {
                \draw[->,line width=\ArrowLineWidth, >={Stealth[length=\arrowlengthh, width=\arrowwidthh]}, color=white, opacity=0, shorten <= -0.25mm] (dyad\i\j)  to[out=\PathOutAngle, in=\PathInAngle] (dyad\j\k); 
            };
        };
    };

 \tikzmath{\AngleDiagOffset=60;}
    \tikzmath{\MidOutAngle=0-\AngleDiagOffset;\MidInAngle=90+\AngleDiagOffset;}
        \tikzmath{\AngleDiagOffset=60;}
    \tikzmath{\MidOutAngle=0-\AngleDiagOffset;\MidInAngle=90+\AngleDiagOffset;}
    \foreach \i in {1,...,2} {
        \tikzmath{\jMin=int(\i+2);\jMax=int(4);};
        \foreach \j in {\jMin,...,\jMax} {
            \tikzmath{\kMin=int(\i+1);\kMax=int(\j-1);};
            \foreach \k in {\kMin,...,\kMax} {
                \tikzmath{\lMin=int(\j+1);\lMax=int(5);};
                \foreach \l in {\lMin,...,\lMax} {
                    \draw[->,line width=\ArrowLineWidth, >={Stealth[length=\arrowlengthh, width=\arrowwidthh]}, color=white, opacity=0, shorten <= -0.25mm] (dyad\i\j)  to[out=\MidOutAngle, in=\MidInAngle] (dyad\k\l); 
                };
            };
        };
    };
    \foreach [count=\j from 2] \dyadY in {2,...,5} {
        \tikzmath{\iMax=\j-1;}
        \foreach [count=\i from 1] \dyadX in {1,...,\iMax} {
            \node (dyad\i\j) [draw=black,fill=white,inner sep=0cm, regular polygon sides=4, minimum size=\DyadBoxSize, line width=\DyadBoxBorderThickness,font=\fontsize{\DyadTextSize}{\DyadTextSize}\selectfont] at (\xStart+\NodeColDelta*\i-\NodeColDelta,\yStart-\NodeRowDelta*\j+\NodeRowDelta) {$X_{\i\j}^{ }$};
        };
    };
  \end{tikzpicture}
  \horizontalspacingnewdags
    \begin{tikzpicture}[remember picture]
      \draw[draw=\retangleBoxColor, fill=none] (\retangleCoverPosX,\retangleCoverPosY) rectangle (\retangleCoverDimX,\retangleCoverDimY);

    \tikzmath{\xStart=1.0;\yStart=11.0;}
    \foreach [count=\j from 2] \dyadY in {2,...,5} {
        \tikzmath{\iMax=\j-1;}
        \foreach [count=\i from 1] \dyadX in {1,...,\iMax} {
            \node (dyad\i\j) [draw=white,fill=none,inner sep=0cm, regular polygon sides=4, minimum size=\DyadBoxSize, line width=\DyadBoxBorderThickness] at (\xStart+\NodeColDelta*\i-\NodeColDelta,\yStart-\NodeRowDelta*\j+\NodeRowDelta) { };
        };
    };

    \tikzmath{\AngleDiagOffset=45;}
    \tikzmath{\NearOutAngle=180+\AngleDiagOffset;\NearInAngle=\AngleDiagOffset;}
    \foreach \i in {2,...,3} {
        \tikzmath{\jMin=int(\i+1);\jMax=int(4);};
        \foreach \j in {\jMin,...,\jMax} {
            \tikzmath{\kMin=int(1);\kMax=int(\i-1);};
            \foreach \k in {\kMin,...,\kMax} {
                \tikzmath{\lMin=int(\j+1);\lMax=int(5);};
                \foreach \l in {\lMin,...,\lMax} {
                    \draw[->,line width=\ArrowLineWidth, >={Stealth[length=\arrowlengthh, width=\arrowwidthh]}, color=NonLocalInteriorColor, opacity=0,  shorten <= -0.25mm] (dyad\i\j)  to[out=\NearOutAngle, in=\NearInAngle] (dyad\k\l); 
                };
            };
        };
    };

    \tikzmath{\AngleDiagOffset=0;}
    \tikzmath{\FarOutAngle=0-\AngleDiagOffset;\FarInAngle=90+\AngleDiagOffset;}
    \foreach \i in {1,...,2} {
        \tikzmath{\jMin=int(\i+1);\jMax=int(3);};
        \foreach \j in {\jMin,...,\jMax} {
            \tikzmath{\kMin=int(\j+1);\kMax=int(5-1);};
            \foreach \k in {\kMin,...,\kMax} {
                \tikzmath{\lMin=int(\k+1);\lMax=int(5);};
                \foreach \l in {\lMin,...,\lMax} {
                    \draw[->,line width=\ArrowLineWidth, >={Stealth[length=\arrowlengthh, width=\arrowwidthh]}, color=NonLocalExteriorColor,opacity=0,  shorten <= -0.25mm] (dyad\i\j)  to[out=\FarOutAngle, in=\FarInAngle] (dyad\k\l); 
                };
            };
        };
    };
    \tikzmath{\AngleDiagOffset=30;}
    \tikzmath{\HubOutAngle=270-\AngleDiagOffset;\HubInAngle=90+\AngleDiagOffset;}
    \foreach \i in {1,...,3} {
        \tikzmath{\jMin=int(\i+1);\jMax=4;};
        \foreach \j in {\jMin,...,\jMax} {
            \tikzmath{\kMin=int(\j+1);\kMax=5;};
            \foreach \k in {\kMin,...,\kMax} {
                \draw[->,line width=\ArrowLineWidth, >={Stealth[length=\arrowlengthh, width=\arrowwidthh]}, color=HubColor, opacity=0, shorten <= -0.25mm] (dyad\i\j)  to[out=\HubOutAngle, in=\HubInAngle] (dyad\i\k); 
            };
        };
    };
    \tikzmath{\AngleDiagOffset=30;}
    \tikzmath{\ForwardOutAngle=-\AngleDiagOffset;\ForwardInAngle=180+\AngleDiagOffset;}
    \foreach \i in {3,...,5} {
        \tikzmath{\jMin=int(1);\jMax=int(\i-2);};
        \foreach \j in {\jMin,...,\jMax} {
            \tikzmath{\kMin=int(\j+1);\kMax=int(\i-1);};
            \foreach \k in {\kMin,...,\kMax} {
                \draw[->,line width=\ArrowLineWidth, >={Stealth[length=\arrowlengthh, width=\arrowwidthh]}, color=ForwardColor,opacity=0,  shorten <= -0.25mm] (dyad\j\i)  to[out=\ForwardOutAngle, in=\ForwardInAngle] (dyad\k\i); 
            };
        };
    };

    \tikzmath{\AngleDiagOffset=30;}
    \tikzmath{\PathOutAngle=270+\AngleDiagOffset;\PathInAngle=180-\AngleDiagOffset;}
    \foreach \i in {1,...,3} {
        \tikzmath{\jMin=int(\i+1);\jMax=int(4);};
        \foreach \j in {\jMin,...,\jMax} {
            \tikzmath{\kMin=int(\j+1);\kMax=int(5);};
            \foreach \k in {\kMin,...,\kMax} {
                \draw[->,line width=\ArrowLineWidth, >={Stealth[length=\arrowlengthh, width=\arrowwidthh]}, color=TransitiveColor, opacity=0, shorten <= -0.25mm] (dyad\i\j)  to[out=\PathOutAngle, in=\PathInAngle] (dyad\j\k); 
            };
        };
    };

 \tikzmath{\AngleDiagOffset=60;}
    \tikzmath{\MidOutAngle=0-\AngleDiagOffset;\MidInAngle=90+\AngleDiagOffset;}
        \tikzmath{\AngleDiagOffset=60;}
    \tikzmath{\MidOutAngle=0-\AngleDiagOffset;\MidInAngle=90+\AngleDiagOffset;}
    \foreach \i in {1,...,2} {
        \tikzmath{\jMin=int(\i+2);\jMax=int(4);};
        \foreach \j in {\jMin,...,\jMax} {
            \tikzmath{\kMin=int(\i+1);\kMax=int(\j-1);};
            \foreach \k in {\kMin,...,\kMax} {
                \tikzmath{\lMin=int(\j+1);\lMax=int(5);};
                \foreach \l in {\lMin,...,\lMax} {
                    \draw[->,line width=\ArrowLineWidth, >={Stealth[length=\arrowlengthh, width=\arrowwidthh]}, color=NonLocalInterfaceColor, shorten <= -0.25mm] (dyad\i\j)  to[out=\MidOutAngle, in=\MidInAngle] (dyad\k\l); 
                };
            };
        };
    };

    \foreach [count=\j from 2] \dyadY in {2,...,5} {
        \tikzmath{\iMax=\j-1;}
        \foreach [count=\i from 1] \dyadX in {1,...,\iMax} {
            \node (dyad\i\j) [draw=black,fill=white,inner sep=0cm, regular polygon sides=4, minimum size=\DyadBoxSize, line width=\DyadBoxBorderThickness,font=\fontsize{\DyadTextSize}{\DyadTextSize}\selectfont] at (\xStart+\NodeColDelta*\i-\NodeColDelta,\yStart-\NodeRowDelta*\j+\NodeRowDelta) {$X_{\i\j}^{ }$};
        };
    };
  \end{tikzpicture}\\ \verticalspacingnewdags
  \begin{tikzpicture}[remember picture]
      \draw[draw=\retangleBoxColor, fill=none] (\retangleCoverPosX,\retangleCoverPosY) rectangle (\retangleCoverDimX,\retangleCoverDimY);

    \tikzmath{\xStart=1.0;\yStart=11.0;}
    \foreach [count=\j from 2] \dyadY in {2,...,5} {
        \tikzmath{\iMax=\j-1;}
        \foreach [count=\i from 1] \dyadX in {1,...,\iMax} {
            \node (dyad\i\j) [draw=white,fill=none,inner sep=0cm, regular polygon sides=4, minimum size=\DyadBoxSize, line width=\DyadBoxBorderThickness] at (\xStart+\NodeColDelta*\i-\NodeColDelta,\yStart-\NodeRowDelta*\j+\NodeRowDelta) { };
        };
    };

    \tikzmath{\AngleDiagOffset=45;}
    \tikzmath{\NearOutAngle=180+\AngleDiagOffset;\NearInAngle=\AngleDiagOffset;}
    \foreach \i in {2,...,3} {
        \tikzmath{\jMin=int(\i+1);\jMax=int(4);};
        \foreach \j in {\jMin,...,\jMax} {
            \tikzmath{\kMin=int(1);\kMax=int(\i-1);};
            \foreach \k in {\kMin,...,\kMax} {
                \tikzmath{\lMin=int(\j+1);\lMax=int(5);};
                \foreach \l in {\lMin,...,\lMax} {
                    \draw[->,line width=\ArrowLineWidth, >={Stealth[length=\arrowlengthh, width=\arrowwidthh]}, color=NonLocalInteriorColor, shorten <= -0.25mm] (dyad\i\j)  to[out=\NearOutAngle, in=\NearInAngle] (dyad\k\l); 
                };
            };
        };
    };

    \tikzmath{\AngleDiagOffset=0;}
    \tikzmath{\FarOutAngle=0-\AngleDiagOffset;\FarInAngle=90+\AngleDiagOffset;}
    \foreach \i in {1,...,2} {
        \tikzmath{\jMin=int(\i+1);\jMax=int(3);};
        \foreach \j in {\jMin,...,\jMax} {
            \tikzmath{\kMin=int(\j+1);\kMax=int(5-1);};
            \foreach \k in {\kMin,...,\kMax} {
                \tikzmath{\lMin=int(\k+1);\lMax=int(5);};
                \foreach \l in {\lMin,...,\lMax} {
                    \draw[->,line width=\ArrowLineWidth, >={Stealth[length=\arrowlengthh, width=\arrowwidthh]}, color=NonLocalExteriorColor, opacity=0, shorten <= -0.25mm] (dyad\i\j)  to[out=\FarOutAngle, in=\FarInAngle] (dyad\k\l); 
                };
            };
        };
    };
    \tikzmath{\AngleDiagOffset=30;}
    \tikzmath{\HubOutAngle=270-\AngleDiagOffset;\HubInAngle=90+\AngleDiagOffset;}
    \foreach \i in {1,...,3} {
        \tikzmath{\jMin=int(\i+1);\jMax=4;};
        \foreach \j in {\jMin,...,\jMax} {
            \tikzmath{\kMin=int(\j+1);\kMax=5;};
            \foreach \k in {\kMin,...,\kMax} {
                \draw[->,line width=\ArrowLineWidth, >={Stealth[length=\arrowlengthh, width=\arrowwidthh]}, color=HubColor, opacity=0, shorten <= -0.25mm] (dyad\i\j)  to[out=\HubOutAngle, in=\HubInAngle] (dyad\i\k); 
            };
        };
    };
    \tikzmath{\AngleDiagOffset=30;}
    \tikzmath{\ForwardOutAngle=-\AngleDiagOffset;\ForwardInAngle=180+\AngleDiagOffset;}
    \foreach \i in {3,...,5} {
        \tikzmath{\jMin=int(1);\jMax=int(\i-2);};
        \foreach \j in {\jMin,...,\jMax} {
            \tikzmath{\kMin=int(\j+1);\kMax=int(\i-1);};
            \foreach \k in {\kMin,...,\kMax} {
                \draw[->,line width=\ArrowLineWidth, >={Stealth[length=\arrowlengthh, width=\arrowwidthh]}, color=ForwardColor,opacity=0,  shorten <= -0.25mm] (dyad\j\i)  to[out=\ForwardOutAngle, in=\ForwardInAngle] (dyad\k\i); 
            };
        };
    };
    \tikzmath{\AngleDiagOffset=30;}
    \tikzmath{\PathOutAngle=270+\AngleDiagOffset;\PathInAngle=180-\AngleDiagOffset;}
    \foreach \i in {1,...,3} {
        \tikzmath{\jMin=int(\i+1);\jMax=int(4);};
        \foreach \j in {\jMin,...,\jMax} {
            \tikzmath{\kMin=int(\j+1);\kMax=int(5);};
            \foreach \k in {\kMin,...,\kMax} {
                \draw[->,line width=\ArrowLineWidth, >={Stealth[length=\arrowlengthh, width=\arrowwidthh]}, color=TransitiveColor,opacity=0,  shorten <= -0.25mm] (dyad\i\j)  to[out=\PathOutAngle, in=\PathInAngle] (dyad\j\k); 
            };
        };
    };

 \tikzmath{\AngleDiagOffset=60;}
    \tikzmath{\MidOutAngle=0-\AngleDiagOffset;\MidInAngle=90+\AngleDiagOffset;}
        \tikzmath{\AngleDiagOffset=60;}
    \tikzmath{\MidOutAngle=0-\AngleDiagOffset;\MidInAngle=90+\AngleDiagOffset;}
    \foreach \i in {1,...,2} {
        \tikzmath{\jMin=int(\i+2);\jMax=int(4);};
        \foreach \j in {\jMin,...,\jMax} {
            \tikzmath{\kMin=int(\i+1);\kMax=int(\j-1);};
            \foreach \k in {\kMin,...,\kMax} {
                \tikzmath{\lMin=int(\j+1);\lMax=int(5);};
                \foreach \l in {\lMin,...,\lMax} {
                    \draw[->,line width=\ArrowLineWidth, >={Stealth[length=\arrowlengthh, width=\arrowwidthh]}, color=NonLocalInterfaceColor, opacity=0, shorten <= -0.25mm] (dyad\i\j)  to[out=\MidOutAngle, in=\MidInAngle] (dyad\k\l); 
                };
            };
        };
    };

    \foreach [count=\j from 2] \dyadY in {2,...,5} {
        \tikzmath{\iMax=\j-1;}
        \foreach [count=\i from 1] \dyadX in {1,...,\iMax} {
            \node (dyad\i\j) [draw=black,fill=white,inner sep=0cm, regular polygon sides=4, minimum size=\DyadBoxSize, line width=\DyadBoxBorderThickness,font=\fontsize{\DyadTextSize}{\DyadTextSize}\selectfont] at (\xStart+\NodeColDelta*\i-\NodeColDelta,\yStart-\NodeRowDelta*\j+\NodeRowDelta) {$X_{\i\j}^{ }$};
        };
    };
  \end{tikzpicture}
  \horizontalspacingnewdags
    \begin{tikzpicture}[remember picture]
    \draw[draw=\retangleBoxColor, fill=none] (\retangleCoverPosX,\retangleCoverPosY) rectangle (\retangleCoverDimX,\retangleCoverDimY);
    \tikzmath{\xStart=1.0;\yStart=11.0;}
    \foreach [count=\j from 2] \dyadY in {2,...,5} {
        \tikzmath{\iMax=\j-1;}
        \foreach [count=\i from 1] \dyadX in {1,...,\iMax} {
            \node (dyad\i\j) [draw=white,fill=none,inner sep=0cm, regular polygon sides=4, minimum size=\DyadBoxSize, line width=\DyadBoxBorderThickness] at (\xStart+\NodeColDelta*\i-\NodeColDelta,\yStart-\NodeRowDelta*\j+\NodeRowDelta) { };
        };
    };

    \tikzmath{\AngleDiagOffset=45;}
    \tikzmath{\NearOutAngle=180+\AngleDiagOffset;\NearInAngle=\AngleDiagOffset;}
    \foreach \i in {2,...,3} {
        \tikzmath{\jMin=int(\i+1);\jMax=int(4);};
        \foreach \j in {\jMin,...,\jMax} {
            \tikzmath{\kMin=int(1);\kMax=int(\i-1);};
            \foreach \k in {\kMin,...,\kMax} {
                \tikzmath{\lMin=int(\j+1);\lMax=int(5);};
                \foreach \l in {\lMin,...,\lMax} {
                    \draw[->,line width=\ArrowLineWidth, >={Stealth[length=\arrowlengthh, width=\arrowwidthh]}, color=NonLocalInteriorColor, shorten <= -0.25mm] (dyad\i\j)  to[out=\NearOutAngle, in=\NearInAngle] (dyad\k\l); 
                };
            };
        };
    };

    \tikzmath{\AngleDiagOffset=0;}
    \tikzmath{\FarOutAngle=0-\AngleDiagOffset;\FarInAngle=90+\AngleDiagOffset;}
    \foreach \i in {1,...,2} {
        \tikzmath{\jMin=int(\i+1);\jMax=int(3);};
        \foreach \j in {\jMin,...,\jMax} {
            \tikzmath{\kMin=int(\j+1);\kMax=int(5-1);};
            \foreach \k in {\kMin,...,\kMax} {
                \tikzmath{\lMin=int(\k+1);\lMax=int(5);};
                \foreach \l in {\lMin,...,\lMax} {
                    \draw[->,line width=\ArrowLineWidth, >={Stealth[length=\arrowlengthh, width=\arrowwidthh]}, color=NonLocalExteriorColor, shorten <= -0.25mm] (dyad\i\j)  to[out=\FarOutAngle, in=\FarInAngle] (dyad\k\l); 
                };
            };
        };
    };
    \tikzmath{\AngleDiagOffset=30;}
    \tikzmath{\HubOutAngle=270-\AngleDiagOffset;\HubInAngle=90+\AngleDiagOffset;}
    \foreach \i in {1,...,3} {
        \tikzmath{\jMin=int(\i+1);\jMax=4;};
        \foreach \j in {\jMin,...,\jMax} {
            \tikzmath{\kMin=int(\j+1);\kMax=5;};
            \foreach \k in {\kMin,...,\kMax} {
                \draw[->,line width=\ArrowLineWidth, >={Stealth[length=\arrowlengthh, width=\arrowwidthh]}, color=HubColor, opacity=0, shorten <= -0.25mm] (dyad\i\j)  to[out=\HubOutAngle, in=\HubInAngle] (dyad\i\k); 
            };
        };
    };
    \tikzmath{\AngleDiagOffset=30;}
    \tikzmath{\ForwardOutAngle=-\AngleDiagOffset;\ForwardInAngle=180+\AngleDiagOffset;}
    \foreach \i in {3,...,5} {
        \tikzmath{\jMin=int(1);\jMax=int(\i-2);};
        \foreach \j in {\jMin,...,\jMax} {
            \tikzmath{\kMin=int(\j+1);\kMax=int(\i-1);};
            \foreach \k in {\kMin,...,\kMax} {
                \draw[->,line width=\ArrowLineWidth, >={Stealth[length=\arrowlengthh, width=\arrowwidthh]}, color=ForwardColor, opacity=0, shorten <= -0.25mm] (dyad\j\i)  to[out=\ForwardOutAngle, in=\ForwardInAngle] (dyad\k\i); 
            };
        };
    };

    \tikzmath{\AngleDiagOffset=30;}
    \tikzmath{\PathOutAngle=270+\AngleDiagOffset;\PathInAngle=180-\AngleDiagOffset;}
    \foreach \i in {1,...,3} {
        \tikzmath{\jMin=int(\i+1);\jMax=int(4);};
        \foreach \j in {\jMin,...,\jMax} {
            \tikzmath{\kMin=int(\j+1);\kMax=int(5);};
            \foreach \k in {\kMin,...,\kMax} {
                \draw[->,line width=\ArrowLineWidth, >={Stealth[length=\arrowlengthh, width=\arrowwidthh]}, color=TransitiveColor,opacity=0,  shorten <= -0.25mm] (dyad\i\j)  to[out=\PathOutAngle, in=\PathInAngle] (dyad\j\k); 
            };
        };
    };

   \tikzmath{\AngleDiagOffset=60;}
    \tikzmath{\MidOutAngle=0-\AngleDiagOffset;\MidInAngle=90+\AngleDiagOffset;}
        \tikzmath{\AngleDiagOffset=60;}
    \tikzmath{\MidOutAngle=0-\AngleDiagOffset;\MidInAngle=90+\AngleDiagOffset;}
    \foreach \i in {1,...,2} {
        \tikzmath{\jMin=int(\i+2);\jMax=int(4);};
        \foreach \j in {\jMin,...,\jMax} {
            \tikzmath{\kMin=int(\i+1);\kMax=int(\j-1);};
            \foreach \k in {\kMin,...,\kMax} {
                \tikzmath{\lMin=int(\j+1);\lMax=int(5);};
                \foreach \l in {\lMin,...,\lMax} {
                    \draw[->,line width=\ArrowLineWidth, >={Stealth[length=\arrowlengthh, width=\arrowwidthh]}, color=NonLocalInterfaceColor, shorten <= -0.25mm] (dyad\i\j)  to[out=\MidOutAngle, in=\MidInAngle] (dyad\k\l); 
                };
            };
        };
    };

    \foreach [count=\j from 2] \dyadY in {2,...,5} {
        \tikzmath{\iMax=\j-1;}
        \foreach [count=\i from 1] \dyadX in {1,...,\iMax} {
            \node (dyad\i\j) [draw=black,fill=white,inner sep=0cm, regular polygon sides=4, minimum size=\DyadBoxSize, line width=\DyadBoxBorderThickness,font=\fontsize{\DyadTextSize}{\DyadTextSize}\selectfont] at (\xStart+\NodeColDelta*\i-\NodeColDelta,\yStart-\NodeRowDelta*\j+\NodeRowDelta) {$X_{\i\j}^{ }$};
        };
    };
  \end{tikzpicture}
   \spaceaftercaptiontwo
\caption{
\textbf{Causal graphs for each of the types of causal arrows between dyads that do not share nodes. 
}\\ 
Causal \mbox{meta-DAGs} between dyads of a growing network with $5$ nodes that are compatible with network models having the following types of causal arrows: 
{\nonLocalExterior} (\textit{top-left}); {\nonLocalInterface} (\textit{top-right}); {\nonLocalInterior} (\textit{bottom-left}); and the three together  (\textit{bottom-right}).
}
\label{Fig:ExampleGraphicalModelNonLocalDirect}
\end{figure}

\newpage 

\spaceafterfivedyadsfig

\begin{figure}[H]
  \centering
  \begin{tikzpicture}[remember picture]
    \draw[draw=\retangleBoxColor, fill=none] (\retangleCoverPosX,\retangleCoverPosY) rectangle (\retangleCoverDimX,\retangleCoverDimY);
    \tikzmath{\xStart=1.0;\yStart=11.0;}
    \foreach [count=\j from 2] \dyadY in {2,...,5} {
        \tikzmath{\iMax=\j-1;}
        \foreach [count=\i from 1] \dyadX in {1,...,\iMax} {
            \node (dyad\i\j) [draw=white,fill=none,inner sep=0cm, regular polygon sides=4, minimum size=\DyadBoxSize, line width=\DyadBoxBorderThickness] at (\xStart+\NodeColDelta*\i-\NodeColDelta,\yStart-\NodeRowDelta*\j+\NodeRowDelta) { };
        };
    };

    \tikzmath{\AngleDiagOffset=45;}
    \tikzmath{\NearOutAngle=180+\AngleDiagOffset;\NearInAngle=\AngleDiagOffset;}
    \foreach \i in {2,...,3} {
        \tikzmath{\jMin=int(\i+1);\jMax=int(4);};
        \foreach \j in {\jMin,...,\jMax} {
            \tikzmath{\kMin=int(1);\kMax=int(\i-1);};
            \foreach \k in {\kMin,...,\kMax} {
                \tikzmath{\lMin=int(\j+1);\lMax=int(5);};
                \foreach \l in {\lMin,...,\lMax} {
                    \draw[->,line width=\ArrowLineWidth, >={Stealth[length=\arrowlengthh, width=\arrowwidthh]}, color=NonLocalInteriorColor, opacity=0, shorten <= -0.25mm] (dyad\i\j)  to[out=\NearOutAngle, in=\NearInAngle] (dyad\k\l); 
                };
            };
        };
    };

    \tikzmath{\AngleDiagOffset=0;}
    \tikzmath{\FarOutAngle=0-\AngleDiagOffset;\FarInAngle=90+\AngleDiagOffset;}
    \foreach \i in {1,...,2} {
        \tikzmath{\jMin=int(\i+1);\jMax=int(3);};
        \foreach \j in {\jMin,...,\jMax} {
            \tikzmath{\kMin=int(\j+1);\kMax=int(5-1);};
            \foreach \k in {\kMin,...,\kMax} {
                \tikzmath{\lMin=int(\k+1);\lMax=int(5);};
                \foreach \l in {\lMin,...,\lMax} {
                    \draw[->,line width=\ArrowLineWidth, >={Stealth[length=\arrowlengthh, width=\arrowwidthh]}, color=NonLocalExteriorColor, opacity=0, shorten <= -0.25mm] (dyad\i\j)  to[out=\FarOutAngle, in=\FarInAngle] (dyad\k\l); 
                };
            };
        };
    };
    \tikzmath{\AngleDiagOffset=30;}
    \tikzmath{\HubOutAngle=270-\AngleDiagOffset;\HubInAngle=90+\AngleDiagOffset;}
    \foreach \i in {1,...,3} {
        \tikzmath{\jMin=int(\i+1);\jMax=4;};
        \foreach \j in {\jMin,...,\jMax} {
            \tikzmath{\kMin=int(\j+1);\kMax=5;};
            \foreach \k in {\kMin,...,\kMax} {
                \draw[->,line width=\ArrowLineWidth, >={Stealth[length=\arrowlengthh, width=\arrowwidthh]}, color=HubColor, shorten <= -0.25mm] (dyad\i\j)  to[out=\HubOutAngle, in=\HubInAngle] (dyad\i\k); 
            };
        };
    };
    \tikzmath{\AngleDiagOffset=30;}
    \tikzmath{\ForwardOutAngle=-\AngleDiagOffset;\ForwardInAngle=180+\AngleDiagOffset;}
    \foreach \i in {3,...,5} {
        \tikzmath{\jMin=int(1);\jMax=int(\i-2);};
        \foreach \j in {\jMin,...,\jMax} {
            \tikzmath{\kMin=int(\j+1);\kMax=int(\i-1);};
            \foreach \k in {\kMin,...,\kMax} {
                \draw[->,line width=\ArrowLineWidth, >={Stealth[length=\arrowlengthh, width=\arrowwidthh]}, color=ForwardColor, opacity=0, shorten <= -0.25mm] (dyad\j\i)  to[out=\ForwardOutAngle, in=\ForwardInAngle] (dyad\k\i); 
            };
        };
    };
    \tikzmath{\AngleDiagOffset=30;}
    \tikzmath{\PathOutAngle=270+\AngleDiagOffset;\PathInAngle=180-\AngleDiagOffset;}
    \foreach \i in {1,...,3} {
        \tikzmath{\jMin=int(\i+1);\jMax=int(4);};
        \foreach \j in {\jMin,...,\jMax} {
            \tikzmath{\kMin=int(\j+1);\kMax=int(5);};
            \foreach \k in {\kMin,...,\kMax} {
                \draw[->,line width=\ArrowLineWidth, >={Stealth[length=\arrowlengthh, width=\arrowwidthh]}, color=TransitiveColor, shorten <= -0.25mm] (dyad\i\j)  to[out=\PathOutAngle, in=\PathInAngle] (dyad\j\k); 
            };
        };
    };

 \tikzmath{\AngleDiagOffset=60;}
    \tikzmath{\MidOutAngle=0-\AngleDiagOffset;\MidInAngle=90+\AngleDiagOffset;}
        \tikzmath{\AngleDiagOffset=60;}
    \tikzmath{\MidOutAngle=0-\AngleDiagOffset;\MidInAngle=90+\AngleDiagOffset;}
    \foreach \i in {1,...,2} {
        \tikzmath{\jMin=int(\i+2);\jMax=int(4);};
        \foreach \j in {\jMin,...,\jMax} {
            \tikzmath{\kMin=int(\i+1);\kMax=int(\j-1);};
            \foreach \k in {\kMin,...,\kMax} {
                \tikzmath{\lMin=int(\j+1);\lMax=int(5);};
                \foreach \l in {\lMin,...,\lMax} {
                    \draw[->,line width=\ArrowLineWidth, >={Stealth[length=\arrowlengthh, width=\arrowwidthh]}, color=NonLocalInterfaceColor, opacity=0, shorten <= -0.25mm] (dyad\i\j)  to[out=\MidOutAngle, in=\MidInAngle] (dyad\k\l); 
                };
            };
        };
    };
    \foreach [count=\j from 2] \dyadY in {2,...,5} {
        \tikzmath{\iMax=\j-1;}
        \foreach [count=\i from 1] \dyadX in {1,...,\iMax} {
            \node (dyad\i\j) [draw=black,fill=white,inner sep=0cm, regular polygon sides=4, minimum size=\DyadBoxSize, line width=\DyadBoxBorderThickness,font=\fontsize{\DyadTextSize}{\DyadTextSize}\selectfont] at (\xStart+\NodeColDelta*\i-\NodeColDelta,\yStart-\NodeRowDelta*\j+\NodeRowDelta) {$X_{\i\j}^{ }$};
        };
    };
  \end{tikzpicture}
  \horizontalspacingnewdags
    \begin{tikzpicture}[remember picture]
    \draw[draw=\retangleBoxColor, fill=none] (\retangleCoverPosX,\retangleCoverPosY) rectangle (\retangleCoverDimX,\retangleCoverDimY);
    \tikzmath{\xStart=1.0;\yStart=11.0;}
    \foreach [count=\j from 2] \dyadY in {2,...,5} {
        \tikzmath{\iMax=\j-1;}
        \foreach [count=\i from 1] \dyadX in {1,...,\iMax} {
            \node (dyad\i\j) [draw=white,fill=none,inner sep=0cm, regular polygon sides=4, minimum size=\DyadBoxSize, line width=\DyadBoxBorderThickness] at (\xStart+\NodeColDelta*\i-\NodeColDelta,\yStart-\NodeRowDelta*\j+\NodeRowDelta) { };
        };
    };

    \tikzmath{\AngleDiagOffset=45;}
    \tikzmath{\NearOutAngle=180+\AngleDiagOffset;\NearInAngle=\AngleDiagOffset;}
    \foreach \i in {2,...,3} {
        \tikzmath{\jMin=int(\i+1);\jMax=int(4);};
        \foreach \j in {\jMin,...,\jMax} {
            \tikzmath{\kMin=int(1);\kMax=int(\i-1);};
            \foreach \k in {\kMin,...,\kMax} {
                \tikzmath{\lMin=int(\j+1);\lMax=int(5);};
                \foreach \l in {\lMin,...,\lMax} {
                    \draw[->,line width=\ArrowLineWidth, >={Stealth[length=\arrowlengthh, width=\arrowwidthh]}, color=NonLocalInteriorColor, opacity=0, shorten <= -0.25mm] (dyad\i\j)  to[out=\NearOutAngle, in=\NearInAngle] (dyad\k\l); 
                };
            };
        };
    };

    \tikzmath{\AngleDiagOffset=0;}
    \tikzmath{\FarOutAngle=0-\AngleDiagOffset;\FarInAngle=90+\AngleDiagOffset;}
    \foreach \i in {1,...,2} {
        \tikzmath{\jMin=int(\i+1);\jMax=int(3);};
        \foreach \j in {\jMin,...,\jMax} {
            \tikzmath{\kMin=int(\j+1);\kMax=int(5-1);};
            \foreach \k in {\kMin,...,\kMax} {
                \tikzmath{\lMin=int(\k+1);\lMax=int(5);};
                \foreach \l in {\lMin,...,\lMax} {
                    \draw[->,line width=\ArrowLineWidth, >={Stealth[length=\arrowlengthh, width=\arrowwidthh]}, color=NonLocalExteriorColor, opacity=0, shorten <= -0.25mm] (dyad\i\j)  to[out=\FarOutAngle, in=\FarInAngle] (dyad\k\l); 
                };
            };
        };
    };
    \tikzmath{\AngleDiagOffset=30;}
    \tikzmath{\HubOutAngle=270-\AngleDiagOffset;\HubInAngle=90+\AngleDiagOffset;}
    \foreach \i in {1,...,3} {
        \tikzmath{\jMin=int(\i+1);\jMax=4;};
        \foreach \j in {\jMin,...,\jMax} {
            \tikzmath{\kMin=int(\j+1);\kMax=5;};
            \foreach \k in {\kMin,...,\kMax} {
                \draw[->,line width=\ArrowLineWidth, >={Stealth[length=\arrowlengthh, width=\arrowwidthh]}, color=HubColor, shorten <= -0.25mm] (dyad\i\j)  to[out=\HubOutAngle, in=\HubInAngle] (dyad\i\k); 
            };
        };
    };
            \tikzmath{\AngleDiagOffset=30;}
    \tikzmath{\ForwardOutAngle=-\AngleDiagOffset;\ForwardInAngle=180+\AngleDiagOffset;}
    \foreach \i in {3,...,5} {
        \tikzmath{\jMin=int(1);\jMax=int(\i-2);};
        \foreach \j in {\jMin,...,\jMax} {
            \tikzmath{\kMin=int(\j+1);\kMax=int(\i-1);};
            \foreach \k in {\kMin,...,\kMax} {
                \draw[->,line width=\ArrowLineWidth, >={Stealth[length=\arrowlengthh, width=\arrowwidthh]}, color=BackwardColor, shorten <= -0.25mm] (dyad\k\i)  to[out=\ForwardInAngle, in=\ForwardOutAngle] (dyad\j\i); 
            };
        };
    };

    \tikzmath{\AngleDiagOffset=30;}
    \tikzmath{\PathOutAngle=270+\AngleDiagOffset;\PathInAngle=180-\AngleDiagOffset;}
    \foreach \i in {1,...,3} {
        \tikzmath{\jMin=int(\i+1);\jMax=int(4);};
        \foreach \j in {\jMin,...,\jMax} {
            \tikzmath{\kMin=int(\j+1);\kMax=int(5);};
            \foreach \k in {\kMin,...,\kMax} {
                \draw[->,line width=\ArrowLineWidth, >={Stealth[length=\arrowlengthh, width=\arrowwidthh]}, color=TransitiveColor, opacity=0, shorten <= -0.25mm] (dyad\i\j)  to[out=\PathOutAngle, in=\PathInAngle] (dyad\j\k); 
            };
        };
    };

 \tikzmath{\AngleDiagOffset=60;}
    \tikzmath{\MidOutAngle=0-\AngleDiagOffset;\MidInAngle=90+\AngleDiagOffset;}
        \tikzmath{\AngleDiagOffset=60;}
    \tikzmath{\MidOutAngle=0-\AngleDiagOffset;\MidInAngle=90+\AngleDiagOffset;}
    \foreach \i in {1,...,2} {
        \tikzmath{\jMin=int(\i+2);\jMax=int(4);};
        \foreach \j in {\jMin,...,\jMax} {
            \tikzmath{\kMin=int(\i+1);\kMax=int(\j-1);};
            \foreach \k in {\kMin,...,\kMax} {
                \tikzmath{\lMin=int(\j+1);\lMax=int(5);};
                \foreach \l in {\lMin,...,\lMax} {
                    \draw[->,line width=\ArrowLineWidth, >={Stealth[length=\arrowlengthh, width=\arrowwidthh]}, color=NonLocalInterfaceColor, opacity=0, shorten <= -0.25mm] (dyad\i\j)  to[out=\MidOutAngle, in=\MidInAngle] (dyad\k\l); 
                };
            };
        };
    };

    \foreach [count=\j from 2] \dyadY in {2,...,5} {
        \tikzmath{\iMax=\j-1;}
        \foreach [count=\i from 1] \dyadX in {1,...,\iMax} {
            \node (dyad\i\j) [draw=black,fill=white,inner sep=0cm, regular polygon sides=4, minimum size=\DyadBoxSize, line width=\DyadBoxBorderThickness,font=\fontsize{\DyadTextSize}{\DyadTextSize}\selectfont] at (\xStart+\NodeColDelta*\i-\NodeColDelta,\yStart-\NodeRowDelta*\j+\NodeRowDelta) {$X_{\i\j}^{ }$};
        };
    };
  \end{tikzpicture}\\ \verticalspacingnewdags
  \begin{tikzpicture}[remember picture]
    \draw[draw=\retangleBoxColor, fill=none] (\retangleCoverPosX,\retangleCoverPosY) rectangle (\retangleCoverDimX,\retangleCoverDimY);
    \tikzmath{\xStart=1.0;\yStart=11.0;}
    \foreach [count=\j from 2] \dyadY in {2,...,5} {
        \tikzmath{\iMax=\j-1;}
        \foreach [count=\i from 1] \dyadX in {1,...,\iMax} {
            \node (dyad\i\j) [draw=white,fill=none,inner sep=0cm, regular polygon sides=4, minimum size=\DyadBoxSize, line width=\DyadBoxBorderThickness] at (\xStart+\NodeColDelta*\i-\NodeColDelta,\yStart-\NodeRowDelta*\j+\NodeRowDelta) { };
        };
    };

    \tikzmath{\AngleDiagOffset=45;}
    \tikzmath{\NearOutAngle=180+\AngleDiagOffset;\NearInAngle=\AngleDiagOffset;}
    \foreach \i in {2,...,3} {
        \tikzmath{\jMin=int(\i+1);\jMax=int(4);};
        \foreach \j in {\jMin,...,\jMax} {
            \tikzmath{\kMin=int(1);\kMax=int(\i-1);};
            \foreach \k in {\kMin,...,\kMax} {
                \tikzmath{\lMin=int(\j+1);\lMax=int(5);};
                \foreach \l in {\lMin,...,\lMax} {
                    \draw[->,line width=\ArrowLineWidth, >={Stealth[length=\arrowlengthh, width=\arrowwidthh]}, color=NonLocalInteriorColor, opacity=0, shorten <= -0.25mm] (dyad\i\j)  to[out=\NearOutAngle, in=\NearInAngle] (dyad\k\l); 
                };
            };
        };
    };

    \tikzmath{\AngleDiagOffset=0;}
    \tikzmath{\FarOutAngle=0-\AngleDiagOffset;\FarInAngle=90+\AngleDiagOffset;}
    \foreach \i in {1,...,2} {
        \tikzmath{\jMin=int(\i+1);\jMax=int(3);};
        \foreach \j in {\jMin,...,\jMax} {
            \tikzmath{\kMin=int(\j+1);\kMax=int(5-1);};
            \foreach \k in {\kMin,...,\kMax} {
                \tikzmath{\lMin=int(\k+1);\lMax=int(5);};
                \foreach \l in {\lMin,...,\lMax} {
                    \draw[->,line width=\ArrowLineWidth, >={Stealth[length=\arrowlengthh, width=\arrowwidthh]}, color=NonLocalExteriorColor, opacity=0, shorten <= -0.25mm] (dyad\i\j)  to[out=\FarOutAngle, in=\FarInAngle] (dyad\k\l); 
                };
            };
        };
    };
    \tikzmath{\AngleDiagOffset=30;}
    \tikzmath{\HubOutAngle=270-\AngleDiagOffset;\HubInAngle=90+\AngleDiagOffset;}
    \foreach \i in {1,...,3} {
        \tikzmath{\jMin=int(\i+1);\jMax=4;};
        \foreach \j in {\jMin,...,\jMax} {
            \tikzmath{\kMin=int(\j+1);\kMax=5;};
            \foreach \k in {\kMin,...,\kMax} {
                \draw[->,line width=\ArrowLineWidth, >={Stealth[length=\arrowlengthh, width=\arrowwidthh]}, color=HubColor, opacity=0,shorten <= -0.25mm] (dyad\i\j)  to[out=\HubOutAngle, in=\HubInAngle] (dyad\i\k); 
            };
        };
    };
    \tikzmath{\AngleDiagOffset=30;}
    \tikzmath{\ForwardOutAngle=-\AngleDiagOffset;\ForwardInAngle=180+\AngleDiagOffset;}
    \foreach \i in {3,...,5} {
        \tikzmath{\jMin=int(1);\jMax=int(\i-2);};
        \foreach \j in {\jMin,...,\jMax} {
            \tikzmath{\kMin=int(\j+1);\kMax=int(\i-1);};
            \foreach \k in {\kMin,...,\kMax} {
                \draw[->,line width=\ArrowLineWidth, >={Stealth[length=\arrowlengthh, width=\arrowwidthh]}, color=ForwardColor, shorten <= -0.25mm] (dyad\j\i)  to[out=\ForwardOutAngle, in=\ForwardInAngle] (dyad\k\i); 
            };
        };
    };
    \tikzmath{\AngleDiagOffset=30;}
    \tikzmath{\PathOutAngle=270+\AngleDiagOffset;\PathInAngle=180-\AngleDiagOffset;}
    \foreach \i in {1,...,3} {
        \tikzmath{\jMin=int(\i+1);\jMax=int(4);};
        \foreach \j in {\jMin,...,\jMax} {
            \tikzmath{\kMin=int(\j+1);\kMax=int(5);};
            \foreach \k in {\kMin,...,\kMax} {
                \draw[->,line width=\ArrowLineWidth, >={Stealth[length=\arrowlengthh, width=\arrowwidthh]}, color=TransitiveColor, shorten <= -0.25mm] (dyad\i\j)  to[out=\PathOutAngle, in=\PathInAngle] (dyad\j\k); 
            };
        };
    };

 \tikzmath{\AngleDiagOffset=60;}
    \tikzmath{\MidOutAngle=0-\AngleDiagOffset;\MidInAngle=90+\AngleDiagOffset;}
        \tikzmath{\AngleDiagOffset=60;}
    \tikzmath{\MidOutAngle=0-\AngleDiagOffset;\MidInAngle=90+\AngleDiagOffset;}
    \foreach \i in {1,...,2} {
        \tikzmath{\jMin=int(\i+2);\jMax=int(4);};
        \foreach \j in {\jMin,...,\jMax} {
            \tikzmath{\kMin=int(\i+1);\kMax=int(\j-1);};
            \foreach \k in {\kMin,...,\kMax} {
                \tikzmath{\lMin=int(\j+1);\lMax=int(5);};
                \foreach \l in {\lMin,...,\lMax} {
                    \draw[->,line width=\ArrowLineWidth, >={Stealth[length=\arrowlengthh, width=\arrowwidthh]}, color=NonLocalInterfaceColor,opacity=0, shorten <= -0.25mm] (dyad\i\j)  to[out=\MidOutAngle, in=\MidInAngle] (dyad\k\l); 
                };
            };
        };
    };

    \foreach [count=\j from 2] \dyadY in {2,...,5} {
        \tikzmath{\iMax=\j-1;}
        \foreach [count=\i from 1] \dyadX in {1,...,\iMax} {
            \node (dyad\i\j) [draw=black,fill=white,inner sep=0cm, regular polygon sides=4, minimum size=\DyadBoxSize, line width=\DyadBoxBorderThickness,font=\fontsize{\DyadTextSize}{\DyadTextSize}\selectfont] at (\xStart+\NodeColDelta*\i-\NodeColDelta,\yStart-\NodeRowDelta*\j+\NodeRowDelta) {$X_{\i\j}^{ }$};
        };
    };
  \end{tikzpicture}
  \horizontalspacingnewdags
    \begin{tikzpicture}[remember picture]
    \draw[draw=\retangleBoxColor, fill=none] (\retangleCoverPosX,\retangleCoverPosY) rectangle (\retangleCoverDimX,\retangleCoverDimY);
    \tikzmath{\xStart=1.0;\yStart=11.0;}
    \foreach [count=\j from 2] \dyadY in {2,...,5} {
        \tikzmath{\iMax=\j-1;}
        \foreach [count=\i from 1] \dyadX in {1,...,\iMax} {
            \node (dyad\i\j) [draw=white,fill=none,inner sep=0cm, regular polygon sides=4, minimum size=\DyadBoxSize, line width=\DyadBoxBorderThickness] at (\xStart+\NodeColDelta*\i-\NodeColDelta,\yStart-\NodeRowDelta*\j+\NodeRowDelta) { };
        };
    };

    \tikzmath{\AngleDiagOffset=45;}
    \tikzmath{\NearOutAngle=180+\AngleDiagOffset;\NearInAngle=\AngleDiagOffset;}
    \foreach \i in {2,...,3} {
        \tikzmath{\jMin=int(\i+1);\jMax=int(4);};
        \foreach \j in {\jMin,...,\jMax} {
            \tikzmath{\kMin=int(1);\kMax=int(\i-1);};
            \foreach \k in {\kMin,...,\kMax} {
                \tikzmath{\lMin=int(\j+1);\lMax=int(5);};
                \foreach \l in {\lMin,...,\lMax} {
                    \draw[->,line width=\ArrowLineWidth, >={Stealth[length=\arrowlengthh, width=\arrowwidthh]}, color=NonLocalInteriorColor, opacity=0, shorten <= -0.25mm] (dyad\i\j)  to[out=\NearOutAngle, in=\NearInAngle] (dyad\k\l); 
                };
            };
        };
    };

    \tikzmath{\AngleDiagOffset=0;}
    \tikzmath{\FarOutAngle=0-\AngleDiagOffset;\FarInAngle=90+\AngleDiagOffset;}
    \foreach \i in {1,...,2} {
        \tikzmath{\jMin=int(\i+1);\jMax=int(3);};
        \foreach \j in {\jMin,...,\jMax} {
            \tikzmath{\kMin=int(\j+1);\kMax=int(5-1);};
            \foreach \k in {\kMin,...,\kMax} {
                \tikzmath{\lMin=int(\k+1);\lMax=int(5);};
                \foreach \l in {\lMin,...,\lMax} {
                    \draw[->,line width=\ArrowLineWidth, >={Stealth[length=\arrowlengthh, width=\arrowwidthh]}, color=NonLocalExteriorColor, opacity=0, shorten <= -0.25mm] (dyad\i\j)  to[out=\FarOutAngle, in=\FarInAngle] (dyad\k\l); 
                };
            };
        };
    };
    \tikzmath{\AngleDiagOffset=30;}
    \tikzmath{\HubOutAngle=270-\AngleDiagOffset;\HubInAngle=90+\AngleDiagOffset;}
    \foreach \i in {1,...,3} {
        \tikzmath{\jMin=int(\i+1);\jMax=4;};
        \foreach \j in {\jMin,...,\jMax} {
            \tikzmath{\kMin=int(\j+1);\kMax=5;};
            \foreach \k in {\kMin,...,\kMax} {
                \draw[->,line width=\ArrowLineWidth, >={Stealth[length=\arrowlengthh, width=\arrowwidthh]}, color=HubColor, shorten <= -0.25mm] (dyad\i\j)  to[out=\HubOutAngle, in=\HubInAngle] (dyad\i\k); 
            };
        };
    };
    \tikzmath{\AngleDiagOffset=30;}
    \tikzmath{\ForwardOutAngle=-\AngleDiagOffset;\ForwardInAngle=180+\AngleDiagOffset;}
    \foreach \i in {3,...,5} {
        \tikzmath{\jMin=int(1);\jMax=int(\i-2);};
        \foreach \j in {\jMin,...,\jMax} {
            \tikzmath{\kMin=int(\j+1);\kMax=int(\i-1);};
            \foreach \k in {\kMin,...,\kMax} {
                \draw[->,line width=\ArrowLineWidth, >={Stealth[length=\arrowlengthh, width=\arrowwidthh]}, color=ForwardColor, shorten <= -0.25mm] (dyad\j\i)  to[out=\ForwardOutAngle, in=\ForwardInAngle] (dyad\k\i); 
            };
        };
    };

    \tikzmath{\AngleDiagOffset=30;}
    \tikzmath{\PathOutAngle=270+\AngleDiagOffset;\PathInAngle=180-\AngleDiagOffset;}
    \foreach \i in {1,...,3} {
        \tikzmath{\jMin=int(\i+1);\jMax=int(4);};
        \foreach \j in {\jMin,...,\jMax} {
            \tikzmath{\kMin=int(\j+1);\kMax=int(5);};
            \foreach \k in {\kMin,...,\kMax} {
                \draw[->,line width=\ArrowLineWidth, >={Stealth[length=\arrowlengthh, width=\arrowwidthh]}, color=TransitiveColor, shorten <= -0.25mm] (dyad\i\j)  to[out=\PathOutAngle, in=\PathInAngle] (dyad\j\k); 
            };
        };
    };

   \tikzmath{\AngleDiagOffset=60;}
    \tikzmath{\MidOutAngle=0-\AngleDiagOffset;\MidInAngle=90+\AngleDiagOffset;}
        \tikzmath{\AngleDiagOffset=60;}
    \tikzmath{\MidOutAngle=0-\AngleDiagOffset;\MidInAngle=90+\AngleDiagOffset;}
    \foreach \i in {1,...,2} {
        \tikzmath{\jMin=int(\i+2);\jMax=int(4);};
        \foreach \j in {\jMin,...,\jMax} {
            \tikzmath{\kMin=int(\i+1);\kMax=int(\j-1);};
            \foreach \k in {\kMin,...,\kMax} {
                \tikzmath{\lMin=int(\j+1);\lMax=int(5);};
                \foreach \l in {\lMin,...,\lMax} {
                    \draw[->,line width=\ArrowLineWidth, >={Stealth[length=\arrowlengthh, width=\arrowwidthh]}, color=NonLocalInterfaceColor, opacity=0, shorten <= -0.25mm] (dyad\i\j)  to[out=\MidOutAngle, in=\MidInAngle] (dyad\k\l); 
                };
            };
        };
    };

    \foreach [count=\j from 2] \dyadY in {2,...,5} {
        \tikzmath{\iMax=\j-1;}
        \foreach [count=\i from 1] \dyadX in {1,...,\iMax} {
            \node (dyad\i\j) [draw=black,fill=white,inner sep=0cm, regular polygon sides=4, minimum size=\DyadBoxSize, line width=\DyadBoxBorderThickness,font=\fontsize{\DyadTextSize}{\DyadTextSize}\selectfont] at (\xStart+\NodeColDelta*\i-\NodeColDelta,\yStart-\NodeRowDelta*\j+\NodeRowDelta) {$X_{\i\j}^{ }$};
        };
    };
  \end{tikzpicture}
  \spaceaftercaptionthree
\caption{
\textbf{Causal graphs with multiple types of causal arrows between dyads. 
}\\ 
Causal \mbox{meta-DAGs} between dyads of a growing network with $5$ nodes 
that are compatible with network models having the following types of causal arrows: 
{\hub} and {\transitive} (\textit{top-left}); {\hub} and {\backward} (\textit{top-right}); {\transitive} and {\forward} (\textit{bottom-
left}); and {\hub}, {\transitive}, 
and {\forward} (\textit{bottom-right}). 
}
\label{Fig:ExampleGraphicalModelLocalDirectSeveral}
\end{figure}




\spacestartappendixsection
\newpage
\section{Hasse Diagram with the 21 Invariant Causal Models}
\label{appendixFigHasse}

\potentiallyImprove{There are 21 causal meta-DAGs with finite ancestral sets whose interventional structure are invariant to node deletion and contraction (Theorem~\ref{thm:21InvariantCausalModels},  \Cref{sec:deletion,sec:poset}).  
They form a partially ordered set (poset), with the partial order relation given by inclusion of the seven types causal arrows.}  

\begin{figure}
\begin{center}
\input{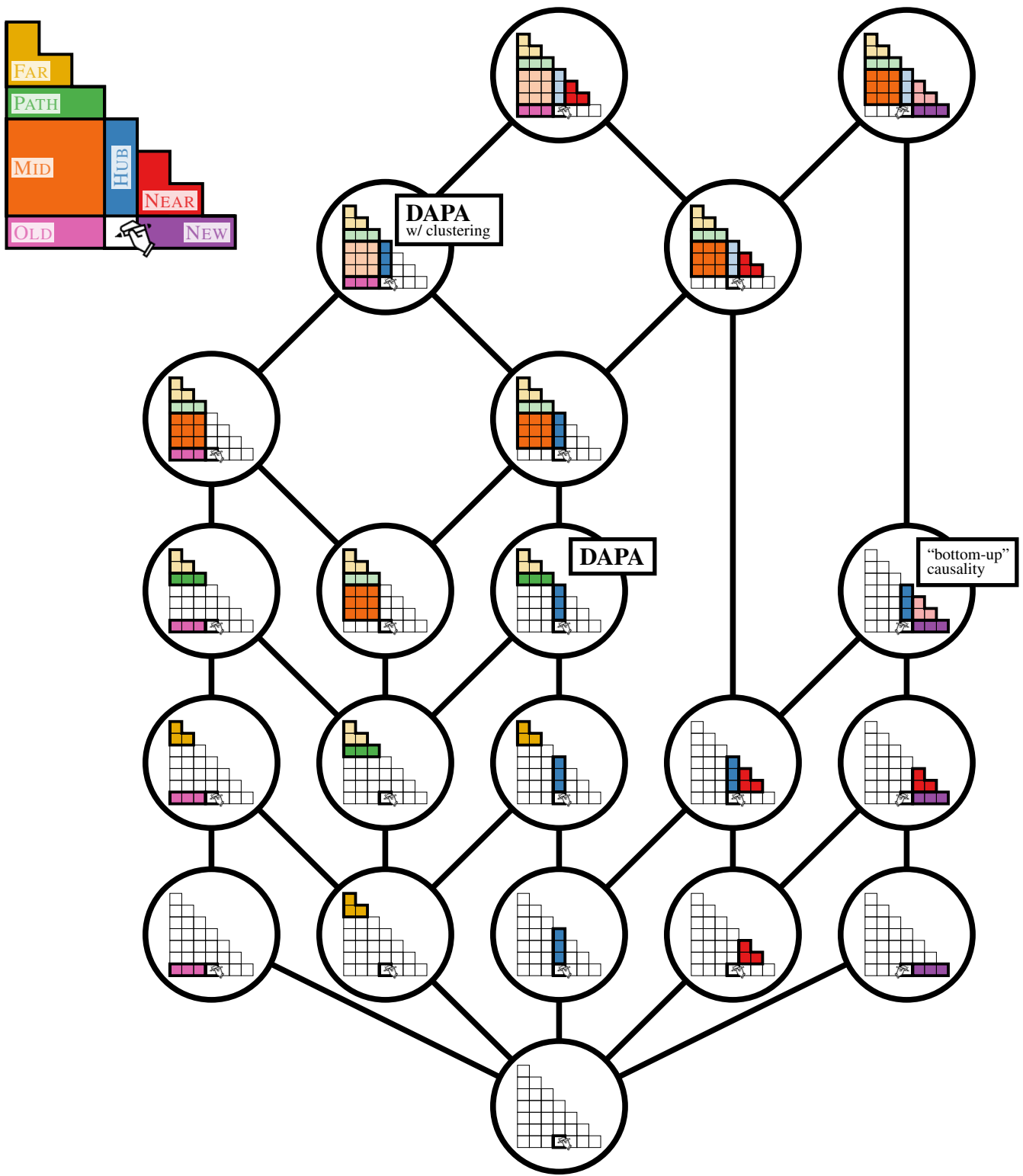}
\end{center}
\caption{
\textbf{Hasse diagram of the 21 transitively-closed deletion-invariant causal meta-DAGs with finite ancestral sets.}\\
}
\label{fig:PosetMetaDAG}
\end{figure}

\potentiallyImprove{Figure~\ref{fig:PosetMetaDAG} display a Hasse diagram of their poset, 
starting from the bottom of the figure and ordering the columns from left to right, 
the type of causal arrows that define each of the 21 causal meta-DAGs:} 
\begin{itemize}
    \item \textit{Bottom row:} 
    \begin{itemize}
        \item no causal arrows.
    \end{itemize}
    \item \textit{$2^{\text{nd}}$ row:} 
    \begin{itemize}
        \item {\OldName} 
 \item {\FarName}
 \item {\HubName} 
 \item {\NearName} 
\item {\NewName}
    \end{itemize}
\item \textit{$3^{\text{rd}}$ row:} 
\begin{itemize}
    \item {\FarName}/{\OldName} 
\item {\PathName} ({\FarName}) 
\item {\FarName}/{\HubName} 
\item {\HubName}/{\NearName} 
\item {\NearName}/{\NewName}
\end{itemize}
\item \textit{$4^{\text{th}}$ row:} 
\begin{itemize}
    \item {\OldName}/{\PathName} ({\FarName}) 
\item {\MidName} ({\PathName}, {\FarName}) 
\item {\HubName}/{\PathName} ({\FarName}) 
\item {\HubName}/{\NewName} ({\NearName}) 
\end{itemize}
\item \textit{$5^{\text{th}}$  row:} 
\begin{itemize}
    \item  {\OldName}/{\MidName} ({\PathName}/{\FarName}) 
\item {\MidName}/{\HubName} ({\PathName}/{\FarName})
\end{itemize}
\item \textit{$6^{\text{th}}$  row:} 
\begin{itemize}
    \item {\OldName}/{\HubName} ({\MidName}/{\PathName}/{\FarName}) 
\item {\MidName}/{\NearName} ({\PathName}/{\FarName}/{\HubName})
\end{itemize}
\item \textit{Top row:} 
\begin{itemize}
    \item {\OldName}/{\NearName} ({\MidName}/{\PathName}/{\FarName}/{\HubName})
    \item {\MidName}/{\NewName} ({\PathName}/{\FarName}/{\HubName}/{\NearName})
\end{itemize}
\end{itemize}
\potentiallyImprove{where a type of causal arrow in parenthesis indicates that it is implied by the addition of the new type of causal arrow.}

\spacestartappendixsection
\newpage
\section{Proof of Theorems for {\ppaAcron} model}
\label{appendixDAPAProofs}

The Distributed Affine Preferential Attachment ({\ppaAcron}) model (Section~\ref{sec:ourppamodel}) is defined as:
\begin{align}
    x_{ij}^{ } &\sim \text{Bernoulli}\big(p_{ij}^{ }\big) \label{eq:PPAbernoulliAppendix} \\
    p_{ij}^{ } &= \frac{\alphaP + \thetain d_i^{\text{in}} + \thetaout d_i^{\text{out}}}{j-2+\alphaP +\betaP} \label{eq:PPAprobAppendix}\\
    d_i^{\text{in}} &= \sum_{\BoxNodeSS{}=i+1}^{j-1} x_{i\BoxNodeSS{}}^{ }  \quad\quad
    d_i^{\text{out}} = \sum_{\BoxNodeSS{}=1}^{i-1} x_{\BoxNodeSS{}i}^{ } \label{eq:PPAdoutAppendix}    
\end{align}
where \mbox{$x_{ij} = 1$} ($\mbox{$x_{ij} = 0$}$) indicates an edge (no edge) between nodes $i$ and $j$.

\subsection{Proof of Theorem~\ref{thm:PPAPhaseTransition} \thmNamePPAAppendix}
\label{appendixProof3Regimes}

Our {\ppaAcron} model exhibits three qualitatively different asymptotic behaviors for the average degree $\avgdegree$ \\[-12pt] 
\vspace{-6pt}
\def\TempKern{\kern4pt}
\begin{center}
\begin{tabular}{llr}
    \vphantom{\Bigg(}constant:      &\TempKern$\displaystyle\frac{2\kern1pt\alphaP}{1-\big(\thetain+\thetaout\big)}$     &\TempKern$0<\thetain+\thetaout<1$ \\
    \vphantom{\Bigg(}logarithmic:   &\TempKern$2\kern1pt\alphaP \log\big(n\big) + C$                            &\TempKern$\thetain+\thetaout=1$ \\
    \vphantom{\Bigg(}polynomial:    &\TempKern$C \times n^{\thetain+\thetaout-1}$                     &\TempKern$1<\thetain+\thetaout<2$
\end{tabular}
\end{center}

\novo{We now derive the expressions for the average degree in the three regimes, 
starting with the constant regime.} 
\novo{To simplify notation on our derivations, 
we will use $n$ for both the step/iteration of the model and the number of nodes.  
\potentiallyRemove{(Indeed, although this model is parallelizable, one can always simulate it sequentially with a single node arriving at each step.)}
}
 
\def\logTIGHT{\log\kern-1pt}
\def\timesLOOSE{\kern1pt\times\kern1pt}
\def\ConstOne{C_1^{ }}
\def\ConstTwo{C_2^{ }}
\subsubsection{Constant degree regime: \protect{\textrm{$0\leq\thetain +\thetaout <1$}}}
In this regime, the average degree converges to a constant: 
\begin{equation}
    \avgdegree = \frac{2\kern1pt\alphaP}{1-(\thetain+\thetaout)}  + \littleO{1} \label{eq:PPAavgdegreeSumThetaLess1}
\end{equation}

After node $n$ has decided all its connections, 
the total number of edges in the network is
\begin{align}
    E(n) = \sum_{i=1}^{n} \din_i = \sum_{i=1}^n \dout_i
\end{align}
The expected number of edges added at each step
is
\begin{align}
     \big\langle E(n+1) - E(n)  \big\rangle 
    &= \sum_{i=1}^{n} \frac{\alphaP + \thetain \din_i + \thetaout \dout_i}{n + \alphaP + \betaP - 1} \nonumber \\
    &= \frac{\alphaP n +\big(\thetain + \thetaout\big) E(n)}{n + \alphaP+ \betaP - 1} \label{eq:EDotOfN}
\end{align}

We make an ansatz of constant average degree 
\begin{align}
    \big\langle E(n) \big\rangle = \ConstOne \timesLOOSE n + g(n) \label{eq:ConstantDegreeAnsatz}
\end{align}
\potentiallyImprove{where ${g(n) = \littleO{n}}$ is subdominant.}  

We will first solve for $\ConstOne$ to obtain the asymptotic average degree ${\langle d \rangle = 2\ConstOne}$,
then we will verify our assumption that ${g(n) = \littleO{n}}$.  

Taking the expectation of \eqref{eq:EDotOfN}, 
\begin{align}
    \big\langle E(n+1) \big\rangle - \big\langle E(n) \big\rangle &= \frac{\alphaP n +\big(\thetain + \thetaout\big) \big\langle E(n) \big\rangle}{n + \alphaP + \betaP - 1} \label{eq:ConstantDegreeAfterExpectation}
\end{align}
using our ansatz \eqref{eq:ConstantDegreeAnsatz},
\begin{align}
     g(n+1) - g(n) + \underbrace{\ConstOne = \Bigg( \alphaP + \big(\thetain + \thetaout\big) \bigg(\ConstOne}_{\text{equate these constant terms}} + \frac{g(n)}{n} \bigg) \Bigg) \timesLOOSE \frac{n}{n + \alphaP + \betaP - 1} \label{eq:ConstantDegreeAfterAnsatz}
\end{align}
and equating the \potentiallyImprove{$\bigTh{1}$ (constant) terms} to solve for $\ConstOne$, 
\begin{align}
    \ConstOne &= \alphaP + \big(\thetain + \thetaout\big) \ConstOne \\
    &= \frac{\alphaP}{1-\big(\thetain + \thetaout\big)} \label{eq:ConstantDegreeSolveForC}
\end{align}
we obtain the asymptotic average degree claimed in \eqref{eq:PPAavgdegreeSumThetaLess1}.  

To check verify our ansatz ${g(n) = \littleO{n}}$, substitute \eqref{eq:ConstantDegreeSolveForC} into \eqref{eq:ConstantDegreeAfterAnsatz}
\begin{align}
    \underbrace{g(n+1) - g(n) = \frac{\big(\thetain + \thetaout\big) g(n)}{n}}_{\text{equate these dominant terms}} - \frac{\alphaP+\betaP-1}{n} + \biggerO{\tfrac{g(n)}{n^2}}
\end{align}
to conclude that ${g(n) = \bigO{n^{\thetain+\thetaout}}}$.  

So our ansatz ${g(n) = \littleO{n}}$ is valid for ${\thetain+\thetaout<1}$, 
which incidentally is precisely when equation~\eqref{eq:PPAavgdegreeSumThetaLess1} is physically meaningful.  


\subsubsection{Logarithmic degree regime: \textrm{\protect$\thetain + \thetaout=1$}}

As ${\thetain + \thetaout \rightarrow 1}$, the average degree predicted by \eqref{eq:PPAavgdegreeSumThetaLess1} diverges, 
and when ${\thetain + \thetaout = 1}$, the average degree is no longer bounded.  
Instead, it grows logarithmically in $n$:
\begin{equation}
    \avgdegree = 2\kern1pt\alphaP\logTIGHT\big(n\big) + C + \littleO{1} \label{eq:PPAavgdegreeSumThetaEqual1}
\end{equation}

Similar to equation~\eqref{eq:ConstantDegreeAnsatz},
we make an ansatz but now of logarithmic average degree 
\begin{align}
    \big\langle E(n) \big\rangle = \ConstOne \timesLOOSE n \log n + g(n) \label{eq:LogDegreeAnsatz}
\end{align}
where ${g(n) = \littleO{n \log n}}$ is subdominant.  

\potentiallyImprove{The change in} ${n \log n}$ is approximately
\begin{align}
\big(n+1\big)\logTIGHT\big(n+1\big) - n\logTIGHT\big(n\big)=\logTIGHT\big(n\big) + 1 + \biggerO{\tfrac{1}{n^2}}.  
\end{align}
Substituting \potentiallyImprove{this} into equation~\eqref{eq:ConstantDegreeAfterExpectation}, 
\begin{align}
     g(n+1) - g(n) + \underbrace{\ConstOne \timesLOOSE \bigg( \logTIGHT\big(n) + 1 \bigg) = \Bigg( \alphaP + \big(\thetain + \thetaout\big) \bigg(\ConstOne\kern1pt\logTIGHT\big(n\big)}_{\text{equate these logarithmic and constant terms}} + \frac{g(n)}{n} \bigg) \Bigg) \timesLOOSE \bigg(1+\biggerO{\tfrac{1}{n}}\bigg)
     \label{eq:LogDegreeAfterAnsatz}
\end{align}
Equating the logarithmic terms, we recover the condition that ${\thetain + \thetaout = 1}$.  
And equating the constant terms, 
\novo{we solve for the constant ${\ConstOne=\alphaP}$.} 

To verify our ansatz ${g(n) = \littleO{n \log n}}$, we consider the {lower-order} terms: 
\begin{align}
    g(n+1) - g(n) = \frac{g(n)}{n} + \biggerO{\tfrac{\log n}{n}}
\end{align}
\potentiallyImprove{The solution can be written} as ${g(n) = C\timesLOOSE n + f(n)}$, where ${f(n) = \littleO{n}}$.  

Note that $C$ is not determined by the asymptotic balance.  
This suggests that different instances of networks generated with the same parameters may limit to different values of $C$.  
This is similar to the situation with the classic P\'olya urn \citep{eggenberger1923statistik}, 
\potentiallyImprove{where any asymptotic ratio is equally stable, resulting in a distribution over these ratios for any fixed set of parameters \citep{mahmoud2008polya, pekoz2019polya}.}

\subsubsection{Sub-linear degree regime: \protect{$1<\thetain +\thetaout<2$}}

When ${1<\thetain + \thetaout < 2}$, the average degree grows as a (sublinear) power of $n$:
\begin{equation}
    \avgdegree \propto C \kern1pt n^{\rho}, \qquad \rho=\thetain +\thetaout-1\label{eq:PPAavgdegreeSumThetaGreater1}
\end{equation}

Again, we make the appropriate ansatz: 
\begin{align}
    \big\langle E(n) \big\rangle = \ConstOne \timesLOOSE n^{1+\rho}_{ } + g(n) \label{eq:SublinearDegreeAnsatz}
\end{align}
where ${g(n) = \littleO{n^{1+\rho}_{ }}}$ is subdominant.  

\potentiallyImprove{The change in ${n^{1+\rho}_{ }}$ is approximately}
\begin{align}
\big(n+1\big)^{1+\rho}_{ } - n^{1+\rho}_{ }=\big(1+\rho\big)n^{\rho}_{ } + \biggerO{n^{\rho-1}_{ }}.  
\end{align}

Substituting this into equation~\eqref{eq:ConstantDegreeAfterExpectation}, 
\begin{align}
     g(n+1) - g(n) + \underbrace{\ConstOne \timesLOOSE \big(1+\rho\big)n^{\rho}_{ } = \Bigg( \alphaP + \big(\thetain + \thetaout\big) \bigg(\ConstOne n^{\rho}_{ }}_{\text{equate these $\bigO{n^{\rho}}$ terms}} + \frac{g(n)}{n} \bigg) \Bigg) \timesLOOSE \bigg(1+\biggerO{\tfrac{1}{n}}\bigg)
     \label{eq:SublinearDegreeAfterAnsatz}
\end{align}
and equating the dominant $\biggerO{\tfrac{1}{n}}$ terms, 
\novo{we solve for the exponent ${\rho=\thetain + \thetaout -1}$.}  
%

Note that this does not fix $\ConstOne$.  
To see why, let us attempt to verify our ansatz ${g(n) = \littleO{n \log n}}$.  
We again equate the remaining {lower-order} terms: 
\begin{align}
    g(n+1) - g(n) = \big(1+\rho\big) \frac{g(n)}{n} + \biggerO{n^{\rho - 1}_{ }}
\end{align}
In this case, it seems as though our ansatz is not verified, with $g(n)$ being the same order as \potentiallyImprove{the ``dominant'' part of the solution:} ${g(n) = \ConstTwo\timesLOOSE n^{1+\rho}_{ } + \littleO{n^{1+\rho}_{ }}}$.  
However, this is in fact not a contradiction --- this is the asymptotic analysis telling us that the original \potentiallyImprove{constant $C$ itself} is not determined.  

Recapitulating the sequence of results:\\[-12pt]
\begin{itemize}
    \item When $\thetain + \thetaout < 1$, the average degree asymptotes to a fixed constant. 
    \item When $\thetain + \thetaout = 1$, the average degree grows logarithmically in $n$, \potentiallyImprove{but} with an arbitrary additive constant.
    \item When $\thetain + \thetaout > 1$, the average degree grows as a sublinear power of $n$, with an arbitrary multiplicative constant.
\end{itemize}
\hfill\qed


\subsection{Proof of Theorem~\ref{thm:PPAPhaseTransitiondegdist} \thmNamePPAAppendixdegdist}
\label{appendix:dapapowerlawproof}

In the {\ppaAcron} model, the asymptotic distribution of node degrees exhibits a power-law tail 
\mbox{$p(d) \propto d^{-\gamma}$}, where the \novo{scaling exponent $\gamma$} depends on either $\thetain$ or $\thetaout$: 
\def\TempKern{\kern30pt}
\begin{align}
    \vphantom{\bigg(}\text{constant:}      &\TempKern\gamma = \frac{1+\thetain}{\thetain}     \kern-2.5cm &0<\thetain+\thetaout\leq1 \\
    \vphantom{\bigg(}\text{polynomial:}    &\TempKern\gamma = \frac{2-\thetaout}{1-\thetaout}        \kern-2.5cm &1\leq\thetain+\thetaout<2 
\end{align}
To show this, we will first characterize the \potentiallyImprove{out-degrees \mbox{$d_j^{\text{out}} = \sum_{i=1}^{j-1} x_{ij}^{ }$} 
(connections that a node $j$ makes with previous nodes).}  
\potentiallyImprove{These out-degrees (and the node arrival time $j$) serve as the initial conditions for the growth of the in-degrees \mbox{$d_j^{\text{in}} = \sum_{k=j+1}^{n} x_{jk}^{ }$} 
(connections that node $j$ makes with later nodes).}

\subsubsection{The out-degrees (connections with previous nodes)}

\potentiallyImprove{The out-degrees do \textit{not} exhibit a power law in this model.}    
\potentiallyImprove{This is because the outcomes of these $x_{ij}^{ }$ (\mbox{$1\leq i < j$}) are conditionally independent given the previous entries $x_{ab}^{ }$ (\mbox{$1\leq a < b < j$}), 
and the sum of independent Bernoulli variables does not exhibit a power-law distribution.}   

Denote these (conditionally) independent probabilities as $p_{ij}^{ }$.  
\potentiallyImprove{In our proof of the sparsity of the model above,} 
we used the fact that the sum of these probabilities \mbox{\smash{$\sum_{i=1}^{j-1}p_{ij}^{ }$}} is the expected change in edges at each step $j$.  
\potentiallyImprove{Moreover, as the network grows, this sum does not change much between each step.}  
Thus, the expected out-degree \mbox{\smash{$d_j^{\text{out}} = \sum_{i=1}^{j-1} x_{ij}^{ }$}} of node $j$ is approximately 
\vspace{-12pt}
\def\TempKern{\kern30pt}
\begin{align}
    \vphantom{\Bigg(}\text{constant:}      &\TempKern\big\langle d_j^{\text{out}} \big\rangle \sim \displaystyle\frac{\alphaP}{1-\big(\thetain+\thetaout\big)}     \kern-2.5cm &0<\thetain+\thetaout<1 \label{eq:dOutConstant}\\
    \vphantom{\Bigg(}\text{logarithmic:}   &\TempKern\big\langle d_j^{\text{out}} \big\rangle \sim \alphaP \log\kern-1pt \big(j\big) + C              \kern-2.5cm &\thetain+\thetaout=1 \label{eq:dOutLogarithmic}\\
    \vphantom{\Bigg(}\text{polynomial:}    &\TempKern\big\langle d_j^{\text{out}} \big\rangle \sim C \times j^{\thetain+\thetaout-1}        \kern-2.5cm &1<\thetain+\thetaout<2 \label{eq:dOutPolynomial}
\end{align}
with a variance upper-bounded by this average \potentiallyImprove{(as they are the sum of independent Bernoulli random variables).}  

\subsubsection{The in-degrees (connections with later nodes)}

For a node $j$ with a given out-degree $d_j^{\text{out}}$, 
the expected in-degree grows according to the difference equation: 
\begin{align}
    \big\langle d_j^{\text{in}}\big\rangle(n+1) - \big\langle d_j^{\text{in}}\big\rangle(n) = \frac{\alphaP + \thetaout d_j^{\text{out}} + \thetain \big\langle d_j^{\text{in}}\big\rangle(n)}{n + \alphaP + \betaP - 1}  \qquad \big\langle d_j^{\text{in}}\big\rangle(j) = 0
\end{align}
 \potentiallyImprove{As this evolution does not depend on the outcome of any other edges in the network,} 
we can write the solution in closed form:
\begin{align}
    \big\langle d_j^{\text{in}}\big\rangle(n) = \frac{\alphaP + \thetaout d_j^{\text{out}}}{\thetain} \Bigg( \frac{\Gamma\big( \alphaP + \betaP + j - 1 \big)}{\Gamma\big( \alphaP + \betaP + \thetain + j - 1 \big)} \frac{\Gamma\big( \alphaP + \betaP + \thetain + n - 1 \big)}{\Gamma\big( \alphaP + \betaP +  n - 1 \big)} - 1\Bigg)
\end{align}
\novo{where is the gamma function.}

For \mbox{$1\ll j < n$}, 
\potentiallyImprove{the ratios of gamma functions can be approximated, 
and the expected degree as a function of $j$ and $n$ is} 
\begin{align}
    \big\langle d_j^{\text{ }}\big\rangle(n) \approx \frac{\alphaP + \thetaout \big\langle d_j^{\text{out}}\big\rangle}{\thetain} \Bigg( \bigg(\frac{n}{j}\bigg)^{\kern-2pt\thetain} - 1\Bigg) + \big\langle d_j^{\text{out}}\big\rangle \label{eq:diTotalAsAFunctionOfN}
\end{align}

\potentiallyImprove{We can extract the power law of the degree distribution from the dependence of the expected degree on the node index $j$.  
Since $\big\langle d_j^{\text{ }}\big\rangle$ is monotonically decreasing in $j$, 
the probability density will be proportional to the reciprocal of the magnitude of the derivative with respect to $j$:}
\begin{align}
    p\big(d\big) \propto \Big| \tfrac{d}{dj} \big\langle d_j^{\text{ }}\big\rangle \Big|^{-1} 
\end{align}
For a power-law degree distribution \mbox{$p\big(d\big) \propto d^{-\gamma}$}, 
the exponent is the change in \mbox{$\log\kern-1pt\big(p(d)\big)$} with respect to \mbox{$\log\kern-2pt\big(d\big)$}:
\begin{align}
    \gamma = \frac{\tfrac{d}{dj}\log\kern-1pt\Big| \tfrac{d}{dj} \big\langle d_j^{\text{ }}\big\rangle \Big| }{\tfrac{d}{dj} \log\kern-1pt\big\langle d_j^{\text{ }}\big\rangle} = \frac{\big\langle d_j^{\text{ }}\big\rangle \tfrac{d^2}{dj^2}\big\langle d_j^{\text{ }}\big\rangle}{\big(\tfrac{d}{dj}\big\langle d_j^{\text{ }}\big\rangle\big)^{\kern-1pt 2}} \label{eq:ScalingExponentDerivatives}
\end{align}
Substitute the expressions for \mbox{$\big\langle d_j^{\text{out}}\big\rangle$} from equations~\eqref{eq:dOutConstant} and \eqref{eq:dOutPolynomial} into equation~\eqref{eq:diTotalAsAFunctionOfN}.  
\potentiallyImprove{When \mbox{$1\ll j \ll n$},} the dominant term has the following scalings:
\begin{align}
    \vphantom{\bigg(}\text{constant:}      &\TempKern\big\langle d_j^{\text{ }}\big\rangle(n) \propto n^{\thetain}_{ } j^{-\thetain}_{ }     \kern-2.5cm &0<\thetain+\thetaout<1 \\
    \vphantom{\bigg(}\text{polynomial:}    &\TempKern\big\langle d_j^{\text{ }}\big\rangle(n) \propto n^{\thetain}_{ } j^{\thetaout - 1}_{ }        \kern-2.5cm &1<\thetain+\thetaout<2 
\end{align}
Substituting these into equation~\eqref{eq:ScalingExponentDerivatives}, we obtain the following power-law exponents: 
\begin{align}
    \vphantom{\bigg(}\text{constant:}      &\TempKern\gamma = \frac{1+\thetain}{\thetain}     \kern-2.5cm &0<\thetain+\thetaout\leq1 \\
    \vphantom{\bigg(}\text{polynomial:}    &\TempKern\gamma = \frac{2-\thetaout}{1-\thetaout}        \kern-2.5cm &1\leq\thetain+\thetaout<2 
\end{align}
Note that both expressions give the same scaling when \mbox{$\thetain+\thetaout=1$} for the intermediate logarithmic regime.  \\
\hfill\qed


\end{document}